 \documentclass[11pt]{article}

 \pdfoutput=1   

\usepackage{graphicx}
\usepackage{amsmath}

\usepackage{enumitem}

\usepackage[francais,english]{babel}

\usepackage{float}
\usepackage{ifthen}
\usepackage{color}

\usepackage{hyperref}

\begin{document}





\title{{\bf The available-enthalpy (flow-exergy) cycle. \\
    Part-II: applications to idealized baroclinic waves.}}

\author{by Pascal Marquet. {\it CNRM/GMAP. M\'et\'eo-France, Toulouse, France.} \\
 Email: pascal.marquet@meteo.fr}

\date{\today}

\maketitle


\vspace*{-10mm}

\begin{center}
{\em Copy of a CNRM-Note submitted in two parts in April 2001 to the
 \underline{Quarterly Journal of the Royal Meteorological Society}.} \\
{\em Published in Vol.129, Issue 593, Part-I (2445--2466) Part-II (2467--2494), July 2003, Part B.} \\
  Part-I: \url{http://onlinelibrary.wiley.com/doi/10.1256/qj.01.62/abstract} \\
  Part-II: \url{http://onlinelibrary.wiley.com/doi/10.1256/qj.01.63/abstract} \\
Comments and corrections are added in footnotes. 
\end{center}
\vspace{1mm}



\vspace*{-2mm}

\begin{abstract}
The local available-enthalpy cycle proposed in Part~I 
of this paper is applied to document energetics 
of three numerical simulations, representing life cycles of idealized 
baroclinic waves. 
An improved temporal numerical scheme 
defined in Part~I is used in this study, together with 
the Arpege-IFS model using a T42 triangular truncation. 
A  $45\,{}^{\circ}$N and $200$~hPa dry unstable jet is constructed with 
the most unstable mode at zonal wave number 8. Energetic 
impacts of both horizontal and vertical diffusion
schemes are determined separately. \\
\hspace*{4mm} The role of ageostrophic winds within the Ekman layer 
is investigated, leading to an explanation for 
large observed values for the dissipation terms
and to a new formulation of the potential-energy conversions. 
The magnitudes of these new conversion terms are 
compared with those of the usual barotropic and
baroclinic conversions. 
A new version for the available-enthalpy 
cycle is proposed. It is suitable for open 
systems and it includes explicitly the potential-energy
component as a transitional reservoir.
Finally, some results from Intensive Observing Period 15 
of the Fronts and Atlantic Storm-Track EXperiment 
(FASTEX) are 
compared with those from the idealized diabatic experiment.
\end{abstract}

 \section{\Large \underline{Introduction}.} 
 \label{section_1}

In the first part of this paper (Marquet,
2003, hereafter referred to as Part~I),
a local and exact available-enthalpy cycle
 has been proposed.
It is designed to clear up the difficulties 
encountered with previous limited-area
applications and, on the global stage, it
must lead to the generally accepted
results, including 
conventional baroclinic and barotropic 
instabilities.

The main local results of Part~I are briefly recalled in 
section 2 and other global results are derived. 
Adiabatic and diabatic
simulations of a life cycle of idealized baroclinic 
waves are described in section 3. The properties 
of the new cycle are examined in section 4,
based on global and local applications.
The temporal scheme defined in Part~I 
is used with a time interval of 3 hours. 
It is explained
how Ekman dissipation and ageostrophic
circulations can account for observed large values 
for dissipation and conversion terms with potential 
energy in the boundary layer. A new version for the 
available-enthalpy cycle, suitable for open 
systems, is proposed in section 5. It includes 
the potential-energy component as a transitional 
reservoir in the Lorenz cycle, located  between 
available-enthalpy and kinetic-energy reservoirs.
Some results from Intensive Observing Period (IOP15) 
 of Fronts and Atlantic Storm-Track EXperiment (FASTEX) 
are compared with the diabatic idealized experiment
in section 6, with final conclusions presented 
in section 7.

 \section{\Large \underline{The limited area available enthalpy cycle}.} 
 \label{section_2}

It is explained in Part~I that 
the budget equations for the available enthalpy 
are derived 
by computing the time derivation of the six components
$a_S$, $a_Z$, $a_E$, $k_S$, $k_Z$ and $k_E$
(all symbols are defined in Appendix-A of Part~I). 
As a result,
the available-enthalpy cycle is given 
by (36) and by Fig.~5(b) in Part~I.

The global version of (36) in Part~I is given by
(\ref{eq:cycle3a})-(\ref{eq:cycle3c}). These equations for
$A_h$, $K$ and the total energy TOT $\:= A_h+K$ are 
obtained when the whole limited-area domain is considered and 
by integration from the top pressure $p_t$ 
to the bottom value $p_b$.
\begin{eqnarray}
    {\partial}_t ( A_h )  \! \!  & = &  \! \! 
            \; - \; B(A_h)                \mbox{\hspace{0.10 cm}}
            \; - \; ( C_S + C_Z + C_E )
            \; - \; B(A_p)                \mbox{\hspace{3mm}}
            \; + \; G \; ,
    \label{eq:cycle3a}  \\
    {\partial}_t ( K )    \!  & = &  \! \!
            \; - \; B(K)                  \mbox{\hspace{0.20 cm}}
            \; + \; ( C_S + C_Z + C_E ) \mbox{\hspace{0.05 cm}}
            \; - \; B({\phi})             \mbox{\hspace{5mm}}
            \; - \; D \; ,
    \label{eq:cycle3b} \\
    {\partial}_t ( TOT )    \!  & = &  \! \!
            \; - \; [\: B(A_h) + B(K) + B(A_p) +  B({\phi}) \:]
            \; + \; G  \mbox{\hspace{0mm}}
            \; - \; D \; .
    \label{eq:cycle3c}
\end{eqnarray}

Boundary terms $B(A_h)$, $B(K)$, $B(A_p)$ and $B({\phi})$ all vanish 
for $z_b=0$, $p_b=$constant, $p_t=0$ and for a domain surrounding the whole 
Earth. These assumptions have been retained in the study of Lorenz (1955,
hereafter L55)
where the total energy is a constant for a pure adiabatic model
if $G-D$ is equal to zero. However $G$ and $D$ are
generally non-zero in numerical simulations because 
the total energy is not rigorously conserved. Indeed,
horizontal spectral schemes are applied on a truncated spectral 
space, leading to approximations and errors. There are other errors
due to time or vertical differencing schemes and
interpolation methods.
The validation of adiabatic and frictionless versions
of the cycle (36) of Part~I will be obtained 
by taking $G=D$ and if the residuals are both 
small in comparison with the physical tendencies generated 
by diabatic processes (i.e. $|G|<\epsilon$ and 
$|D|<\epsilon$).

 \section{\Large \underline{Numerical simulations of baroclinic waves}.}
 \label{section_3}

The concepts of local energetics will be 
illustrated by a study of available-enthalpy diagnostics 
for life-cycle experiments of idealized baroclinic waves.
Energy computations every $3$~h will be undertaken for three 
different numerical simulations with the same
basic state used as a common starting point.
This basic state is obtained by suppressing
orography and by setting the humidity to a very small 
value (but non zero to avoid numerical problems in 
the physics package). The constant surface pressure is equal 
to $1013.25$~hPa. The French 
Arpege\footnote{\color{blue} 
The name ``Arpege'' means ``Action de Recherche Petite 
Echelle et Grande Echelle'' in French (this means ``Large-scale
and Small-scale Research Program'').
It is the French counterpart of the ECMWF-IFS model. 
It share the same dynamic core since 1988, but with a specific 
physics package. Arpege is often used with a variable  
grid mesh, based on the stretched and titled pole
grid option.}
numerical model (Courtier {\it et al.}, 1991) is used with a 
triangular truncation at total wave number $42$, with $31$ 
irregularly spaced hybrid levels and an Eulerian semi-implicit
leapfrog time step of $900$~s. 

The first adiabatic experiment (EXP-A) is subject only to
weak numerical dissipation or generation created by
the Asselin filter, truncation errors or
approximate time or vertical differencing schemes.
There are no explicit horizontal 
and vertical diffusion schemes. This 
adiabatic and frictionless simulation will serve as a 
validation for the new energetic analysis, leading to
a quasi-conservative total energy and giving 
expected small values for dissipation and generation
terms expressed as residuals (including the 
numerical errors). Nevertheless, the last days 
of this simulation are somewhat unrealistic due to
an accumulation of small-scale noisy features 
associated with high total wavenumber. As a consequence 
only the growing stage will be considered for (EXP-A),
namely for the first $10$~days.

The second experiment (EXP-H) is also adiabatic 
and frictionless but with an additional
${\nabla}^6$ horizontal diffusion 
scheme to avoid the accumulation of energy
in the high wave number. The stage of decay of
the baroclinic wave is more realistic than
for EXP-A. The e-folding time of the  horizontal 
diffusion scheme  is $12$~h for the vorticity 
and temperature, $4$~h for the divergence.
\begin{table}[htb]
\caption{\it \small 
The mixing length $L$ for EXP-HV.
Values are given for height $0$ to
$4000$~m. There is a ``$L=0.4 z$'' law
close to the surface, with a maximum
values $L \approx 45$~m at $z = 700$~m, 
with reduction towards the asymptotic value
$L=9$~m in the higher troposphere and 
above.
\label{TabMIXLENGTH}}
\vspace*{2mm}
\centerline
{\begin{tabular}{|c|ccccccccccccc|}
\hline 
 \hline 
Height & 0 & 5 & 50 & 100 &  250 &  500 &  750 & 1000 & 1500 & 2000 & 3000 & 4000 &\\ 
\hline 
 L-mix  & 0 & 2 & 15 &  24 &  37  &  44  &  45  &   43 &   38 &   32 &   24 &   19 &\\ 
\hline 
\hline 
\end{tabular}}
\end{table}

The third diabatic experiment (EXP-HV) 
includes the same ${\nabla}^6$ horizontal diffusion
as for EXP-H, but with a vertical diffusion scheme 
added. As a result, ageostrophic circulations appear 
in the planetary boundary layer and a
crude representation of friction is obtained
inducing large dissipation and generation terms.
The vertical diffusion scheme 
is based on a local exchange coefficient method described
in Louis (1979) and Louis {\it et al.} (1981), with a 
uniform roughness length of $1$~mm and a prescribed 
vertical profile for the mixing length depicted
in Table~\ref{TabMIXLENGTH}.

      \subsection{The dry basic state.} 
      \label{subsection_3.1}

\begin{figure}[t]
\centering
\includegraphics[width=0.54\linewidth,angle=0,clip=true]{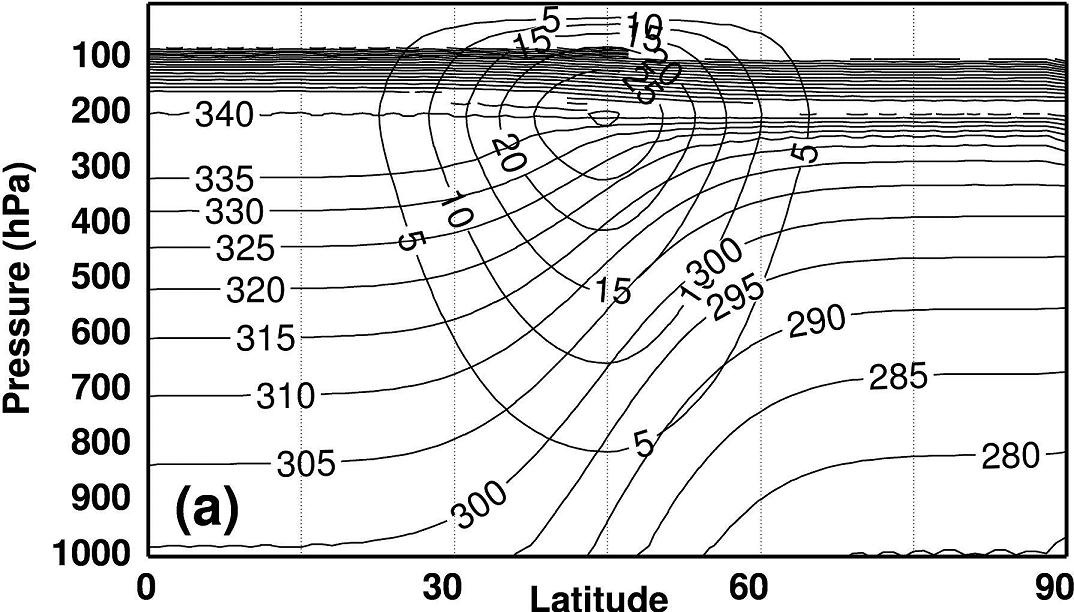}
\includegraphics[width=0.45\linewidth,angle=0,clip=true]{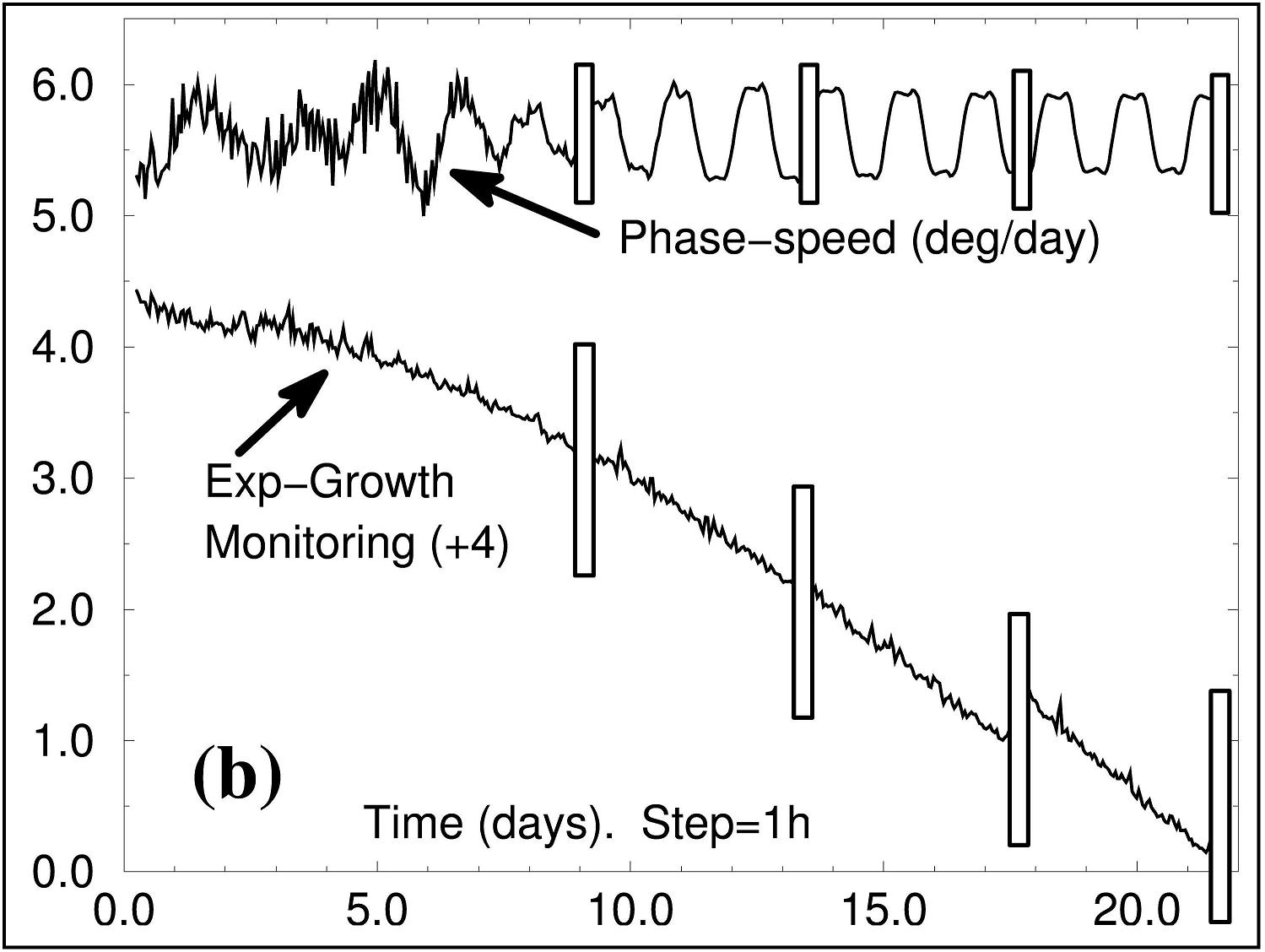}
\caption{\it \small 
(a): latitude-height section for the basic state
showing potential temperature with an interval of $5$~K
and zonal velocity interval of $5$~m~s${}^{-1}$.
(b): the diagnostics of Simmons and Hoskins (1976)
for the $22$~days simulation of the selection 
of the most unstable mode at zonal wave number $8$.
The phase speed is depicted in the upper part in degrees/day
and the relative change in growth rate from one hour to 
the next is depicted on the lower part (going from $0$ to
$-4$, with $+4$ unit added for ease of plotting).
There are four interruptions at about $9.2$,
$13.5$, $17.7$ and $21.5$~days, located at times when the 
perturbation amplitude of the mode is reduced when it became 
too large (see Thorncroft and Hoskins, 1990). The improved
definition of (\ref{AppB:defJETug}) in the Appendix has
no influence on the oscillations observed for the phase speed
after day $5$ of the simulation. An explanation is still to be
discovered.
\label{FigUTZON}}
\end{figure}

The dry basic zonal flow is constructed following 
the approach chosen by C. Freydier when the variable 
resolution version of Arpege-IFS model (Courtier {\it 
et al.}, 1991) was validated. This has
been performed by taking an analytic formulation for 
the zonally symmetric temperature $T(\varphi, p)$. 
The geopotential $\phi(\varphi, p)$ is then computed 
for each pressure level by integrating the hydrostatic 
equation with $\phi =0$ at the surface. The balanced
zonal flow $u(\varphi, p)$ is finally obtained by setting
the zonal and meridional wind tendencies to zero.
The meridional wind is also set to zero
and all the variables ($T,\phi,u$) are zonally 
symmetric (see the Appendix). The result for 
the basic state is depicted on Fig.~\ref{FigUTZON}(a)
where a baroclinic zone is centred at $45\,{}^{\circ}$N, with a 
stationary jet of $30$~m~s$^{-1}$ at $200$~hPa. 

Three simulations EXP-A, EXP-H and EXP-HV are made 
in order to investigate the life cycle of baroclinic waves.
The French Arpege model is used with the same initial 
state for the three simulations. It is a
zonal basic flow with a superimposed most 
unstable normal mode at zonal wavenumber $8$,
determined by use of the method of
Thorncroft and Hoskins (1990, hereafter TH90).
A small random initial perturbation is added to the 
surface pressure of the basic flow. The nonlinear growth 
of the perturbation amplitude is followed during the 
integration of the model until it is exponential to within a 
specified accuracy. The relative changes in growth rate and the 
phase speed are computed every hour according to Simmons and 
Hoskins (1976). The results are shown in Fig.~\ref{FigUTZON}(b)
where the relative change in growth rate from one hour to 
another reaches $10^{-4}$ at day $22$ and where the phase 
speed is close to $5.6$~degrees~(day)${}^{-1}$ after averaging 
time oscillations. Note that, according to TH90, the perturbation
amplitudes are reduced together for $p_s$, $u$, $v$ and $T$ 
by a common factor when they become too large.
At the end of the simulation, the mode is normalized
so that the surface-pressure perturbation is $1$~hPa,
with a common global damping for $p_s$, $u$, $v$ and $T$.

       \subsection{Baroclinic developments} 
      \label{subsection_3.2}

\begin{figure}[t]
\centering
\includegraphics[width=0.49\linewidth,angle=0,clip=true]{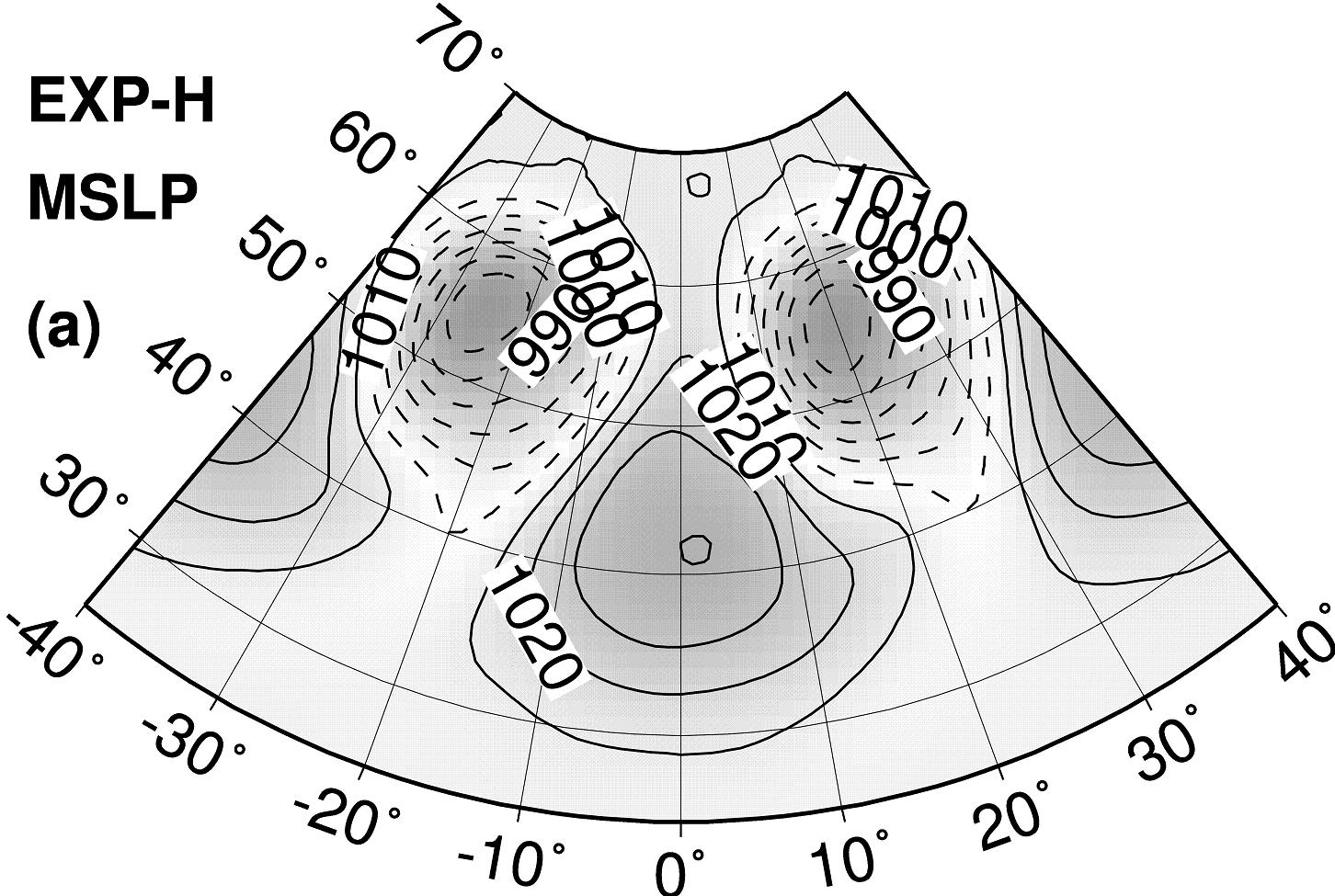}
\includegraphics[width=0.49\linewidth,angle=0,clip=true]{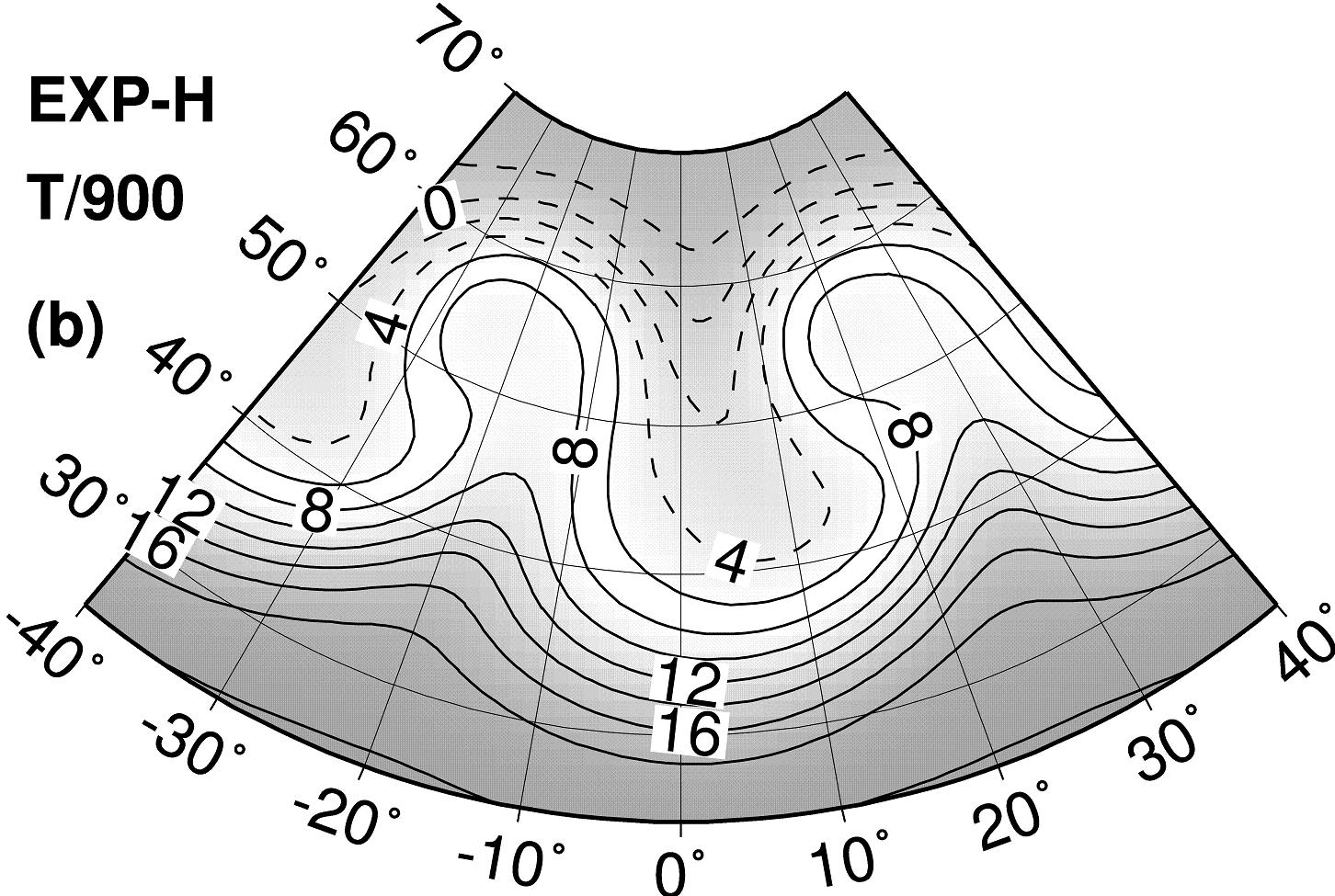}\\
\includegraphics[width=0.49\linewidth,angle=0,clip=true]{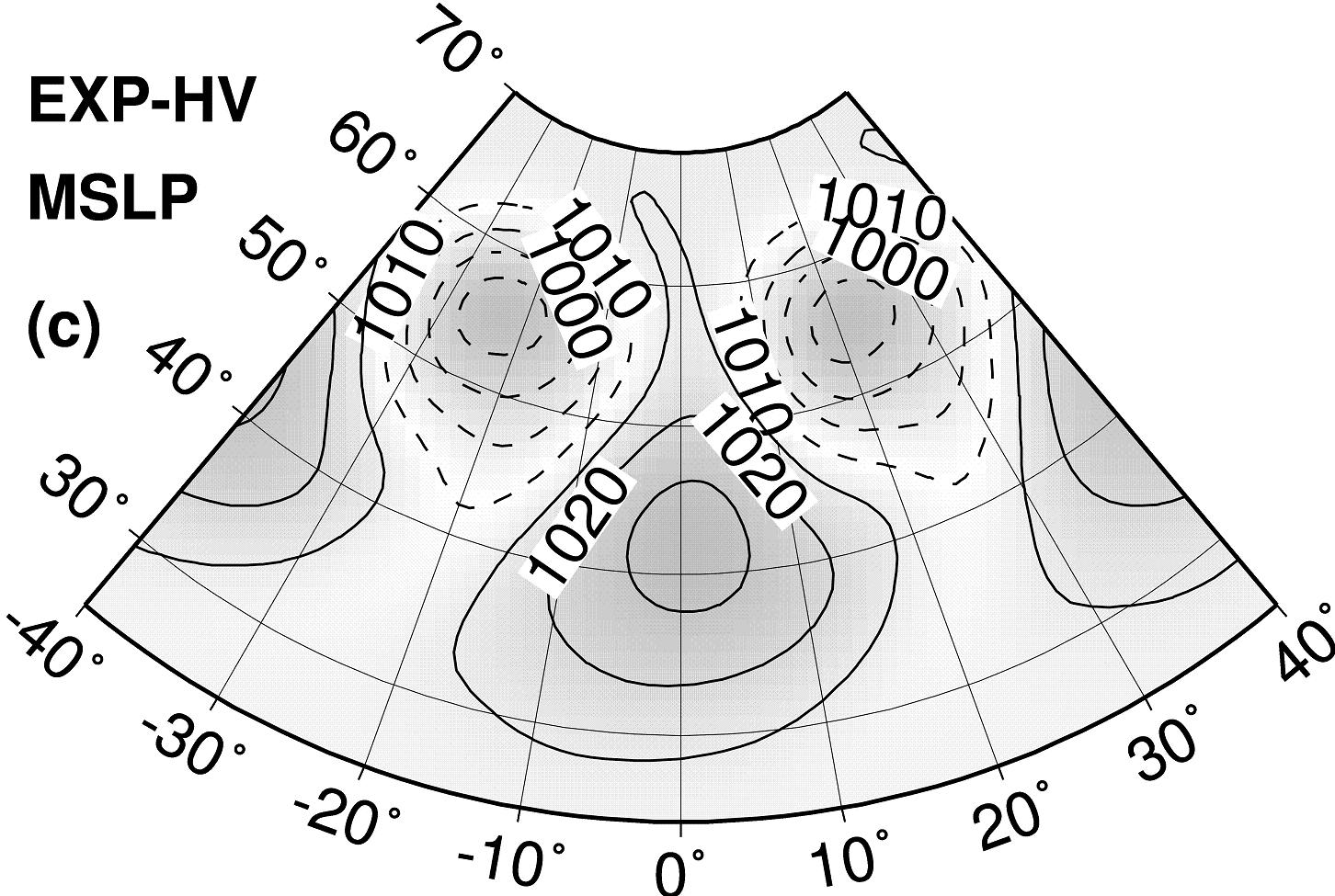}
\includegraphics[width=0.49\linewidth,angle=0,clip=true]{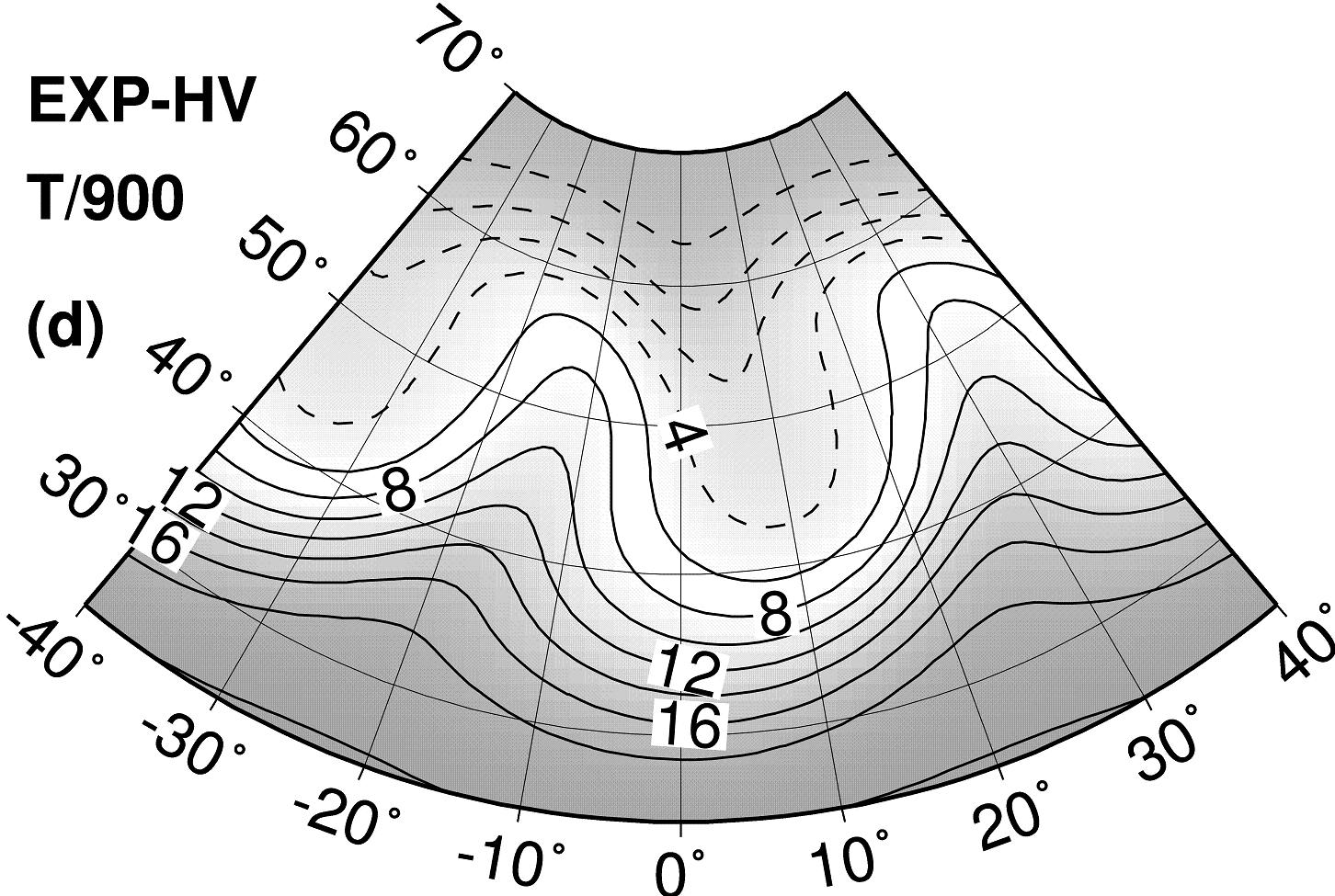}
\vspace*{-2mm}
\caption{\it \small 
Horizontal sections at day $7$ in the life 
cycle for the second and third experiments:
 (a) and (b) for EXP-H and (c) and (d) for
 EXP-HV. 
Sections shown are for:
 (a) and (c) mean sea level pressure at interval of $5$~hPa;
 (b) and (d) temperature at $900$~hPa at interval of $2$~K.
\label{FigPMERT900}}
\end{figure}

The surface pressure and temperature at $900$~hPa 
at day $7$ are shown in Figs.~\ref{FigPMERT900}  
(a) and (b) for the adiabatic simulation EXP-H. The 
same fields are presented in Figs.~\ref{FigPMERT900} (c) and (d) for the 
diabatic simulation EXP-HV.

For EXP-H, the pressure trough and cold front are in good 
agreement with the equivalent results depicted in Fig.~1 of 
Simmons and Hoskins (1978) when they studied a similar T42
adiabatic simulation with a wave-number-6 perturbation of a 
$45\,{}^{\circ}$N jet. 
At day $7$, the surface pressure lows are about $982$~hPa 
at $55$~N, with high pressure reaching $1031$~hPa at $42\,{}^{\circ}$N.
The temperature planetary waves at $900$~hPa depicted in (b) 
are associated with moderate gradients and, as in Fig.~1 
of Simmons and Hoskins (1978) or in Fig.~6 of Hoskins and 
Simmons (1975), the surface troughs and ridges in 
Fig.~\ref{FigPMERT900}~(a) are in
phase for all latitudes from $25\,{}^{\circ}$N to 
$70\,{}^{\circ}$N with the axes of 
warmest and coldest air in Fig.~\ref{FigPMERT900}~(b), 
respectively. 

The surface pressure field for the experiment EXP-HV 
in Fig.~\ref{FigPMERT900}~(c) is almost identical with the corresponding 
adiabatic version in Fig.~\ref{FigPMERT900}~(a), except there is
a global smoothing of the field when the vertical diffusion 
is activated. The pressure lows increase from $982$ to 
$993$~hPa, with an overall gradient decrease. 

The difference in the temperature field at $900$~hPa between
EXP-H and EXP-HV is more important in the range of
latitude from $45\,{}^{\circ}$N to $60\,{}^{\circ}$N. 
There is an Eastward
shift of about $10\,{}^{\circ}$ of the warm axes and of the 
occluded part of the front.
The fields are more profoundly modified closer to the 
surface (not shown) where the vertical diffusion scheme
is acting.

 \section{\Large \underline{Energetic analysis of the baroclinic waves}.}
 \label{section_4}

      \subsection{Practical computations of the available 
                  enthalpy cycle.} 
      \label{subsection_4.1}

An output dataset from Arpege is used
for the $27$ post-processed data on pressure levels,
at intervals of $25$~hPa from $1000$ to
$800$~hPa, $50$~hPa from $800$ to $100$~hPa and with the 
four upper levels at $70, 50, 30$ and $10$~hPa. 
Time derivatives and other terms of cycle 
(36) of Part~I are evaluated with meteorological data
available every $3$~h. 

The post-processed fields are 
($u$, $v$, $T$, $\omega$) for all pressure levels and
on the same $1\,{}^{\circ} \times 1\,{}^{\circ}$ 
latitude$\: \times \:$longitude grid. Vertical and
horizontal differencing schemes used in the diagnostic 
package are second-order centred grid point schemes, 
different from those used in Arpege where $\omega$ values are 
located on half-levels and where horizontal derivatives are
computed in spectral space. The reason why diagnostics
have not been computed directly in Arpege with the
same numerical schemes is to allow future studies of different
models with ``pressure levels'' and
``latitude$\times$longitude'' grid structures as
a common starting point for the same diagnostic package. 

The modified time scheme described in Part~I will be
used to solve the generic equations of (36) in
Part~I, denoted by ${\partial}_t (Z) = C$.
The notations $ Z^{(+)}$, $ Z^{(-)}$ 
and $ C^{(0)}$ are used for values of $Z$ and $C$ at time 
$ t_0 + \Delta t$, $ t_0 + \Delta t$ and $t_0$,
respectively. The scheme\footnote{\color{blue} 
This scheme is expressed differently in Part~I: 
$\; C^{(0)} \: + \: [\: C^{(+)} - 2 \, C^{(0)} + C^{(-)} \:]/6$.
The two formulations are clearly equivalent.}
 takes
\vspace{-0.15cm}
\begin{eqnarray}
   \frac{ Z^{(+)} \: - \: Z^{(-)} }
   {2 \; \Delta t}
& \approx &
   \frac{  \:  C^{(+)} \: + \:  4 \, C^{(0)} \: + \:  C^{(-)} \:
   } {6}
\; .  \label{eq:bilanINT}
\end{eqnarray}

For the present ideal case study, the scale-length
is equal to the half-wavelength of the normal mode,
i.e. $360 / 16 = 22.5\,{}^{\circ}$. The advection is obtained from 
day-to-day observations of the mode and the value is
different from the theoretical value of
$5.6\,{}^{\circ}$/day, as shown in 
Fig.~{\ref{FigUTZON}}~(b). The real value is
about $7\,{}^{\circ}$/day. For a time interval 
of $3$~h, the critical time interval is equal 
to  $25/7 \approx 3.2$~{days} and the value of $3$~h 
is thus far below the corresponding limit of 
$0.8\times3.2=2.6$~{days} required for an accurate application 
of (\ref{eq:bilanINT}) with an error lower than 
$0.2$~\% (see Table~1 in Part~I).

      \subsection{Global results (adiabatic and diabatic 
                  simulations).} 
      \label{subsection_4.2}

\begin{figure}[t]
\centering
\includegraphics[width=0.49\linewidth,angle=0,clip=true]{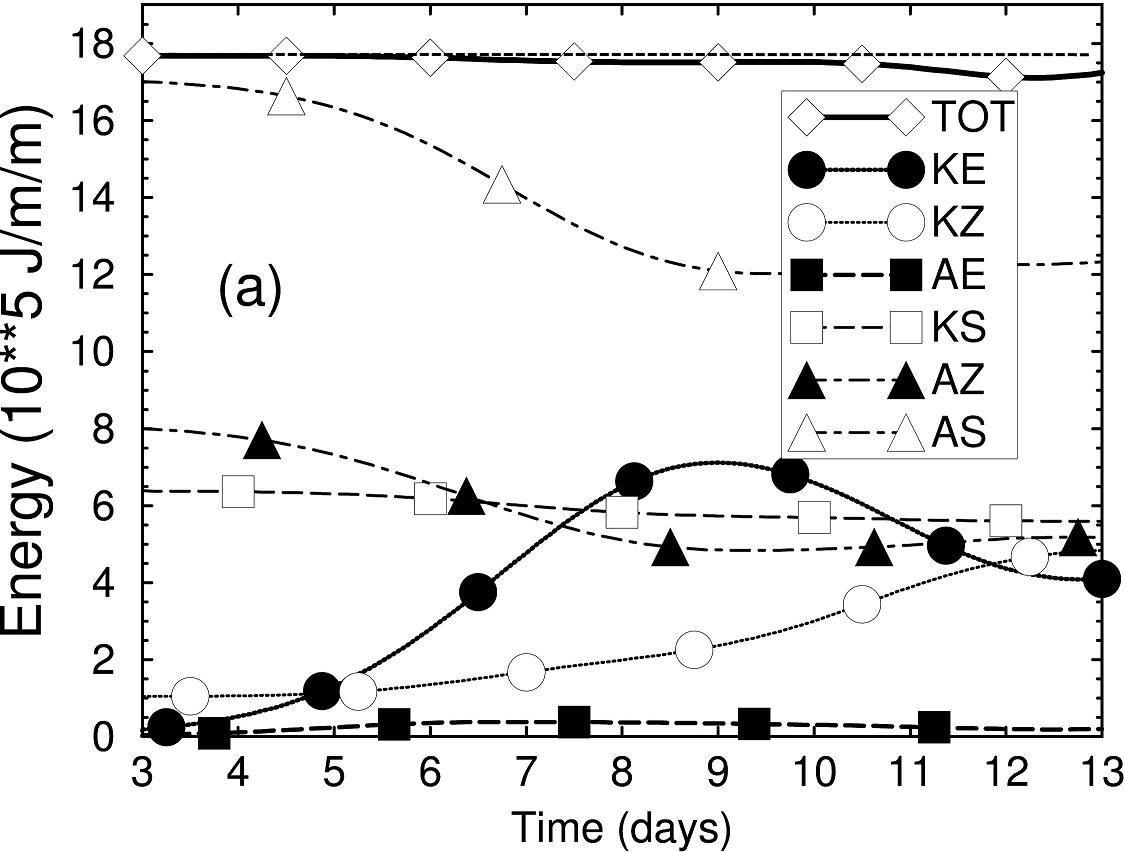}
\includegraphics[width=0.49\linewidth,angle=0,clip=true]{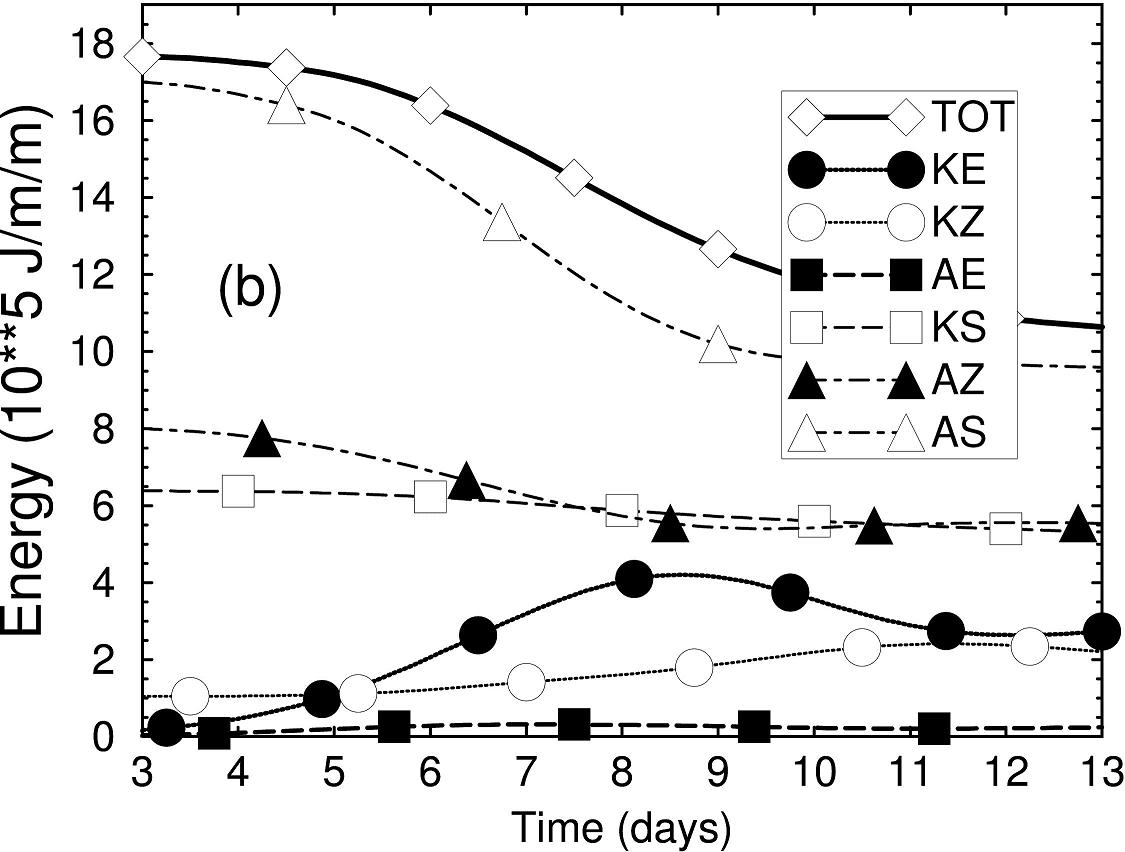}\\
\includegraphics[width=0.49\linewidth,angle=0,clip=true]{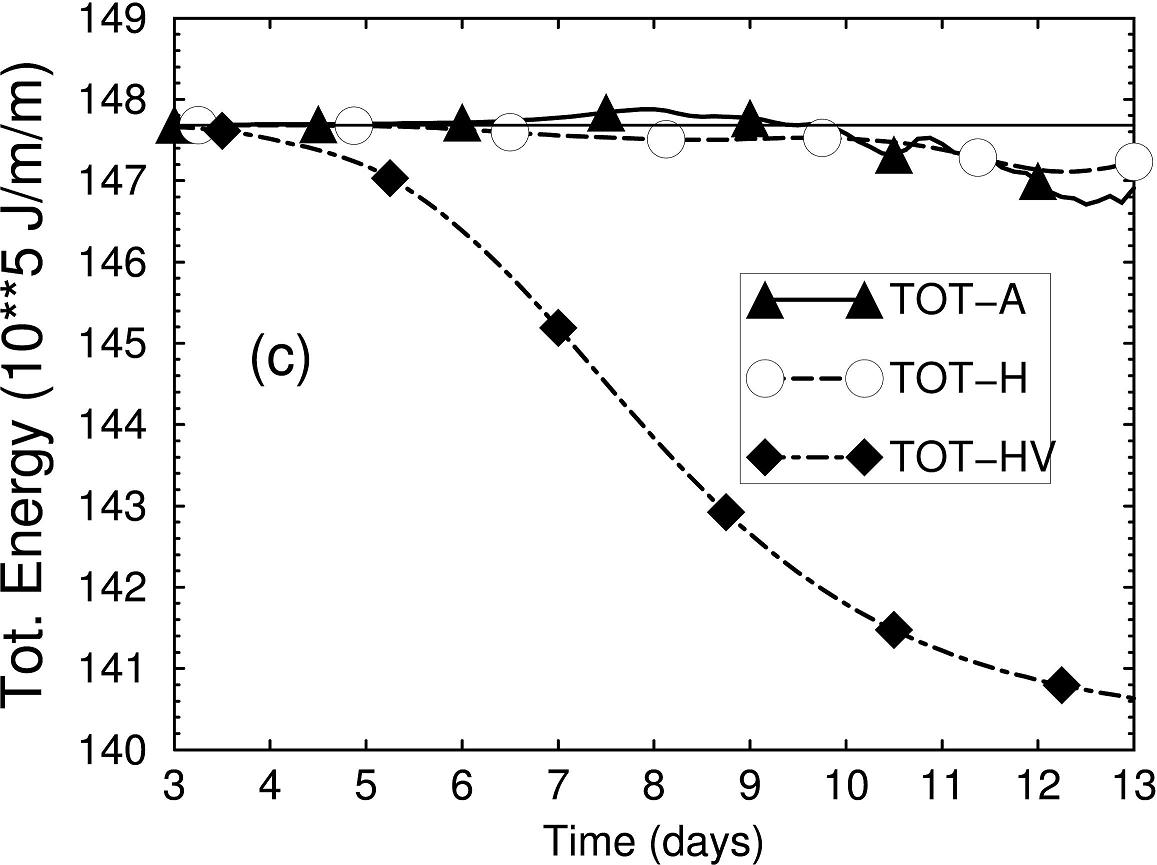}
\includegraphics[width=0.49\linewidth,angle=0,clip=true]{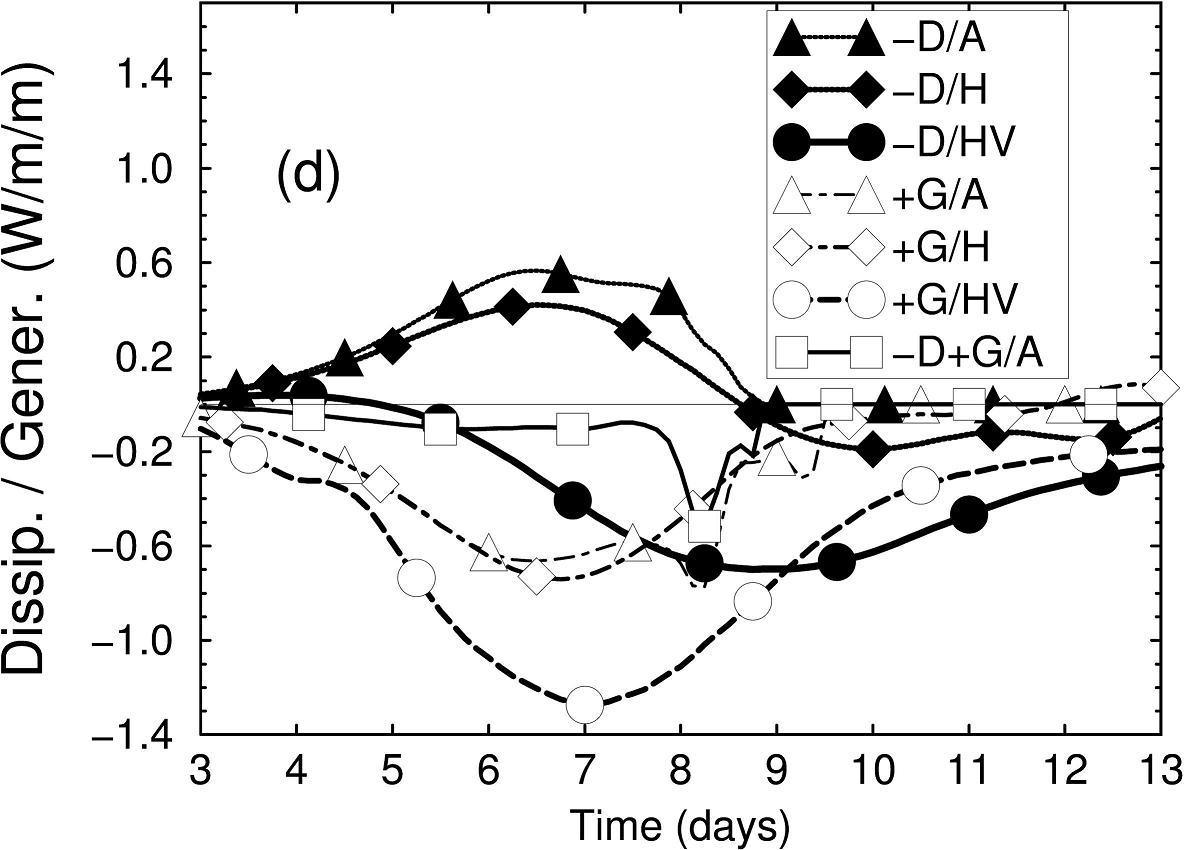}
\vspace*{-2mm}
\caption{\it \small 
Evolution of global energy components of the basic 
life cycle, from day $3$ to day $13$.
(a) Energy components for the simulation EXP-H with the
    weak ${\nabla}^6$ horizontal diffusion. Energy are in
   units of 
    $10^{5}$~J~m${}^{-2}$. The stability component $A_S$
    and the total energy TOT$\:=K_E+A_E+K_Z+A_Z+K_S+A_S$
    have had $115$ and $130$ units subtracted for ease 
    of plotting, respectively.
(b) As (a) but for the diabatic simulation EXP-HV with an
     additional vertical diffusion.
(c) Evolution of total energy for
     EXP-A, EXP-H and EXP-HV. Changes in energy 
     are in units of W~m${}^{-2}$.
(d) Evolution of the total dissipation term $-D=-D_E-D_Z-D_S$
    and the total generation term $G=G_E+G_Z+G_S$ for the same
    three simulations as in (c). The sum $-D+G$ is depicted
    with open square symbol only for the adiabatic simulation
    EXP-A. Dissipation and generation of energy are in
    units of  W~m${}^{-2}$.
   See text for explanation.
\label{FigEXP182021}}
\end{figure}

Evolutions of global energy components of the basic 
life cycle are presented in Fig.~{\ref{FigEXP182021}}
for days $3$ to $13$ of the simulations EXP-A, EXP-H and 
EXP-HV. The large computational domain extends vertically from 
$1000$ to $10$~hPa and the horizontal limits are
$25\,{}^{\circ}$N and $65\,{}^{\circ}$N, $0\,{}^{\circ}$E 
and $180\,{}^{\circ}$E. Although it
is not a real global domain (say $90\,{}^{\circ}$S 
to $90\,{}^{\circ}$N and
$180\,{}^{\circ}$W to $180\,{}^{\circ}$E), 
it is clear from
Figs.~\ref{FigPMERT900}(a)--(d) that fields are 
nearly zonally symmetric to north and south
of the computational domain and periodic in longitude 
(there is an even number of planetary waves $m=8$). 
Unused parts of this global domain  
would not have contributed significantly to energetics
of the simulations and it is expected that, for this large computational 
domain, the baroclinic or barotropic signals will 
be enhanced  and be easier to analyse than for
the real global domain.

Global energy components for the simulation EXP-H 
are presented on Fig.~{\ref{FigEXP182021}}~(a).
As expected the total energy TOT is a constant up to day 
$10$ for this adiabatic case. 
The rapid increase
of $KE$ is maximum at day $6.5$ and $KE$ reaches its maximum 
value at day $9$ ($+ 7.$~$10^{5}$~J~m${}^{-2}$). 
There is a slow and less important increase in 
$KZ$ in the growing stage of the mode, followed by a more 
rapid increase between days $9.5$ and $12$ when $KE$ 
start to decrease and when the mode decays. 
The initial increase in $KE$ and $KZ$ up to day $9$ 
is thus obtained at the expense of $AS$ and $AZ$, 
via possible baroclinic processes to be confirmed later 
on. The rapid increase of $KZ$ in the depletion stage
is realized at the expense of $KE$, because only $KZ$
and $KE$ vary in time after day $9$. This transfer
of energy from $KE$ to $KZ$ is usually considered as a
barotropic conversion. As for the changes in the global 
components $AE$ and $KS$, they are less than 
$\pm 1.$~$10^{5}$~J~m${}^{-2}$ and do not give
contributions to the global energetics.
All these global results are similar to those reported
in the ``T42'' study of Pearce (1978, hereafter P78)
 and in the ``T95'' study of TH90, with 
almost the same time-scale and the same intensity as in P78,
whereas in TH90 the growth of $KE$ is more rapid with a maximum 
value for $KE$ reaching $11.$~$10^{5}$~J~m${}^{-2}$ at 
day $6.5$.

Figure~{\ref{FigEXP182021}}~(b) shows the results for 
the diabatic simulation EXP-HV including the vertical
diffusion scheme.
In comparison with EXP-H, the total energy decreases continuously
during the $13$ days of the simulation. The decrease 
in the global static stability component $AS$ 
is more important. It is a direct impact of the vertical mixing, 
leading to a decrease in the low-level vertical gradients of 
temperature (not shown). Another important feature is that 
the increases in $KE$ and $KZ$ are smaller than for EXP-H.
The effect of smoothing the horizontal gradients of the fields,
already mentioned in the comments of Fig.~\ref{FigPMERT900}, is 
confirmed by the energetic analysis and it leads to smaller
values for $KE$ and $KZ$ at each stage of the simulation. 
Changes in $KS$ and $AE$ are small
(less than $\pm 1.$~$10^{5}$~J~m${}^{-2}$), as for EXP-H.

Total energy TOT is a constant for EXP-A in
Fig.~{\ref{FigEXP182021}}~(c), with, however, a 
small and unrealistic increase above the 
initial value during the growing stage of the mode
(from days $6$ to $9$). 
Values of TOT for EXP-H
are also nearly constant but they lead to a more 
realistic continuous and small decrease, indicating that the
explicit ${\nabla}^6$ horizontal diffusion has a small but 
positive global impact on the total energy. The e-folding times 
for horizontal diffusion are short enough to suppress the
noisy features at high zonal wavenumber and 
do not lead to excessive numerical flattening 
of the fields that could induce spurious changes in total energy.
Relative changes of TOT are less than $0.2$~\% up to day $10$ 
for the two adiabatic simulations and absolute changes are less 
than $\pm0.25$~$10^{5}$~J~m${}^{-2}$.
As expected, they are small in comparison with the gain in 
eddy kinetic energy ($+7.$~$10^{5}$~J~m${}^{-2}$) and
they can be compared with the small changes in $AE$ or 
$KS$.

Global dissipation and generation terms are
depicted on Fig.~{\ref{FigEXP182021}}~(d) for the
three simulations and the difference $G-D$ is plotted 
only for EXP-A with the open square symbol. 
It is expected from (\ref{eq:cycle3c}) 
that, for a nearly constant total energy and for zero boundary
fluxes, $G-D$ must be small. Absolute 
values are indeed lower than $0.1$~W~m${}^{-2}$ up to day $8$,
whereas $D$ and $G$ reach $\pm 0.6$~W~m${}^{-2}$ for the
adiabatic cases.
This result is another global validation for the cycle 
(\ref{eq:cycle3c}), indicating that no global terms are missing
and that the numerical computations are accurate enough.
Although the dissipation is positive for EXP-A and EXP-H,
the behaviour for the diabatic simulation EXP-HV is more 
realistic. There is a negative global 
dissipation (dark circle) reaching $- 0.7$~W~m${}^{-2}$ at day $9$ and
a negative generation (open circle) reaching $- 1.3$~W~m${}^{-2}$ at day $7$. 
These negative and large generations and dissipations for EXP-HV 
explain the decreasing of $AS$ in Fig.~{\ref{FigEXP182021}}~(b)
and the moderate increase in $K_Z$ and $K_E$.

      \subsection{Local results for $\overline{k_E}$ 
                  (adiabatic simulation).} 
      \label{subsection_4.3}

\begin{figure}[t]
\centering
\includegraphics[width=0.46\linewidth,angle=0,clip=true]{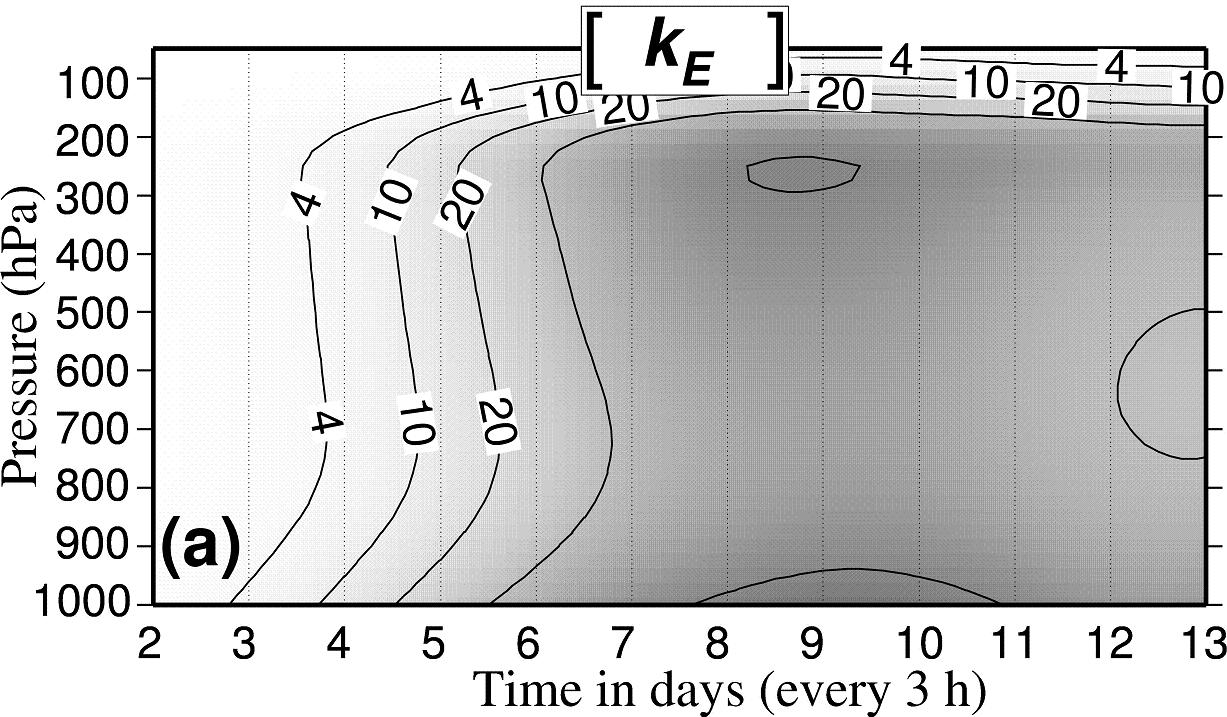}
\includegraphics[width=0.46\linewidth,angle=0,clip=true]{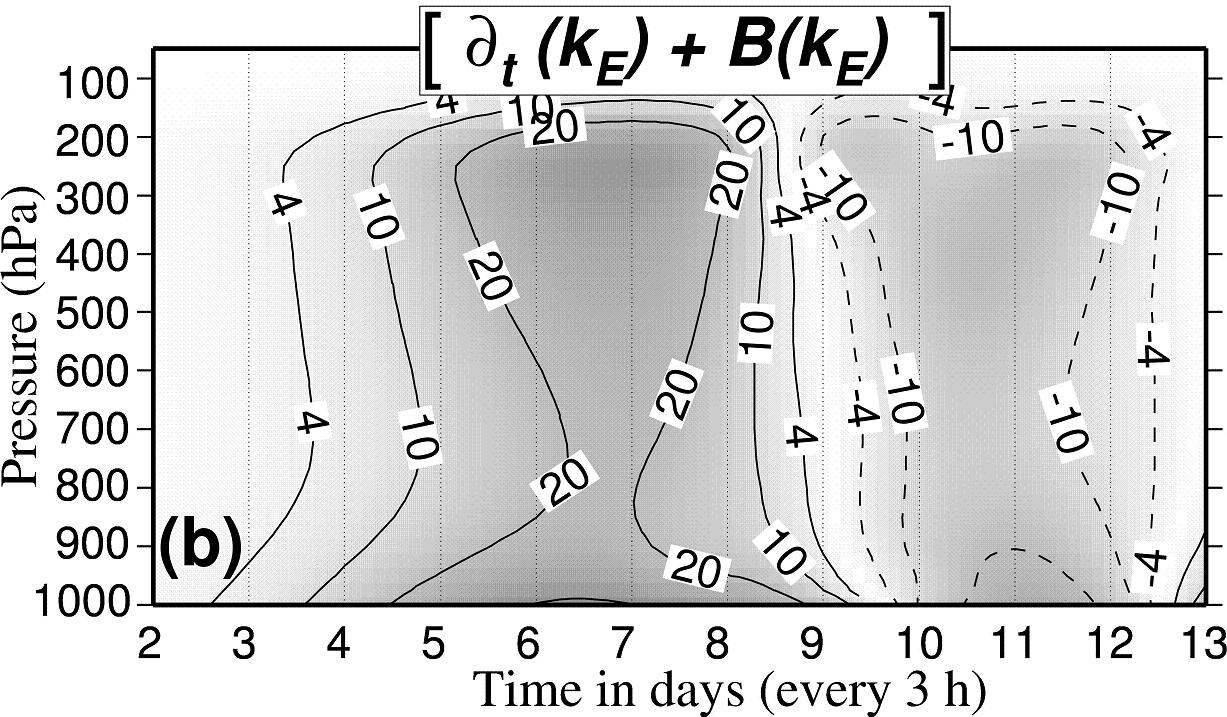}\\
\includegraphics[width=0.46\linewidth,angle=0,clip=true]{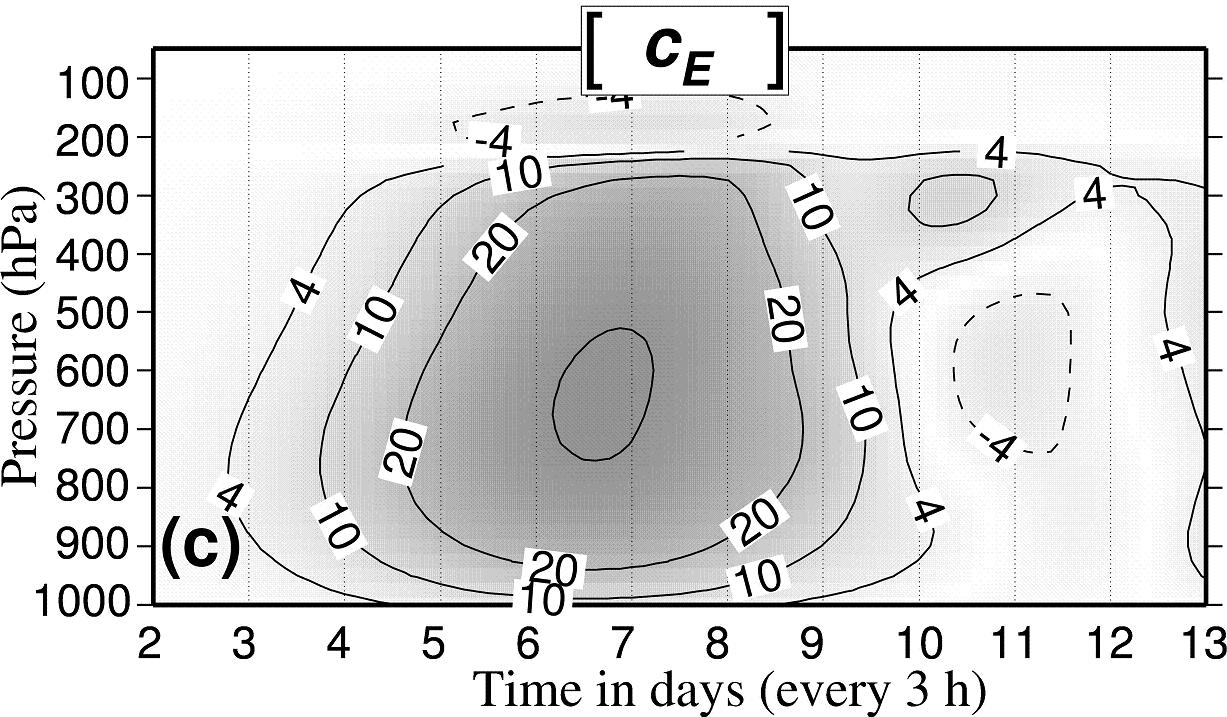}
\includegraphics[width=0.46\linewidth,angle=0,clip=true]{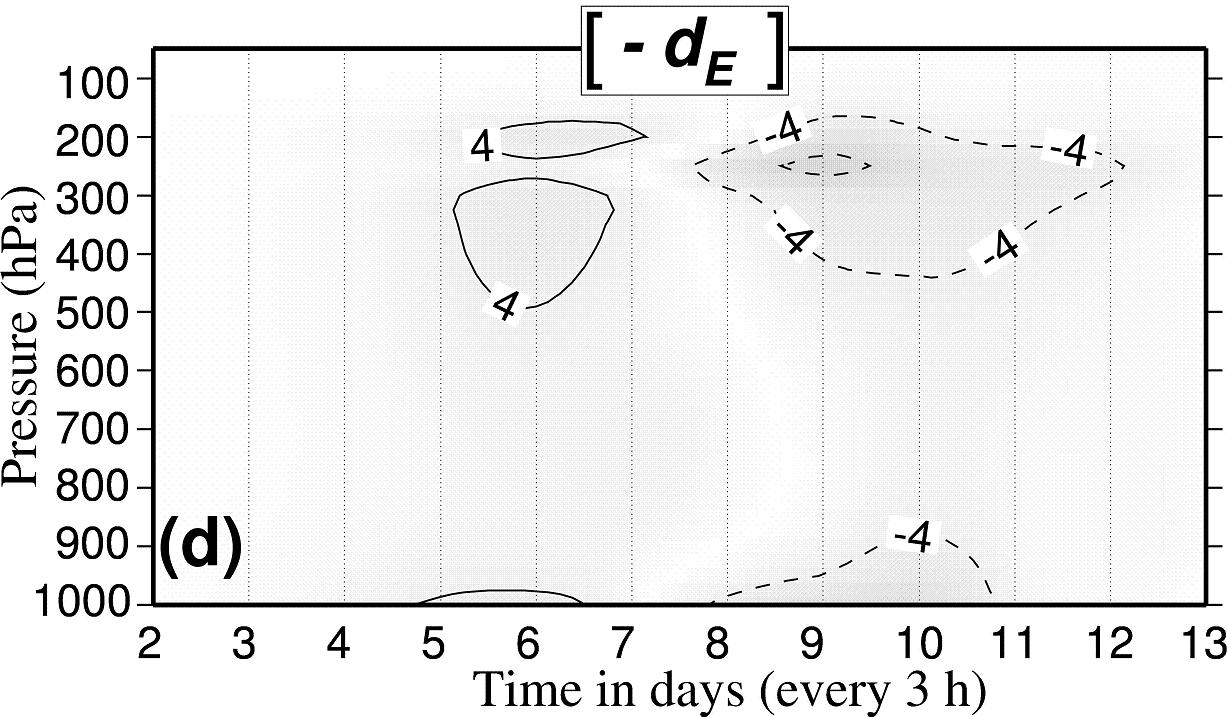}\\
\includegraphics[width=0.46\linewidth,angle=0,clip=true]{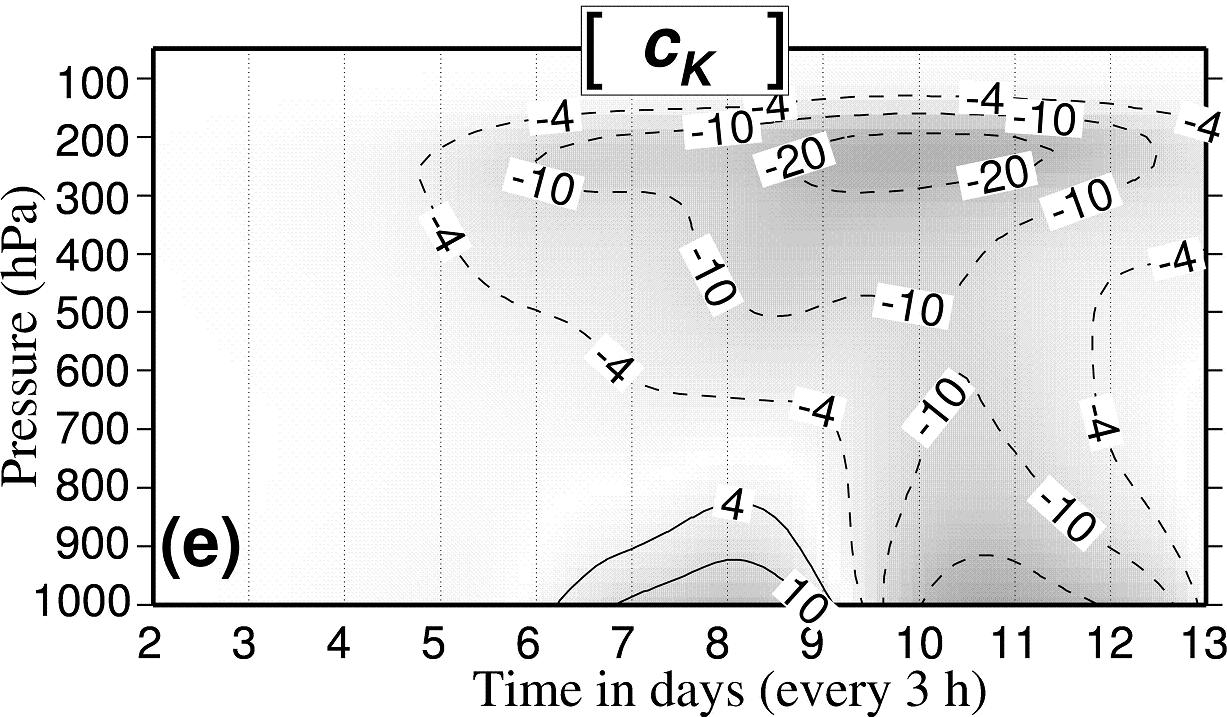}
\includegraphics[width=0.46\linewidth,angle=0,clip=true]{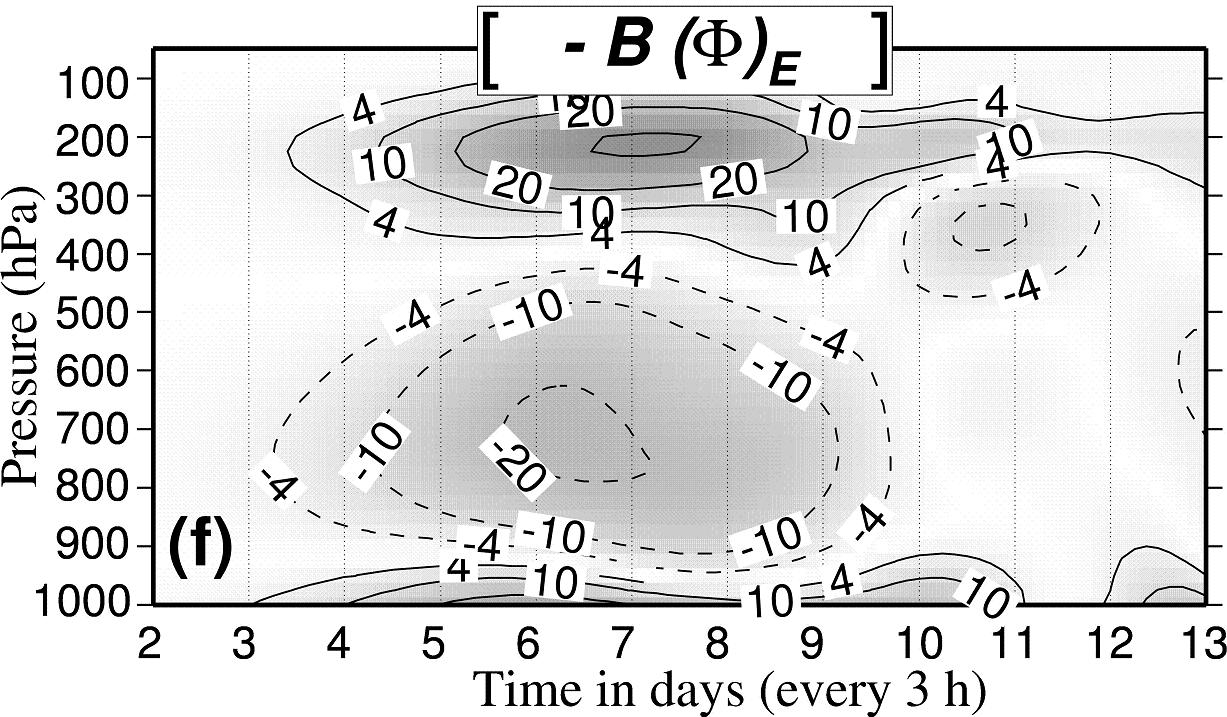}
\vspace*{-2mm}
\caption{\it \small 
Time--pressure diagrams from day $2$ to day $13$
of the adiabatic simulation EXP-H with the ${\nabla}^6$ horizontal 
diffusion and for the eddy kinetic energy reservoir. 
The horizontal domain extends from $25\,{}^{\circ}$N to $65\,{}^{\circ}$N
in latitude and it includes all the longitudes from $0$ to $180\,{}^{\circ}$.
  The isopleths $\pm 4$, $\pm 10$ and $\pm 20$ are annotated,
  followed by the contoured isopleths $\pm 40$, $\pm 100$, 
  $\pm 200$, $\pm 400$ and $\pm 1000$.
(a) The component $\overline{k_E}$ (J~{kg}${}^{-1}$). 
(b) The total budget 
$\overline{ {\partial}_t ( k_E ) } + \overline{ B( k_{E} )}$:
it is the sum of the local time derivative 
plus the divergence of boundary flux. 
(c) The usual baroclinic conversion $\overline{ c_E }$.
(d) The dissipation term $- \overline{d_E}$, expressed 
as a residual of the equation for $\overline{ {\partial}_t ( k_E ) }$
in Eqs.~(36) of Part~I.
(e) The usual barotropic conversion $\overline{ c_K }$.
(f) The direct conversion term $- \overline{{B( \phi )}_E}$,
between $\overline{\phi}$ and $\overline{k_E}$.
Units are $10^{-5}$~W~{kg}${}^{-1}$ for (b)--(f).
See text and Appendix~A of Part~I for explanation of symbols.
\label{FigEXP20}}
\end{figure}

Results for the large computational domain 
($25\,{}^{\circ}$N--$65\,{}^{\circ}$N ; $0\,{}^{\circ}$--$180\,{}^{\circ}$E)
are presented in 
Fig.~{\ref{FigEXP20}} for the eddy kinetic energy 
and for the adiabatic simulation. According to
formulation (36) of the cycle of Part~I, the reservoir
$\overline{k_E}$ is connected to $\overline{k_Z}$ and
$\overline{a_E}$ by the barotropic and baroclinic conversions
$\overline{c_K}$ and $\overline{c_E}$, respectively. 
The intensity of these conversions will be evaluated
with the aid of pressure-time diagrams. The new direct conversion 
with $\overline{\phi}$, denoted by $- \overline{{B( \phi )}_E}$, and 
the dissipation $- \overline{d_E}$ must also be carefully investigated.

There are two maximum values for $\overline{k_E}$ in 
Fig.~{\ref{FigEXP20}}~(a) at day $9$. The first maximum is 
located just below the jet at $250$~hPa and the other one 
is located at the surface. A relative minimum is observed up to
day $10$ in the middle troposphere between $800$ and 
$500$~hPa.

Figures {\ref{FigEXP20}}~(b)--(f) are expressed with a common unit 
of $10^{-5}$~W~{kg}${}^{-1}$. The total change in time
$\overline{ {\partial}_t ( k_E ) } + 
\overline{ B( k_{E} )}$ is depicted in (b) including
the boundary flux term, with observed small values
for $\overline{ B( k_{E} )}$ (not shown). 
The maximum and minimum values are located, 
as for Fig.~{\ref{FigEXP20}}~(a), at the surface and at about $250$~hPa
(at days $6.5$ for the maximum and at day $11$ 
for the minimum).

The equation for $\overline{{\partial}_t ( k_E )}$ 
corresponds in Fig.~{\ref{FigEXP20}}
to: (b)$\:=\:$(c)$\:+\:$(e)$\:+\:$(f)$\:+\:$(d). 
An expected result is that the dissipation in Fig.~{\ref{FigEXP20}}~(d)
for such an adiabatic simulation
should be small in comparison with others terms 
Figs.~{\ref{FigEXP20}}~(b), (c), (e) and (f), for all levels and at
any time. And indeed, the isopleths for $\pm 5$ 
units appear only in Fig.~{\ref{FigEXP20}}~(d)
at the levels and at the moment when
absolute changes in $\overline{k_{E}}$ are maximum in 
Fig.~{\ref{FigEXP20}}~(b). 
Therefore, the dissipation for this adiabatic simulation 
is only due to errors in the numerical 
schemes already mentioned. 
It is demonstrated that, on a local stage, there are 
no missing terms in (36) of Part~I because, in that case,
any forgotten or approximated terms would have contributed
with an opposite sign to the residual $-\overline{d_{E}}$ 
in Fig.~{\ref{FigEXP20}}~(d).

Figure~{\ref{FigEXP20}}~(b) shows a growing stage of the 
eddy kinetic energy from day $4$ to day $9$, associated 
with a large domain of positive values visible from 
the surface up to $150$~hPa, followed by a depletion stage 
corresponding to negative values after day $9$.
These growing and decreasing stages for $\overline{k_E}$ 
are usually associated with baroclinic 
and barotropic developments, respectively.
However, if the baroclinic and barotropic conversions in 
Fig.~{\ref{FigEXP20}}~(c) and (e) have 
he correct sign to partly explain the total 
change of $\overline{k_E}$ in 
Fig.~{\ref{FigEXP20}}~(b), direct comparisons of
Figs.~{\ref{FigEXP20}}~(c) and (b) show that the fields 
are out of phase.

Maximum values observed in the middle troposphere
for $\overline{c_E}$ correspond to minimum 
values in Fig.~{\ref{FigEXP20}}~(b). 
Similarly, maximum values for the 
total change in Fig.~{\ref{FigEXP20}}~(b) 
close to the jet and at the 
surface are associated with small values of 
$\overline{c_E}$.

A possible explanation for the difference between 
Figs.~{\ref{FigEXP20}} (b) and (c) is to consider that the 
term $- \overline{{B( \phi )}_E}$
in Fig.~{\ref{FigEXP20}}~(f) exports the positive 
middle-troposphere baroclinic 
input of energy toward the surface and the jet. This
mechanism has been described in Orlanski and Sheldon, 1995
(hereafter referred to as OS95) as vertical redistributions
of energy via work done by pressure forces. It 
is also in agreement with the fact that Fig.~{\ref{FigEXP20}}~(f) 
is neglected in 
global study of L55 because vertical redistributions correspond
to small or zero integral values over the depth of the atmosphere 
(not shown).

Figure~{\ref{FigEXP20}}~(f) shows a complex pattern 
for the vertical redistribution term
$- \overline{{B( \phi )}_E}$.
Positive values close to the surface are the main
source for the developments of the mode.
It is true for the maximum of $+42$ units observed at day $6.5$ 
in Fig.~{\ref{FigEXP20}}~(b) which is partitioned 
into $+3$ units from $- \overline{d_E}$ 
in Fig.~{\ref{FigEXP20}}~(d), $+5$ units from $\overline{ c_K }$ 
in Fig.~{\ref{FigEXP20}}~(e), $+11$ units from $\overline{ c_E }$ 
in Fig.~{\ref{FigEXP20}}~(c) and $+23$ units from 
$- \overline{{B( \phi )}_E}$ in Fig.~{\ref{FigEXP20}}~(f).
The larger term is the positive vertical redistribution 
Fig.~{\ref{FigEXP20}}~(f)
rather than the baroclinic conversion. 

Negative values in the 
middle troposphere for 
Fig.~{\ref{FigEXP20}}~(f) are almost balanced 
by the baroclinic conversion 
Fig.~{\ref{FigEXP20}}~(c). This baroclinic 
conversion cancels out above the $250$~hPa level
and the positive values between $350$ and $100$~hPa for 
$- \overline{{B( \phi )}_E}$ are the only contribution 
among 
Figs.~{\ref{FigEXP20}}~(c), (d), (e) or (f) that can explain 
the growth of $\overline{k_E}$ for the stratospheric part of
the jet (above the level $250$~hPa). 
Therefore the direct conversion 
between $\overline{\phi}$ and $\overline{k_E}$
plays a major role in energetics of stratospheric 
circulations when $\overline{k_E}$ is considered.

The barotropic conversion $\overline{c_K}$ in 
Fig.~{\ref{FigEXP20}}~(e) partly explains the 
depletion stage of the mode, with negative values
for the jet and in the boundary layer after day $9$. 
However negative values after day $9$ in 
Figs.~{\ref{FigEXP20}}~(d) 
and (f) are two other contributions for the decrease in $\overline{k_E}$.
As for the positive values for $\overline{c_K}$ in the 
boundary layer from day $6$ to day $9$, they could be interpreted 
as a barotropic instability. Nevertheless, this explanation will not
be confirmed for EXP-HV when the vertical diffusion scheme is 
activated. These spurious barotropic instabilities are the result
of a lack of surface dissipation in adiabatic simulations.

      \subsection{Local results for $\overline{k_E}$ 
                  (diabatic simulation).} 
      \label{subsection_4.4}

\begin{figure}[t]
\centering
\includegraphics[width=0.46\linewidth,angle=0,clip=true]{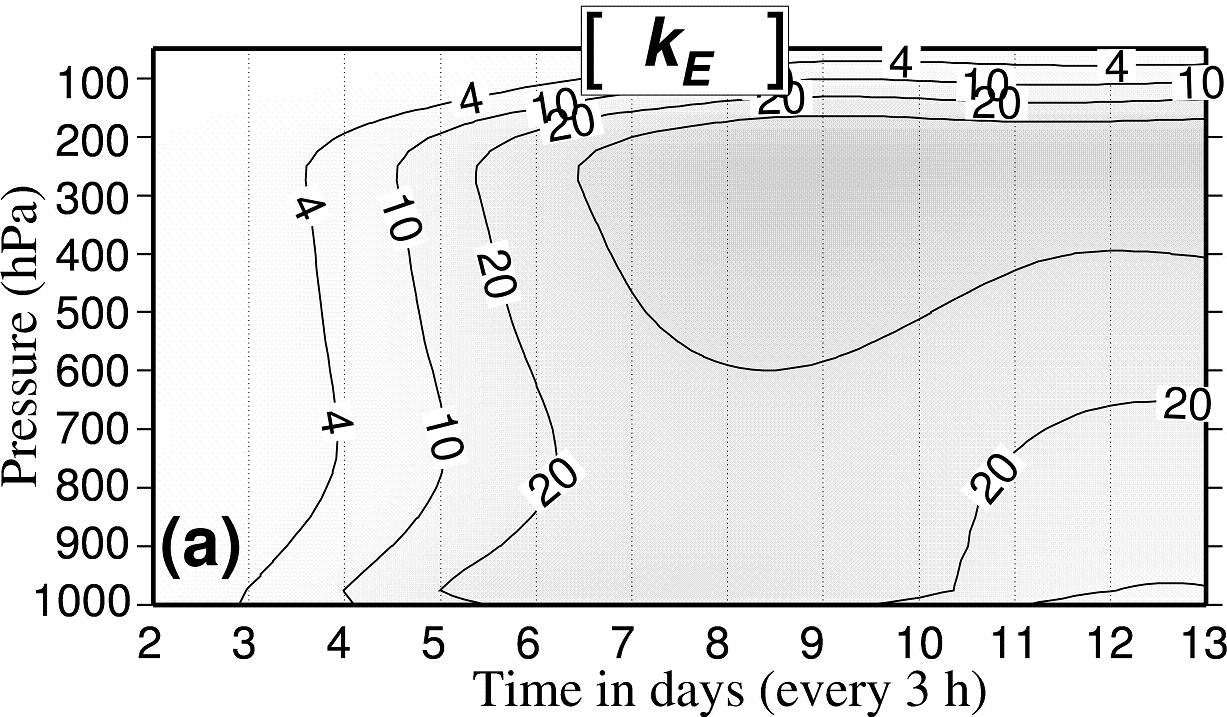}
\includegraphics[width=0.46\linewidth,angle=0,clip=true]{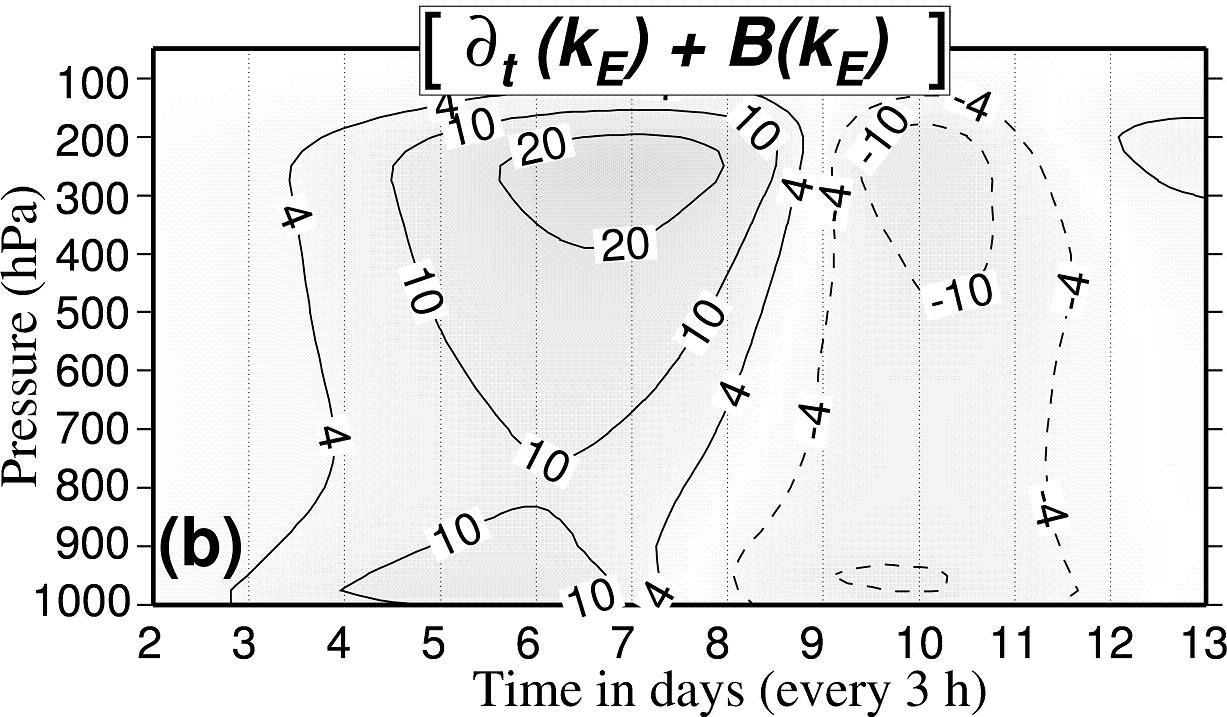}\\
\includegraphics[width=0.46\linewidth,angle=0,clip=true]{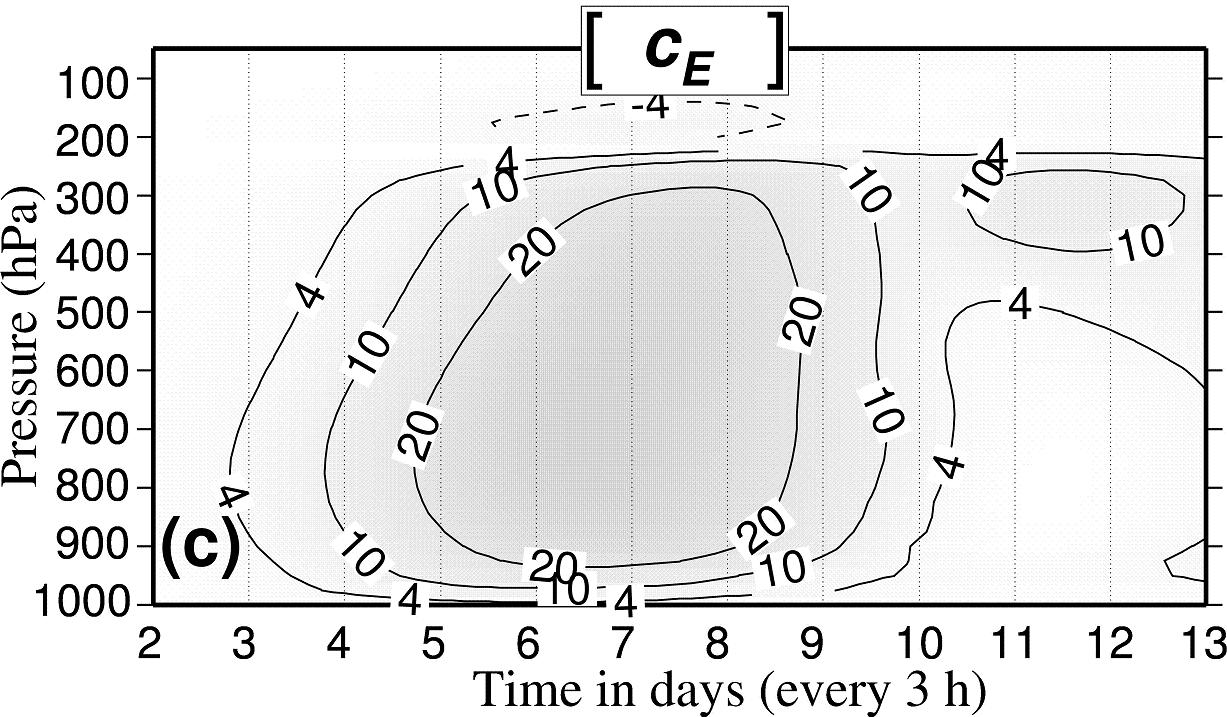}
\includegraphics[width=0.46\linewidth,angle=0,clip=true]{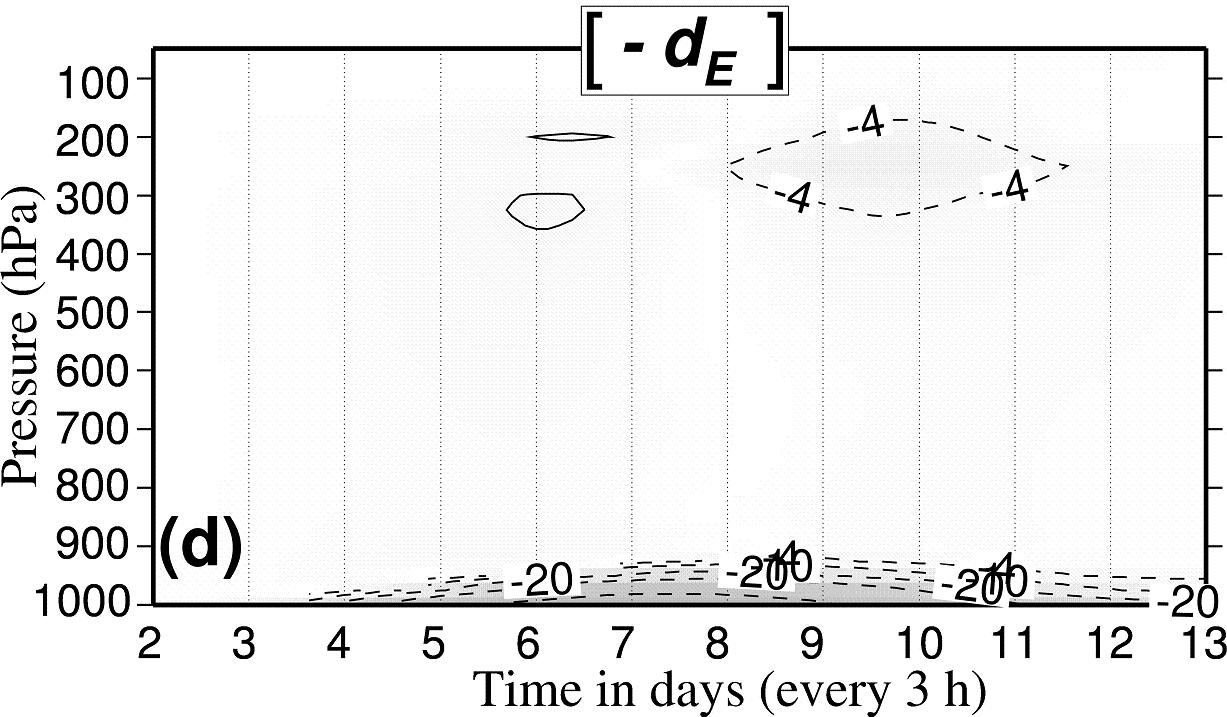}\\
\includegraphics[width=0.46\linewidth,angle=0,clip=true]{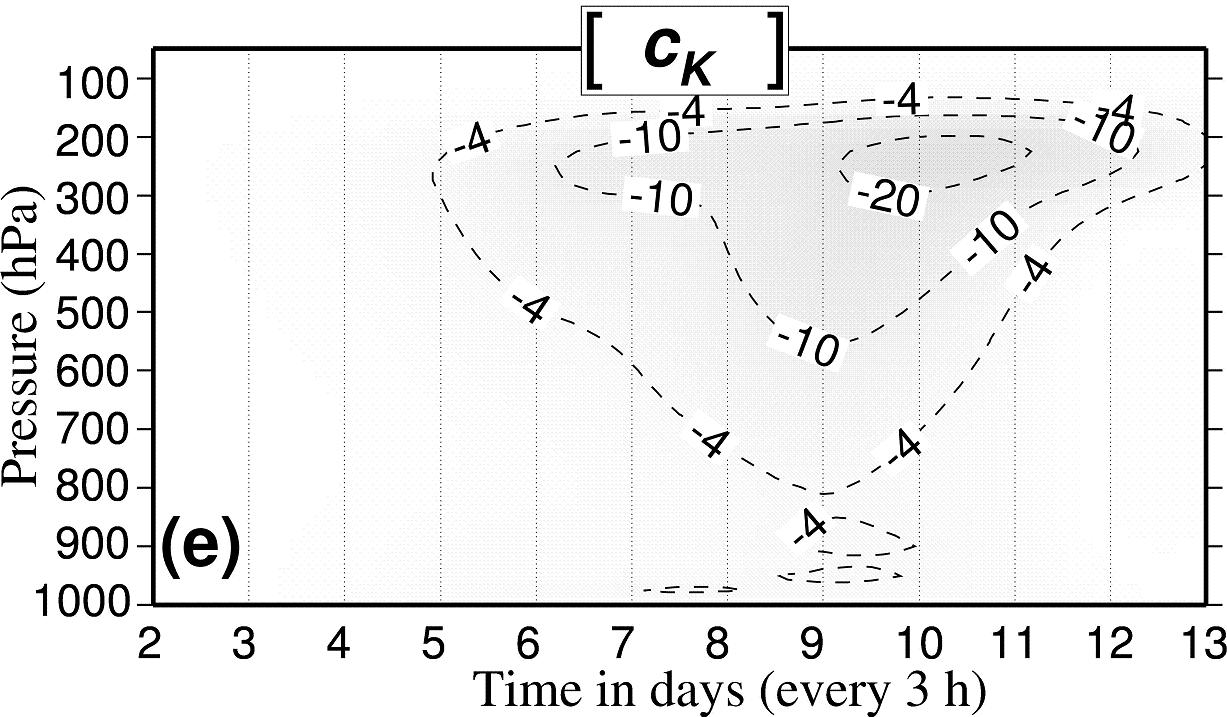}
\includegraphics[width=0.46\linewidth,angle=0,clip=true]{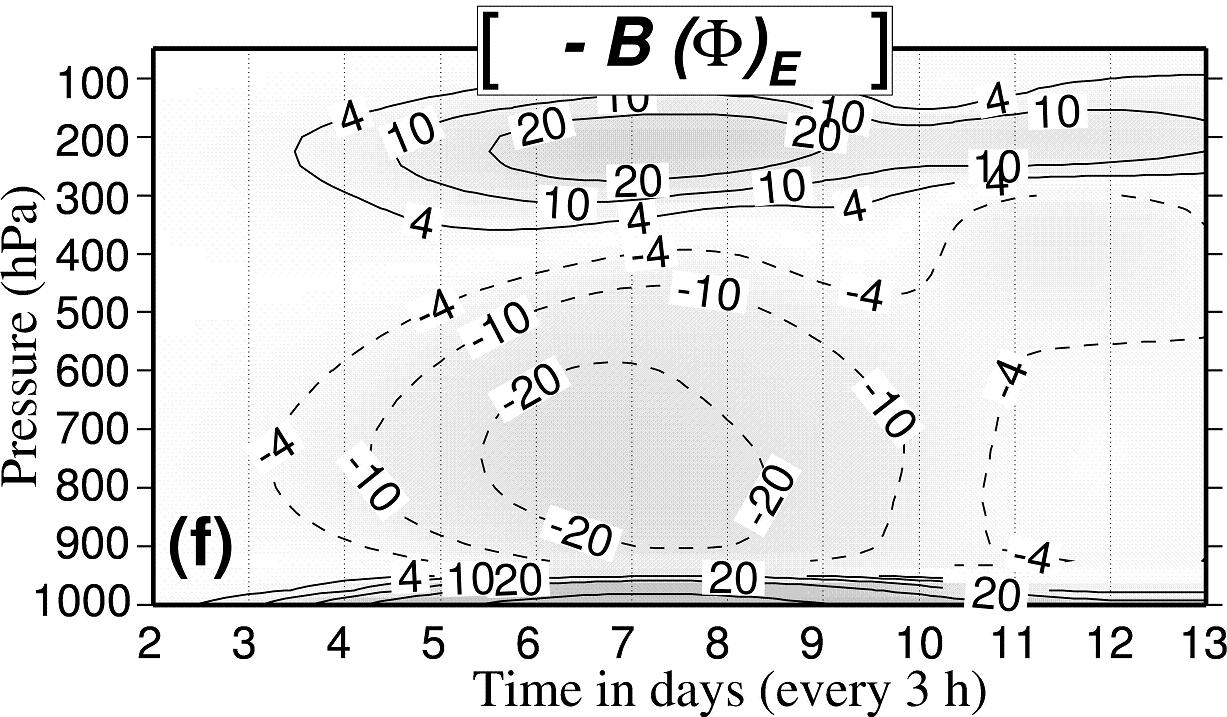}
\vspace*{-2mm}
\caption{\it \small 
As Fig.~{\ref{FigEXP20}} and for 
$\overline{k_E}$ see text), with the same domain,
interval and units, but for the diabatic simulation EXP-HV.
Both the horizontal and the vertical diffusion act 
on wind and temperature. The equation for $\overline{k_E}$
corresponds to (b)$\:=\:$(c)$\:+\:$(e)$\:+\:$(f)$\:+\:$(d). 
The numerous isopleths 
located near the surface in (d) and (f) correspond to real 
large negative values for the dissipation (d) and large 
positive values for the conversion (f).
The annotated isopleths are $\pm 4$, $\pm 10$ and $\pm 20$,
followed by the contoured isopleths $\pm 40$, $\pm 100$, 
$\pm 200$, $\pm 400$ and $\pm 1000$.
\label{FigEXP21a}}
\end{figure}

Results for the diabatic simulation EXP-HV with both
horizontal and vertical diffusions acting together
are presented on Fig.~{\ref{FigEXP21a}} for 
$\overline{k_E}$. When comparing EXP-H and EXP-HV,
the differences are not important for the total change
in Fig.~{\ref{FigEXP21a}}~(a) for the upper troposphere 
(above the $600$~hPa level)
and close to the jet. Differences are more important in the 
boundary layer between $1000$ and $900$~hPa when
they are associated with the vertical diffusion scheme.
The maximum of $\overline{k_E}$ occurs at day $7.5$, two
days ahead in comparison with EXP-H. The maximum growth 
of the mode in Fig.~{\ref{FigEXP21a}}~(b) has decreased from $42$ 
to $17$ units in the boundary layer with respect to 
Fig.~{\ref{FigEXP20}}~(b). A vertical displacement of this maximum 
is observed for the diabatic simulation: it rises from the surface 
to $975$ or $950$~hPa levels (the first diagnostic pressure levels).

Even if general characteristics are left unchanged between
Figs.~{\ref{FigEXP20}}~(c) and {\ref{FigEXP21a}}~(c),
baroclinic conversions decrease for all days and for 
all levels when the vertical diffusion is activated. 
There are large values ($\approx 35$ units) 
in the middle troposphere, with small values close to the surface 
and at the levels above $250$~hPa. As for the barotropic conversion,
comparisons of Figs.~{\ref{FigEXP20}}~(e) and {\ref{FigEXP21a}}~(e)
show large modifications occurring below the level $800$~hPa,
where the unrealistic positive and negative maximum observed at days 
$8$ and $11$ for EXP-H disappear.
The result is a barotropic decrease of the 
mode for the jet after day $9$, but with no equivalent 
barotropic stabilisation close to the surface. 
As a consequence, the decrease 
of the mode in the boundary layer after day $8.5$ in 
Fig.~{\ref{FigEXP21a}}~(b) must be explained
by non-barotropic components, i.e. the dissipation 
$- \overline{d_E}$ and/or the conversion 
$- \overline{{B(\phi)}_E}$.

Although the dissipation is, as expected, weak at all levels for
the adiabatic simulation in Fig.~{\ref{FigEXP20}}~(d), large
negative values are observed in Fig.~{\ref{FigEXP21a}}~(d)
in the boundary layer ($\approx -150$ units).
They correspond to large and expected dissipations created by 
the vertical diffusion scheme and by the surface friction.
As in EXP-H, values for $- \overline{d_E}$ above the level $900$~hPa 
where the vertical diffusion is not active are 
small in comparison 
with those in Figs.~{\ref{FigEXP21a}}~(b), 
(c), (e) and (f). It is a local 
validation for the cycle (36) of Part~I, obtained for the 
diabatic simulation in the free atmosphere.

\begin{figure}[t]
\centering
\includegraphics[width=0.49\linewidth,angle=0,clip=true]{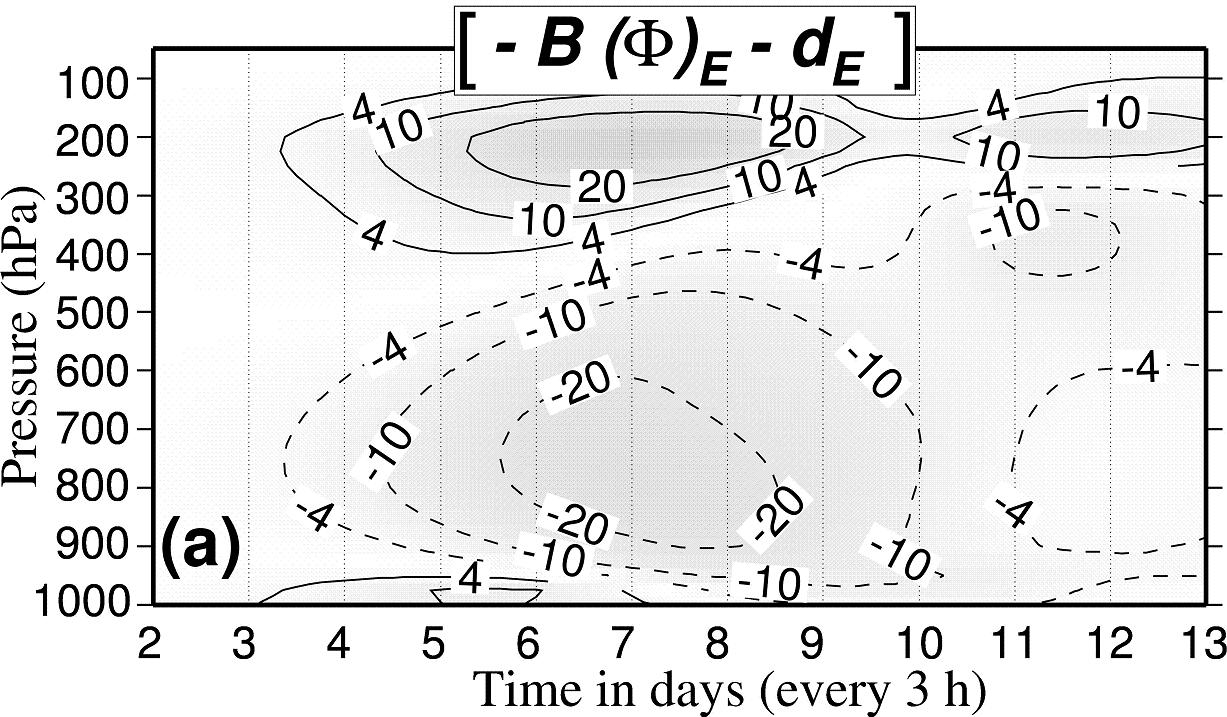}
\includegraphics[width=0.49\linewidth,angle=0,clip=true]{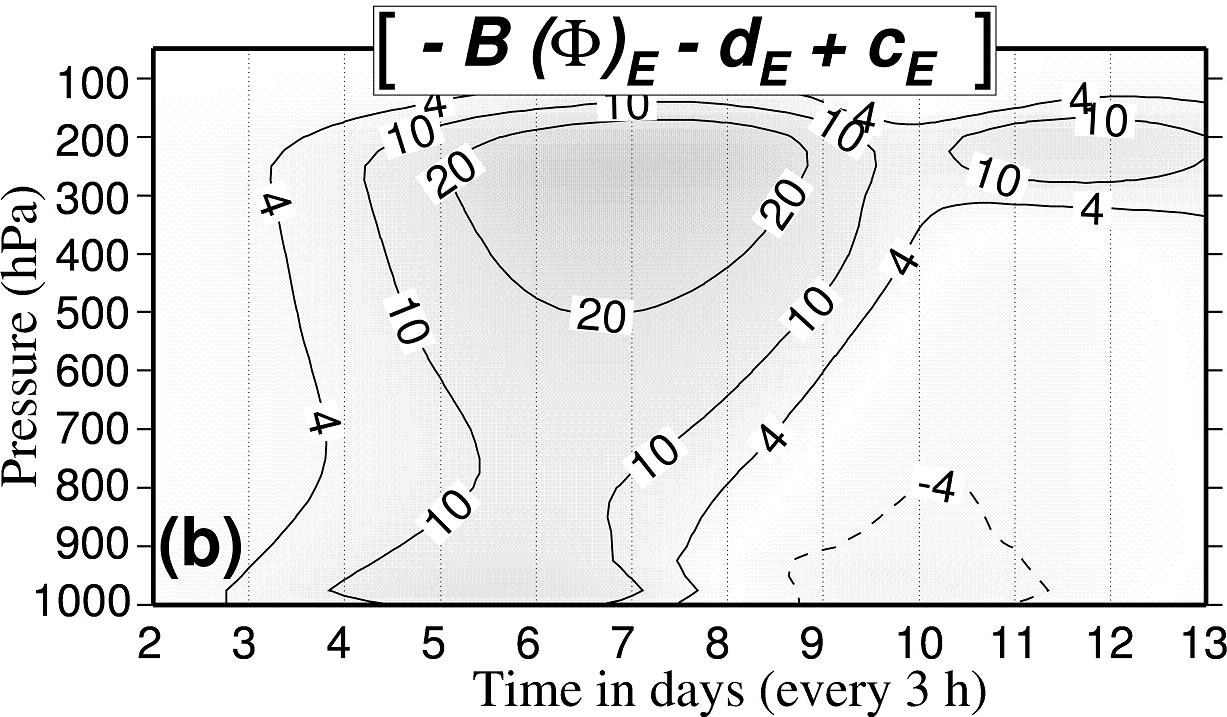}
\vspace*{-2mm}
\caption{\it \small 
Time--pressure diagrams from day $2$ to day $13$
for the same simulation EXP-HV as in Fig.~\ref{FigEXP21a}.
(a) The sum $- \overline{{B( \phi )}_E} - \overline{d_E}$.
(b) The sum $- \overline{{B( \phi )}_E} - \overline{d_E} 
    + \overline{c_E}$.
See text for explanation.
\label{FigEXP21b}}
\end{figure}

The validation of large values for $- \overline{d_E}$ in
Fig.~{\ref{FigEXP21a}}~(d) must be carefully realized.
Indeed, many authors have obtained similar large dissipation
terms computed as residuals and associated with large 
boundary terms $- \overline{{B( \phi )}_E}$
(Muench 1965, Brennan and Vincent 1980, Michaelides 1987). 
They consider these large dissipation terms as
not reliable. However, in this diabatic study, it seems that
large negative values for $- \overline{d_E}$ in the boundary 
layer are balanced by large positive counterparts observed for 
$- \overline{{B( \phi )}_E}$ in Fig.~{\ref{FigEXP21a}}~(f).
The sum of 
Figs.~{\ref{FigEXP21a}}~(d) and (f)
is presented in 
Fig.~{\ref{FigEXP21b}}~(a). The balance between
$- \overline{d_E}$ and $- \overline{{B( \phi )}_E}$ 
is established close to the surface, leading to moderate 
positive values up to day $8$ for the sum ($\leq +12$ units), 
whereas they reach $\pm 150$~units separately.

The equation for $\overline{k_E}$ in 
(36) of Part~I can be further modified 
by adding the baroclinic conversion 
$\overline{c_E}$ to the sum $- \overline{{B( \phi )}_E}
- \overline{d_E}$, to give the non-barotropic part of 
the total change of $\overline{k_E}$
depicted in Fig.~\ref{FigEXP21b}(b).
If the barotropic part in
Fig.~\ref{FigEXP21a}~(e) is easy to interpret,
the non-barotropic part in Fig.~\ref{FigEXP21b}~(b)
is a new combination of terms.
As explained in (49) of Part~I, it is the sum of the
eddy dissipation $- \overline{d_E}$ plus the work of 
eddy part of the pressure force, leading to
\vspace{-0.15cm}
\begin{eqnarray}
   - \: \overline{{B( \phi )}_E} \:
   + \: \overline{c_E} \:
   - \: \overline{d_E}
& = &
   - \: \overline{{({\bf U}_h)}_{\lambda} \: . \: 
   {({\bf \nabla}_{\!p} \, \phi)}_{\lambda}} \:
   - \: \overline{d_E}
\: .  \label{eq:newbaroc}
\end{eqnarray}
The depletion stage of the mode is represented by negative 
values in the boundary layer after day $8$ in
Fig.~\ref{FigEXP21a}~(b). A surprising result is
that this decreasing of the mode is not 
exclusively due to the barotropic conversion 
$\overline{c_E}$ in Fig.~\ref{FigEXP21a}~(e). 
It is also a consequence of the negative values observed 
in the boundary layer for the non-barotropic part
defined by (\ref{eq:newbaroc}).
The maximum of $-12$ units at day $10$ close to
$950$~hPa in Fig.~\ref{FigEXP21a}~(b) is separated into
$-4$~units for $\overline{c_K}$ in Fig.~\ref{FigEXP21a}~(e) 
and $-8$~units in Fig.~\ref{FigEXP21b}~(b). The non-barotropic 
part is thus the largest term.

By defining (\ref{eq:newbaroc}), a cancellation of the baroclinic 
conversion is obtained and, similarly to OS95, divergence 
of ageostrophic geopotential fluxes appear to be the main 
feature in local energetics of baroclinic systems.
Indeed, (\ref{eq:newbaroc}) can be interpreted as an equilibrium 
between the dissipation and the energetic impacts of ageostrophic 
circulations. 
The computations of the eddy terms in (\ref{eq:newbaroc})
are done with a zonally symmetric Coriolis term, leading to
${(f)}_{\lambda}=0$.
The result is that the eddy part of the pressure force 
may be written as 
\begin{equation}
- {({\bf \nabla}_{\!p} \, \phi)}_{\lambda}
\; = \; 
f \:\: {\bf k} \times {({\bf U}_g )}_{\lambda} 
\: .
\nonumber
\end{equation}
The eddy conversion term then becomes
\begin{equation}
- \overline{ {({\bf U}_h)}_{\lambda} . \: 
{({\bf \nabla}_{\!p} \, \phi)}_{\lambda} } \:
\; = \; 
{\bf k} \: . \;  \overline{ \:
             f \: {({\bf U}_g)}_{\lambda}
      \times {({\bf U}_h)}_{\lambda} }
\: .
\nonumber
\end{equation}
Using both
${({\bf U}_h)}_{\lambda} = {({\bf U}_a)}_{\lambda}
+ {({\bf U}_g)}_{\lambda}$ and the property
${({\bf U}_g)}_{\lambda} \times 
{({\bf U}_g)}_{\lambda} \equiv 0$ and without
loss of generality, (\ref{eq:newbaroc})
depicted in Fig.~\ref{FigEXP21b}~(b) can thus be
rewritten as 
\begin{equation}
{\bf k} \: . \; 
\overline{ \: f \: {({\bf U}_g )}_{\lambda}
\times {({\bf U}_a)}_{\lambda} }
\; - \;
\overline{d_E}
\: .
\nonumber
\end{equation}
There is a conversion if
the ageostrophic wind ${({\bf U}_a)}_{\lambda}$ 
is different from zero and
if it not parallel to ${({\bf U}_g)}_{\lambda}$.
However, this formulation differs from the 
vertical redistribution of energy described in
OS95 paper because it is not a divergence of some 
ageostrophic geopotential fluxes, in the form 
$\overline{ {\bf \nabla}_{\!p} {( \, \phi \: {\bf U}_h )}_a }$,
for instance.

      \subsection{Integral results for $\overline{k_E}$ 
                  (diabatic simulation).} 
      \label{subsection_4.5}

\begin{figure}[t]
\centering
\includegraphics[width=0.49\linewidth,angle=0,clip=true]{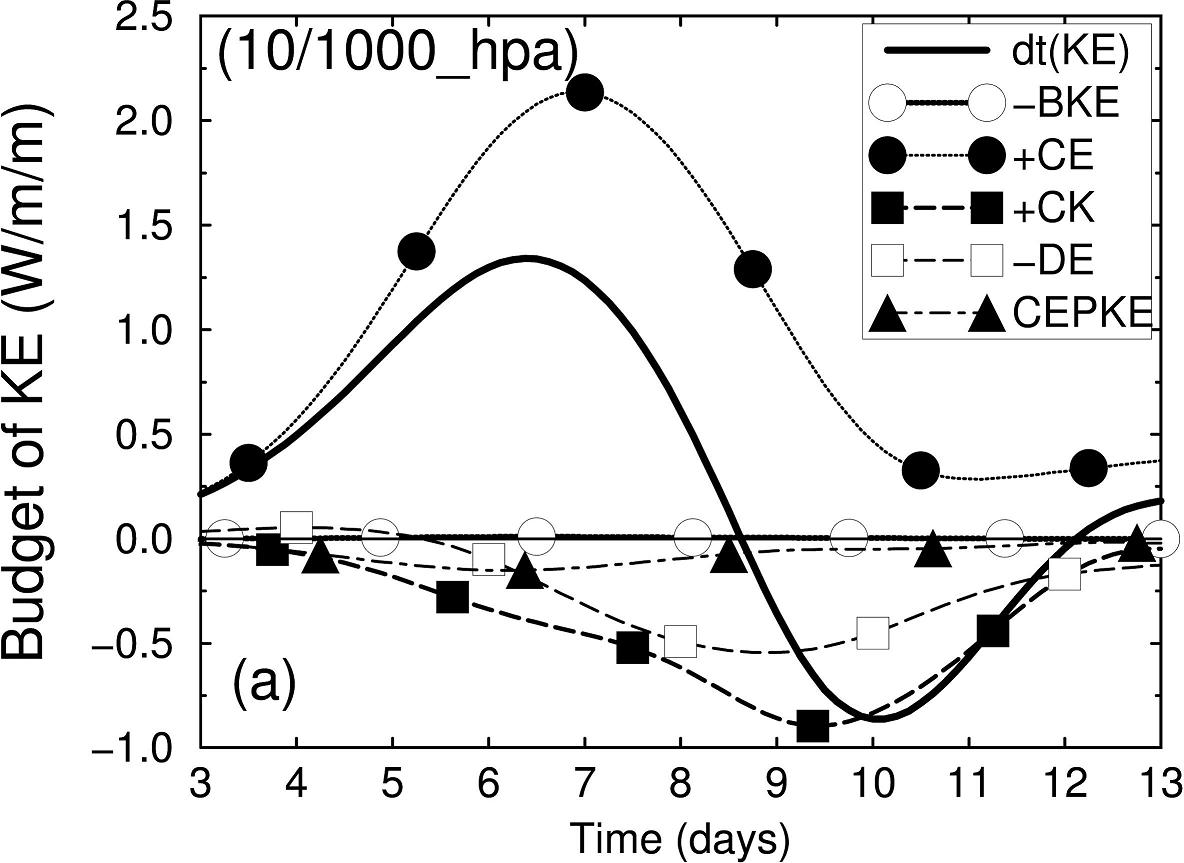}
\includegraphics[width=0.49\linewidth,angle=0,clip=true]{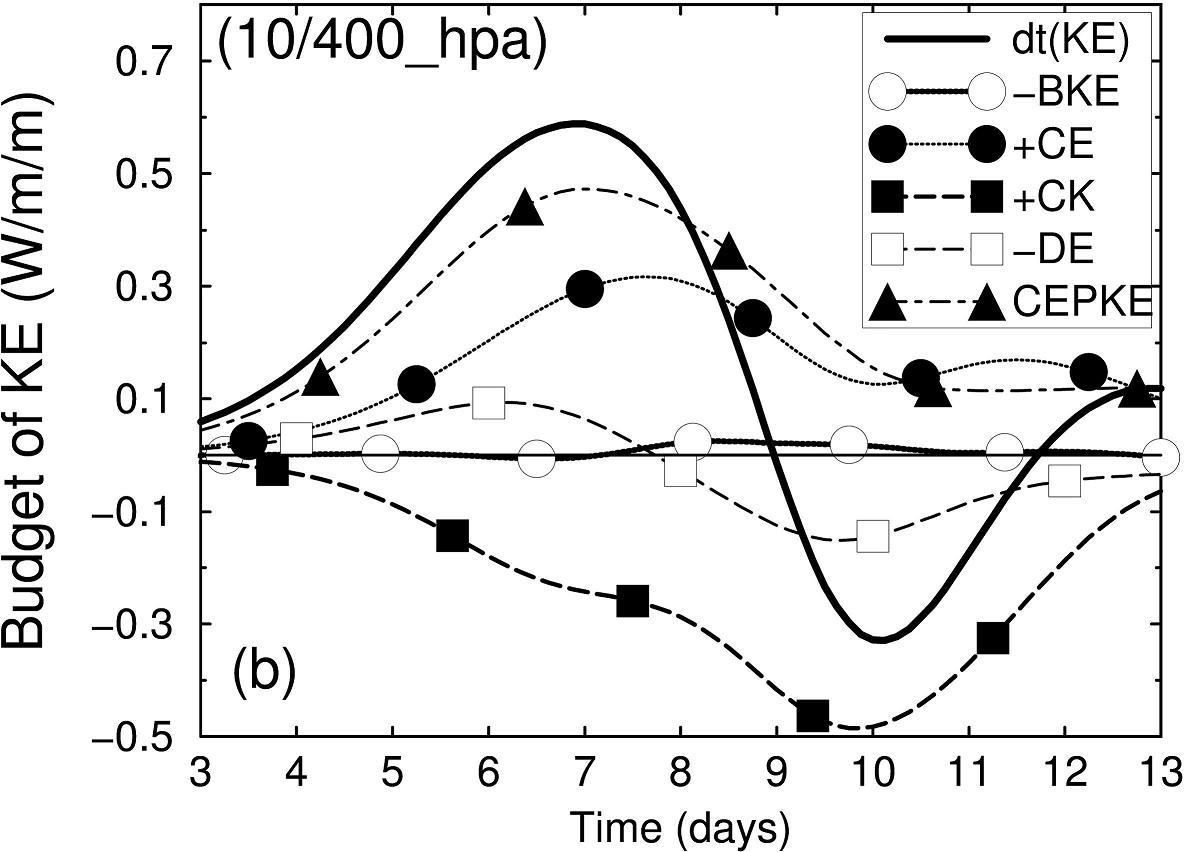}\\
\includegraphics[width=0.49\linewidth,angle=0,clip=true]{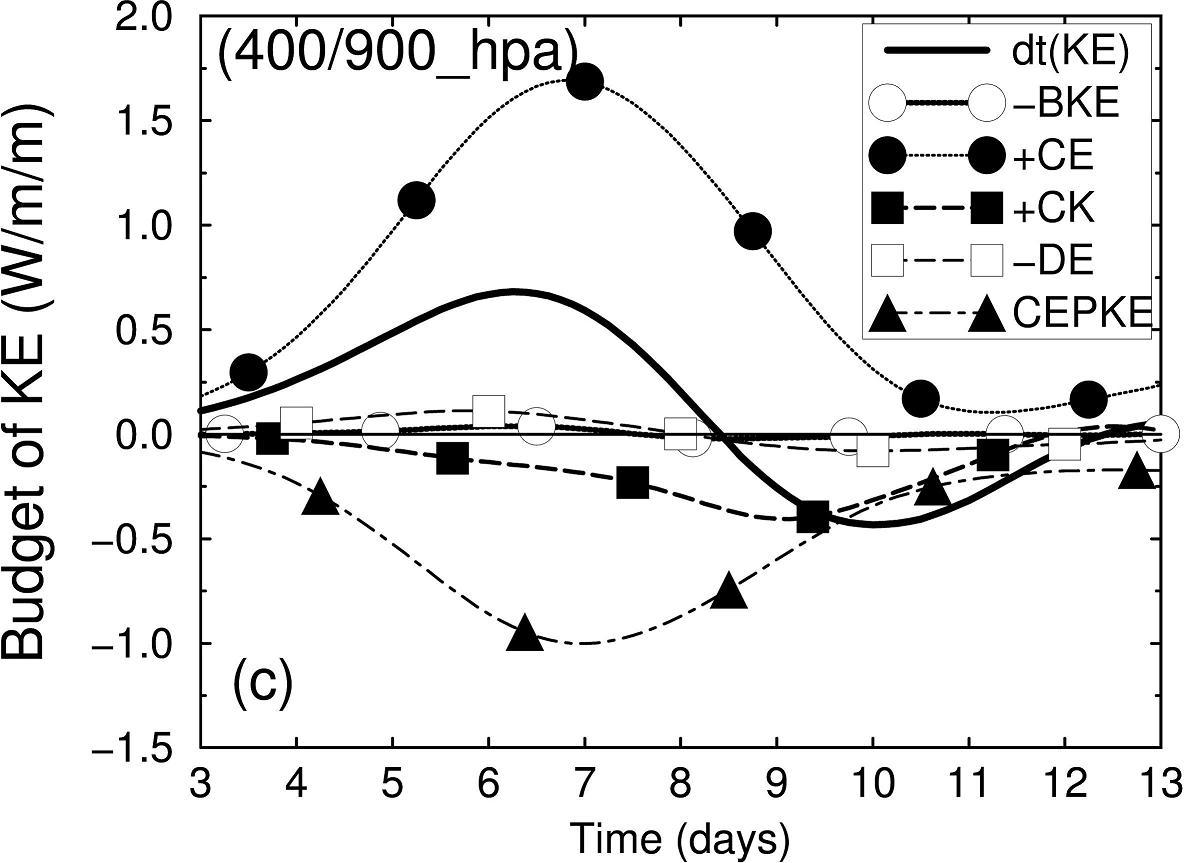}
\includegraphics[width=0.49\linewidth,angle=0,clip=true]{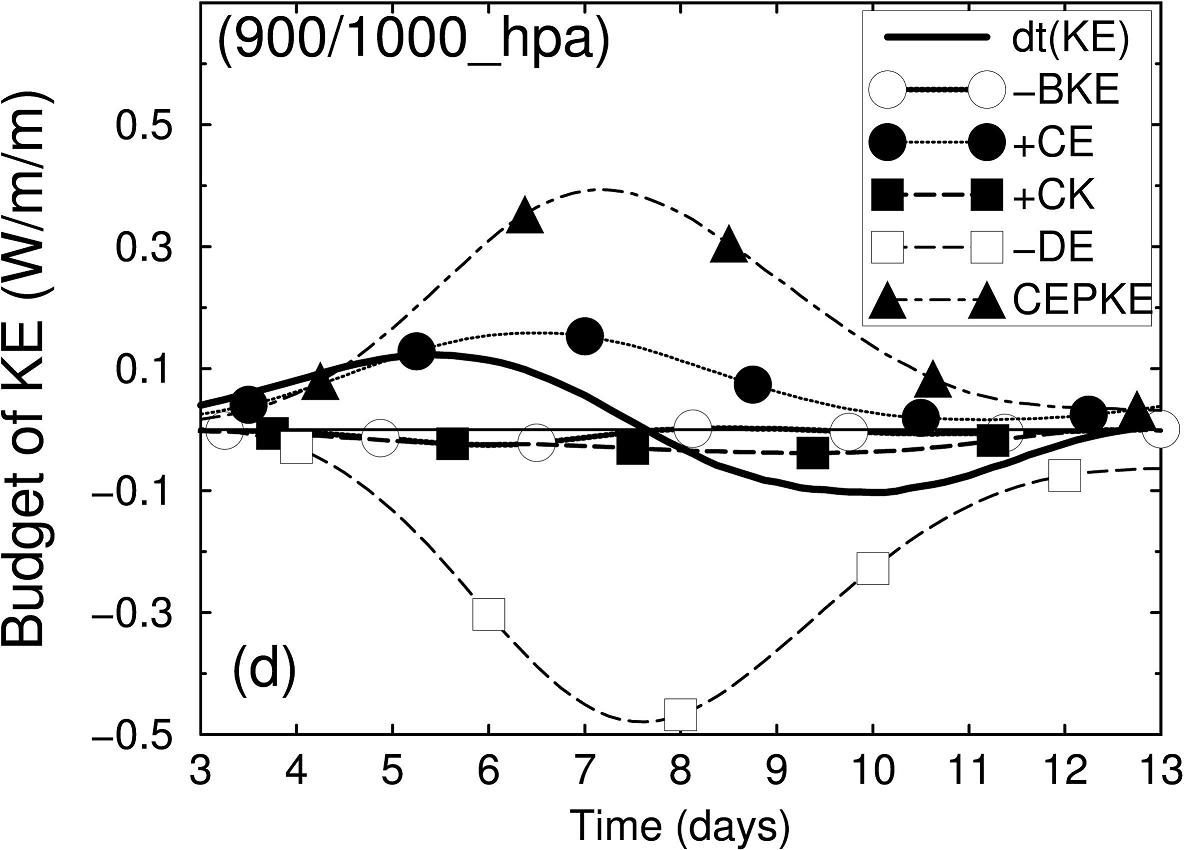}
\vspace*{-2mm}
\caption{\it \small 
Vertical integrals of budget equation 
for $\overline{k_E}$ (W~m${}^{-2}$). 
Terms are 
computed for the diabatic simulation EXP-HV and for four layers.
Tendencies $\overline{{\partial}_t(k_E)}$, denoted by dt(KE)
in the figure (thick solid line);
boundary terms $-\overline{ B( k_{E} )}$,
denoted -BKE (open circles and solid line);
baroclinic conversions $\overline{ c_E }$, 
denoted +CE  (dark circles and thin solid line); 
barotropic conversions $\overline{c_K}$,
denoted CK (dark squares and thick 
dashed line);
dissipation terms $- \overline{d_E}$, denoted -DE
(open squares and thin dashed line);
conversion terms with potential 
energy $- \overline{{B(\phi)}_E}$,
denoted CEPKE (dark triangles and mixed dashed line).
(a) $10$ to $1000$~hPa (global case).
(b) $10$ to $400$~hPa (stratosphere and jet).  
(c) $400$ to $900$~hPa (middle troposphere).  
(d) $900$ to $1000$~hPa (planetary boundary layer).  
\label{FigEXP21c}}
\end{figure}

The five ``time--pressure'' diagrams (Figs.~\ref{FigEXP21a}~(b)--(f))
exhibit complex patterns, with several 
minimum and maximum values located at different levels.
It seems worthwhile, however, to investigate the budget equation 
for $\overline{k_E}$ for three homogeneous layers: (i) the planetary boundary 
layer (from $1000$ to $900$~hPa), (ii) the middle troposphere (from $900$ 
to $400$~hPa) and (iii) the stratosphere and the jet ($400$ to $10$~hPa). 
A fourth global layer will also be considered: the integral from
$1000$ to $10$~hPa corresponding to the global available-enthalpy
cycle.

The global case is presented in Fig.~\ref{FigEXP21c}~(a) for EXP-HV. 
As expected, the increase in $\overline{k_E}$ up to day $8.5$ (positive 
values for the tendencies) is mainly controlled by the positive baroclinic 
conversion, with smaller negative contributions produced by the barotropic 
conversion and the dissipation terms. The boundary term 
$-\overline{ B( k_{E} )}$ is very small and it is not considered 
in the global studies (L55, P78).

The additional global conversion $- \overline{{B(\phi)}_E}$
denoted by $\mbox{\small CEPKE}$ in Fig.~\ref{FigEXP21c}~(a), 
is smaller than other terms (except 
$-\mbox{\small BKE} = -\overline{B(k_{E})}\approx0$). 
It is usually neglected in global studies (L55 and P78).
However it reaches $-0.2$~W~m${}^{-2}$ at day 
$6$, a value equal to half the barotropic 
conversion in the growing stage of the mode. As already mentioned
in Part~I, this term is not a real boundary term, i.e. 
$- \overline{{B(\phi)}_E} \neq  - \overline{B({\phi}_E)}$,
and its vertical integral must not cancel out. 
Locally observed large values of 
$- \overline{{B(\phi)}_E} = \mbox{\small CEPKE}$ in 
Fig.~\ref{FigEXP21a}~(f) and small corresponding global 
values of $\mbox{\small CEPKE}$ 
in Fig.~\ref{FigEXP21c}~(a) lead to a justification for 
the concept of vertical redistribution of energy described in OS95 
(see sections 4.3  and 4.4). As a consequence, $- \overline{{B(\phi)}_E}$ 
should appear both in local and global studies.

The budget of $\overline{k_E}$ for the middle troposphere in 
Fig.~\ref{FigEXP21c}~(c) is almost similar to the global case, 
although observed negative values of $\mbox{\small CEPKE}$
are much larger according to the large negative region in
Fig.~\ref{FigEXP21a}~(f). Another difference is that the dissipation 
is very small, owing to the distance from the jet and from the surface
where eddy diffusion and ageostrophic circulations occur
preferentially. Clearly, the barotropic and the baroclinic terms 
alone cannot explain  the budget of $\overline{k_E}$. In the
middle troposphere, the leading terms are $\overline{c_E}$, 
$- \overline{{B(\phi)}_E}$ and $\overline{c_K}$, in
decreasing order.

In the stratosphere and for the jet, the growing stage of the mode
in Fig.~\ref{FigEXP21c}~(b) correspond to a balance between
positive barotropic and negative baroclinic components. 
The additional vertical redistribution term
$- \overline{{B(\phi)}_E}$ is positive and it
is the largest term up to day $9$. Like the baroclinic
conversion $\overline{c_E}$, it can explain the increase of 
$\overline{k_E}$. The small eddy dissipation 
$-\overline{d_E}$ is positive and then negative, 
reaching $+0.1$~W~m${}^{-2}$ at day $6$ and $-0.15$~W~m${}^{-2}$ 
at day $10$, like for the middle troposphere case.

As expected, the depletion stage of the mode after day $9$ 
in Figs.~\ref{FigEXP21c}~(a) and (b) corresponds
to global negative values for the dissipation and for 
the barotropic conversion. However the decrease of
$\overline{k_E}$ after day $9$ for the middle troposphere 
in Figs.~\ref{FigEXP21c}~(c) is caused by negative values for
barotropic conversion and $- \overline{{B(\phi)}_E}$, with
a very small contribution from eddy dissipation.

In the planetary boundary layer, changes in $\overline{k_E}$ 
in Fig.~\ref{FigEXP21c}~(d) correspond to positive values for
the baroclinic conversion, at least up to day $6$. But the
dominant feature in this diagram is the balance between the 
large terms $- \overline{{B(\phi)}_E}$ and $-\overline{d_E}$. 
Results derived in section 4.4 show that the total balance 
between the three terms is equal 
to ${\bf k} \: . \: \overline{ \: f \: {({\bf U}_g )}_{\lambda} 
\times {({\bf U}_a)}_{\lambda} } - \overline{d_E}$. The 
depletion stage of the mode is due to observed negative values
in the Ekman layer for this total balance, because
the effect of ageostrophic circulations and eddy dissipation
do not exactly compensate each other. 
This is only true for the Ekman spiral.

      \subsection{Ekman dissipation for diabatic simulations.} 
      \label{subsection_4.6}

Explanations for the balance between large
positive and negative values for 
\begin{equation}
- D \; = \; 
{\bf U}_h \, . \: {\bf F}_h
\; \; \; \; \; \; \mbox{and} \; \;  \; \;
C_{a} \;  = \; 
- {\bf U}_h \, . \, {\bf \nabla}_{\!p}(\, \phi \, )
\; , \nonumber
\end{equation}
as observed in the boundary
layer in Figs.~\ref{FigEXP21a}~(d), \ref{FigEXP21a}~(f) 
and \ref{FigEXP21c}~(d), are given in this section.
It is demonstrated that
the two terms have opposite signs with 
exactly the same magnitudes
in the case of an idealized Ekman spiral, in which
case  the non-barotropic part (\ref{eq:newbaroc})
is zero in the boundary layer.

If the horizontal wind and the frictional force
are expressed in the complex form 
\begin{equation}
V = \: u + i \, v
\; \; \; \; \; \; \mbox{and} \; \;  \; \;
F_h = ({F}_h){}_y + i \, ({F}_h){}_y
\; , \nonumber
\end{equation}
the dissipation is equal to the real part 
$-D = {\Re}\mbox{e} \! \left( \overline{V} F_h \right)$
and it can be verified that the
conversion term due to the ageostrophic 
wind is equal to $C_{a} = {\Im}\mbox{m} \! 
\left( \, f \, \overline{V_g} \: V_a \: \right)$,
where the overbars represent complex conjugates.

In complex formulation, the Ekman spiral frictional 
force due to the vertical dissipation scheme is
given by $F_h = i \, f \, V_a$ and for this value
the dissipation due to the
vertical diffusion is exactly compensated by 
the conversion created by ageostrophic secondary 
circulations, or equivalently 
\begin{equation}
D \: = \: C_a 
 \: = \: f \: \left( u_g v_a - v_g u_a \right)
 \: = \: f \; {\bf k} \, . 
         \left( {\bf U}_g \times {\bf U}_a \right)
\; . \nonumber
\end{equation}

The common value for $C_a = D$ can be computed
by using the Ekman spiral for a general surface angle
$\theta$ and for a height of planetary boundary layer
given by $H_{pbl}$. In complex form, the ageostrophic 
wind is written
\begin{eqnarray}
  V_a \; \;  = \; \; V_g \; \: \sqrt{2} \; \: \sin(\theta) \; \:
        \exp\left( - \frac{\pi z }{\Lambda} \right) \; \:
        \exp\left[ \, i \, \pi \, \frac{(H_{pbl} - z )}{\Lambda} \right]
   \: , \label{EKMAN1}  
\end{eqnarray}
where the scale height $\Lambda = H_{pbl} / (\theta  / \pi + 3 / 4)$ 
equals $H_{pbl}$ for the zero surface wind case 
$\theta = \pi/4$; it 
becomes smaller than $H_{pbl}$ for decreasing $\theta$
and increasing surface wind
(ex. $\Lambda \approx 0.917 \, H_{pbl}$ for $\theta=\pi/6 $
$= 30\,{}^{\circ}$). The conversion is computed by using (\ref{EKMAN1}) and
$C_a = {\Im}\mbox{m}  \! \left( \, f \: \overline{V_g} \:  V_a \: \right)$,
to give
\begin{eqnarray}
  C_a \; \;  = \; \; D \; \;  = \; \; 
         f  \; \: {(V_g)}{}^2 \; \: \sqrt{2}  \; \: \sin(\theta)  \; \:
        \exp\left( - \frac{\pi z }{\Lambda} \right) \; \:
        \sin\left[ \, \pi \, \frac{(H_{pbl} - z )}{\Lambda} \right]
   \: . \label{EKMAN2}  
\end{eqnarray}
The maximum value for $C_a$ is obtained by cancelling
the derivative with respect to $Z = z / \Lambda $ 
of the function $f(Z) = \exp\{ - \pi Z \}$ 
$\sin\{ \pi \, (H_{pbl}/\Lambda - Z ) \}$. The height
for this maximum is equal to $z_{max} = \Lambda /4  + 
( H_{pbl} - \Lambda )$. For $\theta = \pi/n$, this
expression reduces to $z_{max} = H_{pbl} / \{ 4 + 3(n-4)/4 \}$.
It gives $z_{max} = 0.25 \: H_{pbl}$ for $n=4$ (latitude $45\,{}^{\circ}$), 
$ z_{max}  \approx 0.182 \: H_{pbl}$ for $n=6$ (latitude $30\,{}^{\circ}$)
and $ z_{max}  \approx 0.143 \: H_{pbl} $ for $n=8$ (latitude $22.5\,{}^{\circ}$).

\begin{figure}[t]
\centering
\includegraphics[width=0.49\linewidth,angle=0,clip=true]{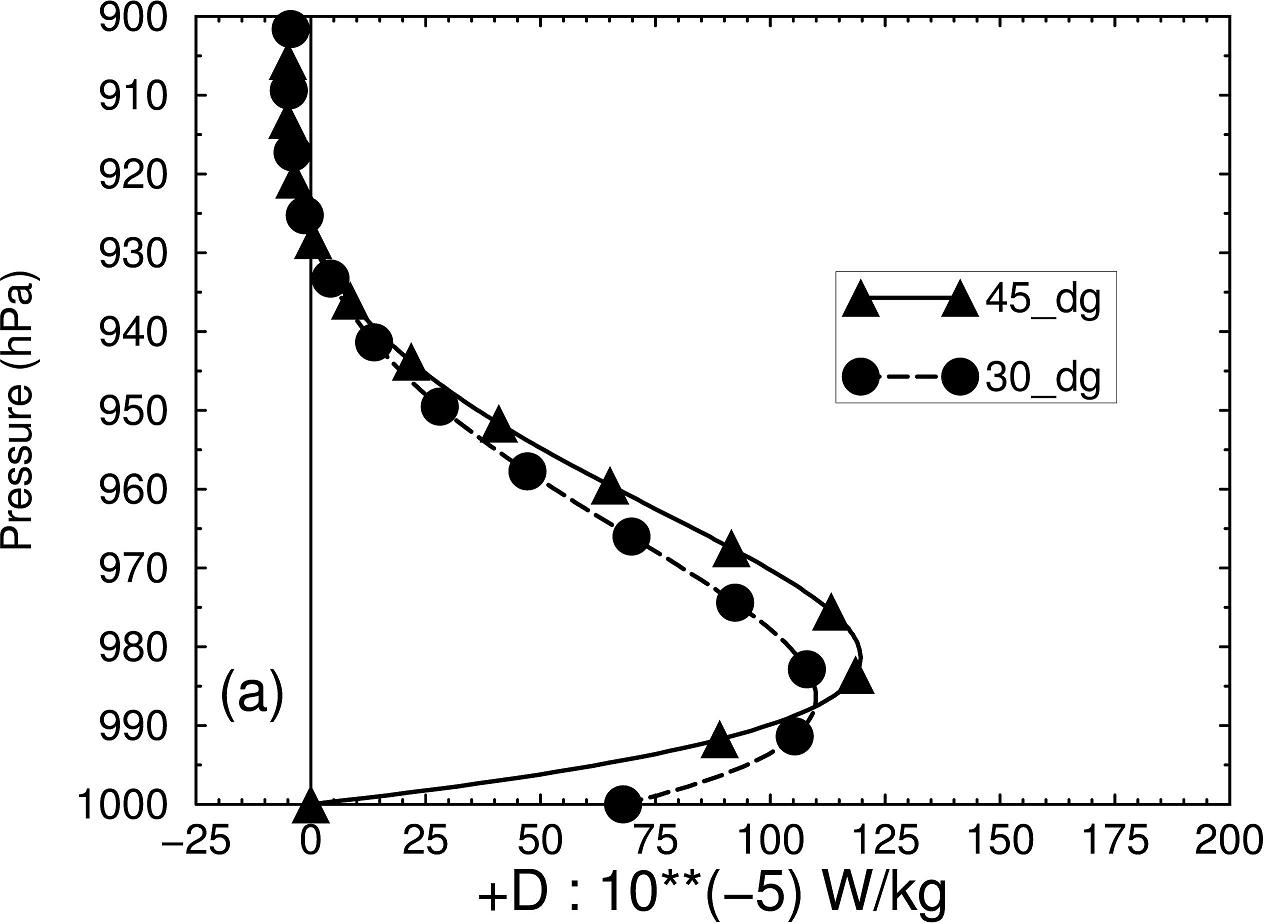}
\includegraphics[width=0.49\linewidth,angle=0,clip=true]{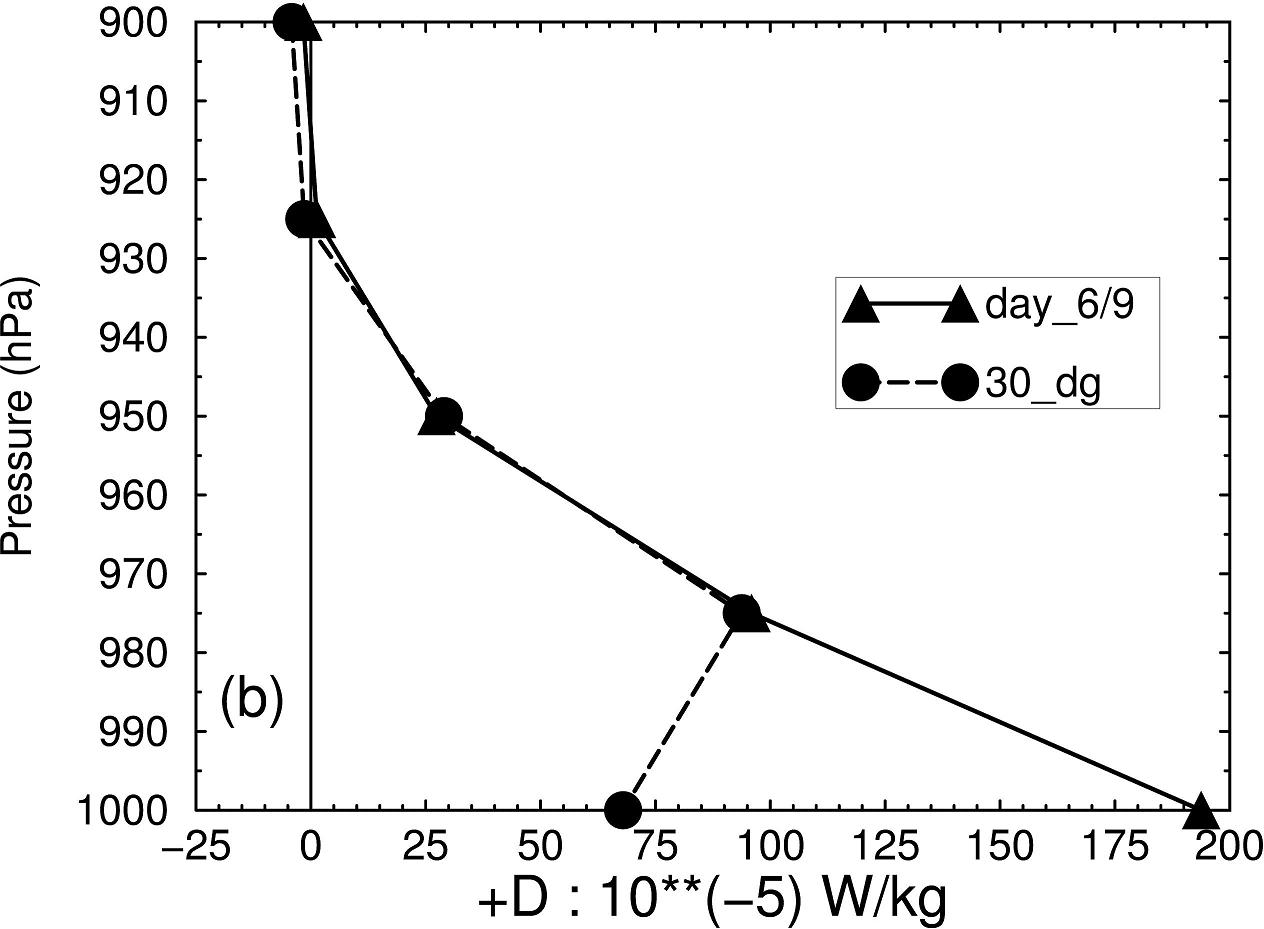}
\vspace*{-2mm}
\caption{\it \small 
Positive values of $+D>0$ for idealized Ekman spirals
and within the boundary layer ($-D$ is the usual 
negative dissipation).
(a) Two theoretical spirals with the surface angle 
    $\theta=45\,{}^{\circ}$ and 
    $\theta=30\,{}^{\circ}$. A refined 
    vertical resolution of $400$ levels is used.
(b) The theoretical spiral with $\theta=30\,{}^{\circ}$
   and the averaged observed values from day $6$ to day $9$ 
   for the diabatic simulation EXP-HV. There are
   four vertical levels above the $1000$~hPa surface
   pressure, corresponding to the interval of $25$~hPa
   between the constant post-processed pressure levels
   ($975$, $950$, $925$ and $900$~hPa).
\label{FigEKMAN21}}
\end{figure}

Examples of different dissipation terms ($+D > 0$)
for idealized Ekman spirals are presented 
in Fig.~\ref{FigEKMAN21}~(a) for two surface 
angles ($\theta=45\,{}^{\circ}$ and $\theta=30\,{}^{\circ}$) and 
with a high vertical resolution. The height of 
the maximum for $+D$ is lower for the case $30\,{}^{\circ}$N, 
in agreement with the values $0.25 \: H_{pbl}$ and 
$0.182 \: H_{pbl}$ for $45\,{}^{\circ}$N and $30\,{}^{\circ}$N,
respectively.

Equation~(\ref{EKMAN2}) is used with
a height of the boundary layer set to
$H_{pbl}=600$~m, corresponding to the pressure $928$~hPa 
where the dissipation cancels out on Fig.~\ref{FigEKMAN21}~(b). 
The Coriolis parameter for an average latitude of 
$45\,{}^{\circ}$N is set to $10{}^{-4} \, \mbox{s}{}^{-1}$ and
the module of the geostrophic wind is set 
to $6\,\mbox{m}\,\mbox{s}{}^{-1}$.

Figure~\ref{FigEKMAN21}~(b) shows the successful fitting 
between the idealized Ekman spiral ($\theta=30\,{}^{\circ}$)
and the averaged observed values from day $6$ to day $9$ 
for the diabatic simulation EXP-HV depicted on 
Fig.~\ref{FigEXP21a}~(d). Dissipations in the free atmosphere 
above the level $928$~hPa are small and $-D$ is
positive as predicted by the Ekman spiral. Below the
level $928$~hPa, the observed and theoretical curves
can be superposed at $950$ and $975$~hPa. However
the fitting cannot be obtained for the level 
$1000$~hPa because this level is subject to several errors.
Firstly the $25$~hPa post-processed level interval is 
coarser than the uneven model levels interval, equal
to $13$ and $24$~hPa for the first ones. As a consequence
the vertical differencing schemes differ. Secondly, there
are interpolations and/or extrapolation of the
fields for the case of surface pressure greater and/or
lower than $1000$~hPa, with the side effect of excluding
or including unrealistic or fictitious mass of atmosphere 
into the energetic budget.

This interpretation of large local values for atmospheric dissipation 
expressed as residuals is an important result of this paper.
It is the ultimate local validation of the available-enthalpy 
cycle (36) of Part~I, giving the proof that
there are no approximate or missing terms. 
However, the real ageostrophic circulations diverge from 
the idealized Ekman spiral case, because the eddy coefficients 
are not independent of height. It is the reason why the sum
(\ref{eq:newbaroc}) depicted in Fig.~\ref{FigEXP21b}(b) 
is not exactly zero and imbalances between
${\bf k} \: . \: \overline{ \: f \: {({\bf U}_g )}_{\lambda}
\times {({\bf U}_a)}_{\lambda} }$ and 
$ - \overline{d_E}$ yield positive values in the 
growing stage of the mode, with observed maximum 
values in the boundary layer ($975$~hPa). As for the
maximum values below the jet in Fig.~\ref{FigEXP21b}~(b), 
they must be interpreted in term of a balance between
effects of ageostrophic circulations and the baroclinic
conversion, because the dissipation is very small in this
region (see Fig.~\ref{FigEXP21a}).

      \subsection{Local results for $\overline{a_E}$ and 
    $\overline{k_S}$ (diabatic simulation).} 
      \label{subsection_4.7}

The equation for $\overline{a_E}$ in the cycle (36)
of Part~I is similar to the equation for $\overline{k_E}$,
but with different conversion terms and
a generation term in place of the
dissipation. There is no term equivalent to
$- \overline{{B( \phi )}_E}$. Observed values
for $\overline{{\partial}_t (a_E)}$, $\overline{B(a_{E})}$ 
and $\overline{g_{E}}$ are small (not shown). They are
less than $\pm 4$ units of $10^{-5}$~W~{kg}${}^{-1}$, for all levels 
and for the whole simulation from day $2$ to $13$.
As a consequence, there is a close balance 
$\overline{c_A} \approx \overline{c_E}$ and
the eddy available enthalpy reservoir $\overline{a_E}$ behaves 
like a catalytic component. There is a direct transfer of energy
from $\overline{a_Z}$ toward $\overline{k_E}$ through 
$\overline{a_E}$, with no change in $\overline{a_E}$.

The analysis of all terms in the equation for 
$\overline{k_S}$ (results not shown) reveal
that the external path ``$\overline{c_S}$'' and
``$- \overline{{B( \phi )}_S}$'' is the main feature
in the troposphere, with large and opposite values
leading to $\overline{c_S} - \overline{{B( \phi )}_S} 
\approx 0$. There is a catalytic behaviour for $\overline{k_S}$,
with a direct transfer of energy via the external path
$\overline{a_S} \leftrightarrow \overline{k_S}
\leftrightarrow \overline{\phi}$.
In the stratosphere, values for 
$\overline{{\partial}_t (k_S)}$, $\overline{B(k_{S})}$,
$\overline{B(k_{cS})}$, $\overline{c_{KS}}$
and $-\overline{d_{S}}$ are very small and
the balance of terms is somewhat different. The decrease of 
$\overline{k_S}$ close to the jet ($-8$ units) is explained 
by negative ageostrophic redistribution terms
$\overline{c_S} - \overline{{B( \phi )}_S} 
\approx -22$ units and by $\overline{B(k_{cS})} 
\approx +16$ units.
\begin{figure}[t]
\centering
\includegraphics[width=0.49\linewidth,angle=0,clip=true]{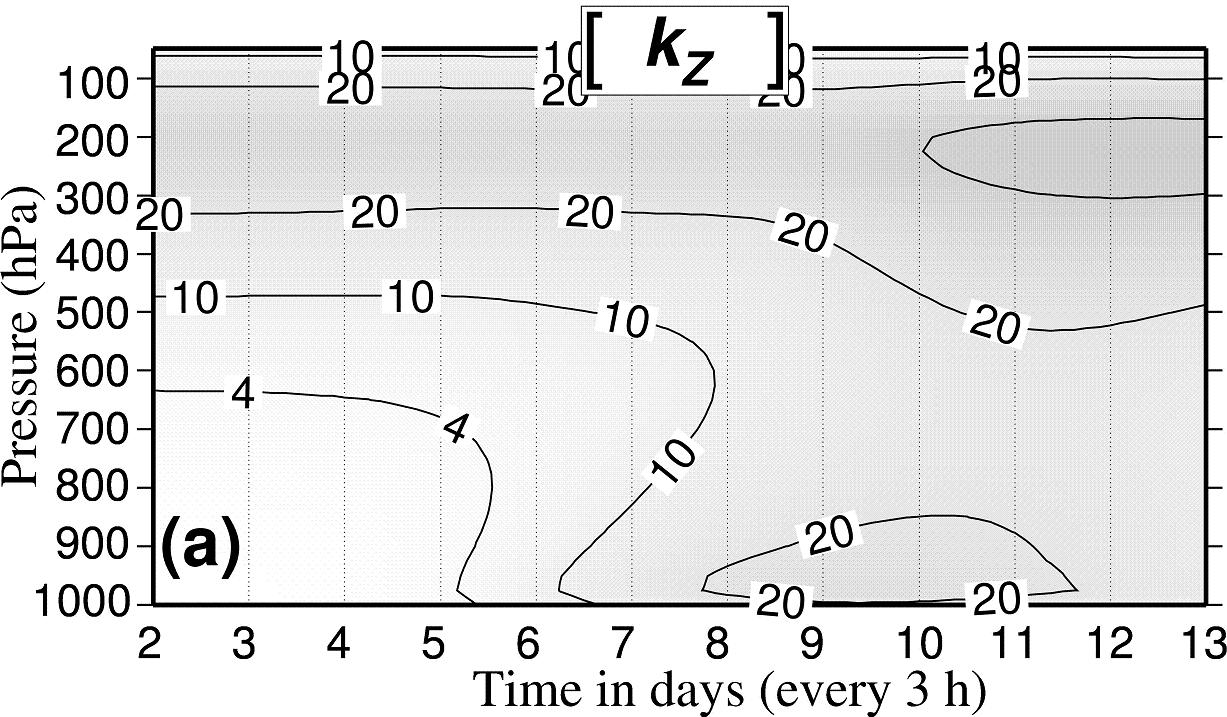}
\includegraphics[width=0.49\linewidth,angle=0,clip=true]{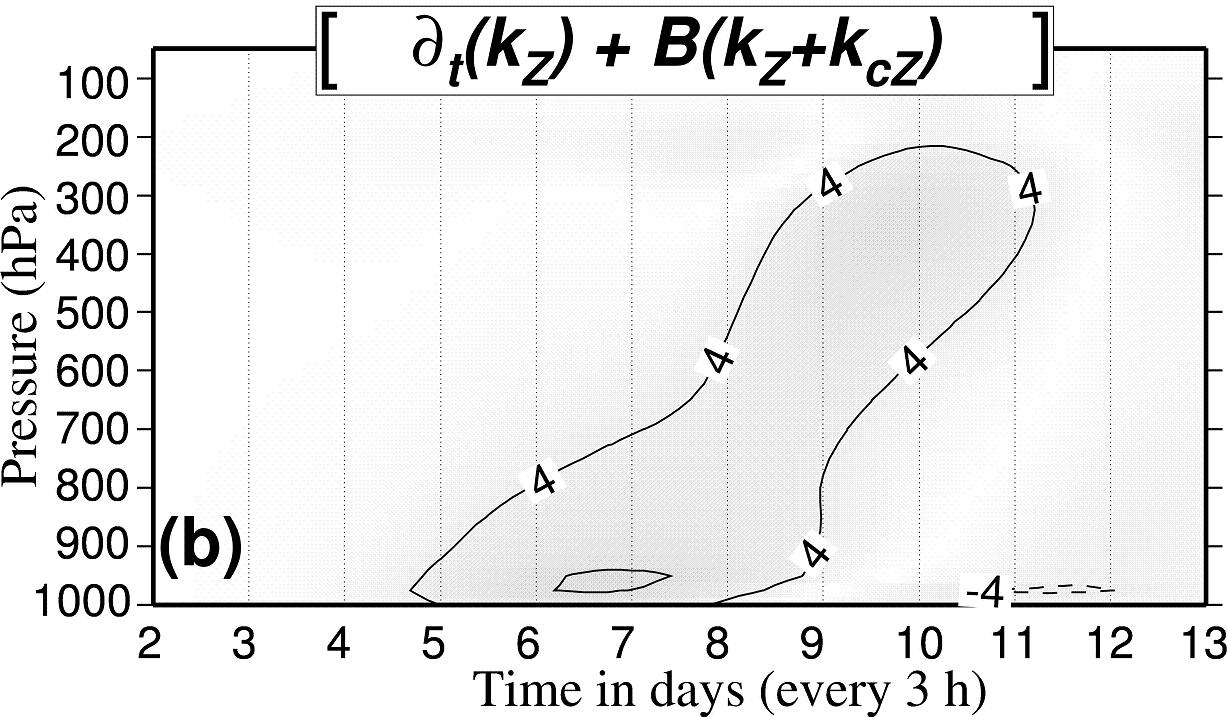}\\
\includegraphics[width=0.49\linewidth,angle=0,clip=true]{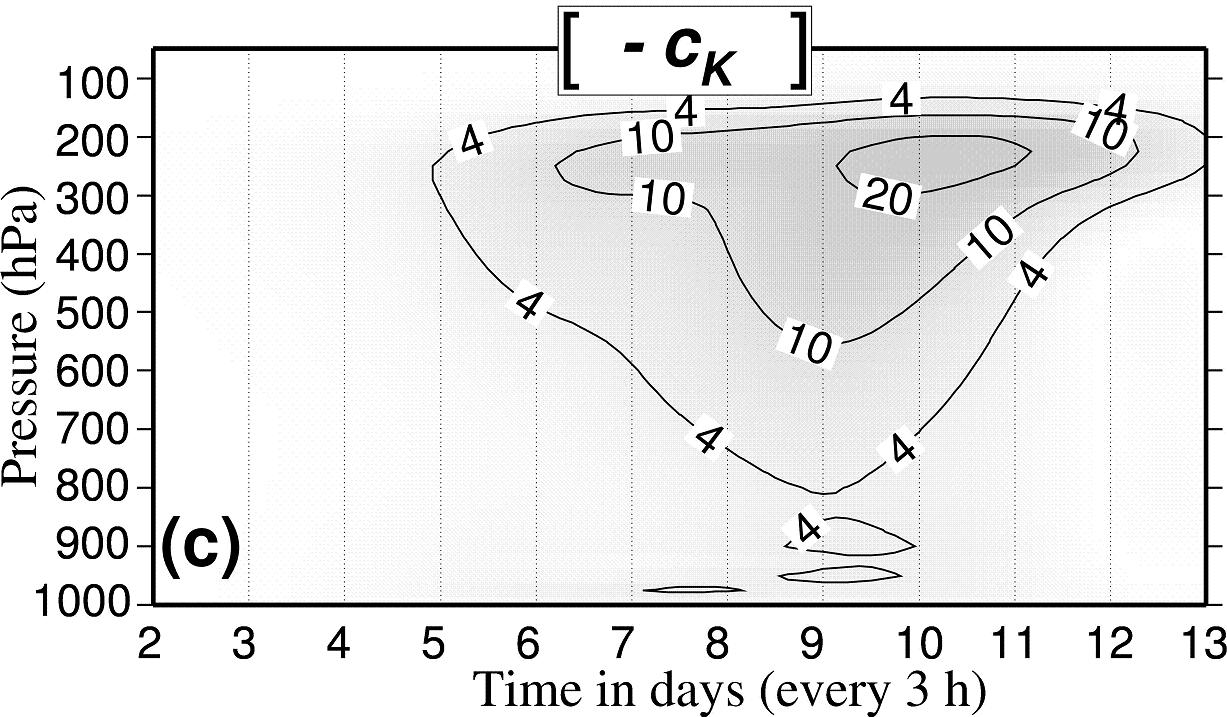}
\includegraphics[width=0.49\linewidth,angle=0,clip=true]{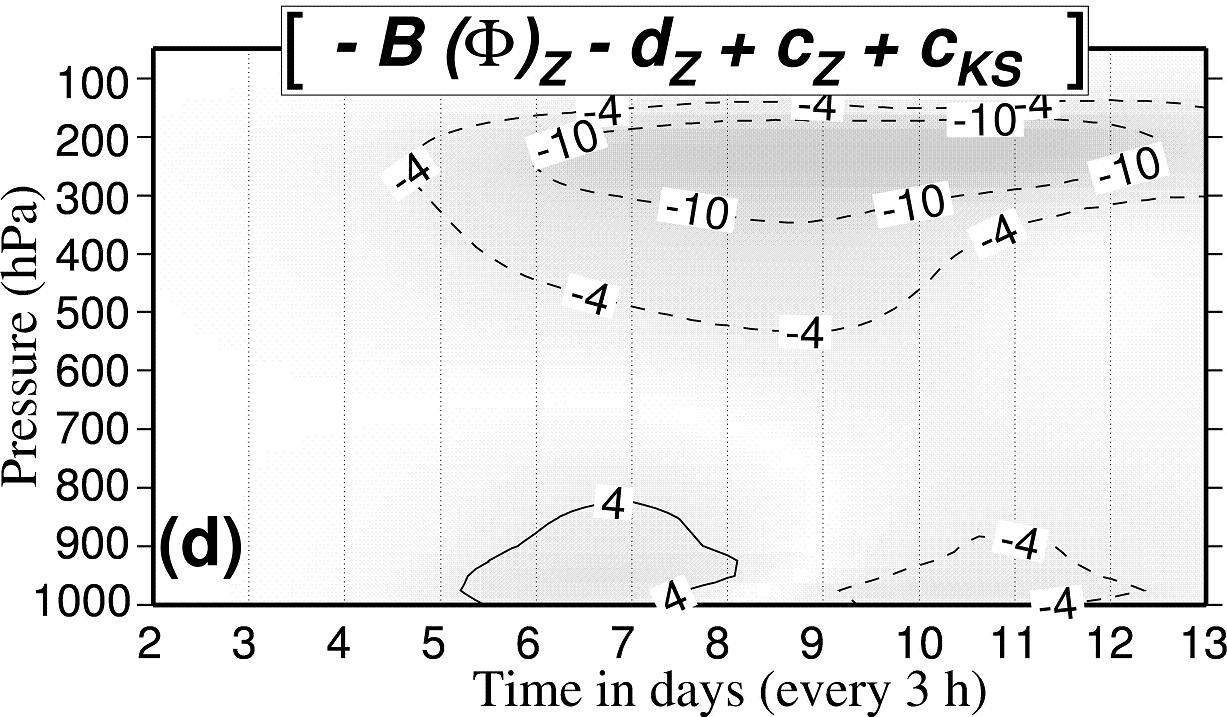}
\vspace*{-2mm}
\caption{\it \small 
As Fig.~\ref{FigEXP21a}, with the same large computational
domain and for the diabatic simulation EXP-HV, but for the 
zonal kinetic energy component $\overline{k_Z}$.
(a) The component $\overline{k_Z}$. 
(b) The total budget 
$\overline{{\partial}_t(k_Z)} + \overline{B(k_{Z}+k_{cZ})}$:
it is the sum of the local time derivative 
plus the divergence of two boundary fluxes.
The budget of $\overline{k_Z}$ writes (b)$\:=\:$(c)$\:+\:$(d).
(c) The usual barotropic conversion $-\overline{ c_K }$.
(d) The non-barotropic part of the budget for $\overline{k_Z}$,
equal to the sum $- \overline{{B( \phi )}_Z} - \overline{d_Z} 
+ \overline{c_Z}+ \overline{c_{KS}}$.
\label{Fig21kz}}
\end{figure}

      \subsection{Local results for $\overline{k_Z}$ 
      (diabatic simulation).} 
      \label{subsection_4.8}

The budget for the zonal component $\overline{k_Z}$ 
is depicted in Fig.~\ref{Fig21kz}~(a). 
Initial values of the zonal wind described 
in Fig.~{\ref{FigUTZON}}~(a) lead to maximum values 
for $\overline{k_Z}$ from $100$ to 
$300$~hPa, with small values in the lower troposphere. 
A comparison 
with the eddy component $\overline{k_Z}$ depicted in 
Fig.~\ref{FigEXP21a}~(a)
shows that maximum values appear $2$ days later in the boundary layer,
when $\overline{k_E}$ has reached its maximum or starts to decrease.
A delay of $3$ to $3.5$ days is also observed close to the 
jet. 
The total change in time of $\overline{k_Z}$ in Fig.~\ref{Fig21kz}~(b)
is computed including the two boundary terms $\overline{B(k_{Z})}$ and
$\overline{B(k_{cZ})}$, with (b)$\: = \:$(c)$\: + \:$(d).
The two divergence terms for Fig.~\ref{Fig21kz}~(b)
 are small with respect
to $\overline{{\partial}_t(k_Z)}$ (not shown). The total change
is maximum at day $7$ in the boundary layer ($950$~hPa) and it is
maximum at day $10$ just below the jet (from $300$ to $400$~hPa).
The delay observed in Fig.~\ref{Fig21kz}~(b) in comparison with
Fig.~\ref{FigEXP21a}~(b) is $1$ day in the boundary layer and $3$ 
days just below the jet.

The barotropic conversion $-\overline{c_K}$ in Fig.~\ref{Fig21kz}~(c)
can partly explain the growth of $\overline{k_Z}$ just below 
the jet, but the positive values for $-\overline{c_K}$ 
are too large and occur too early in the 
simulation.
They are also located at elevations that are too high  
(from $150$ to $300$~hPa).
In the boundary layer, the barotropic conversion cannot explain the 
growth of $\overline{k_Z}$ because there is no significant maximum in 
Fig.~\ref{Fig21kz}~(c) between days $6$ and $8$. 
As a result, the change in zonal kinetic 
energy component is improperly determined by the barotropic conversion 
term.

\begin{figure}[t]
\centering
\includegraphics[width=0.46\linewidth,angle=0,clip=true]{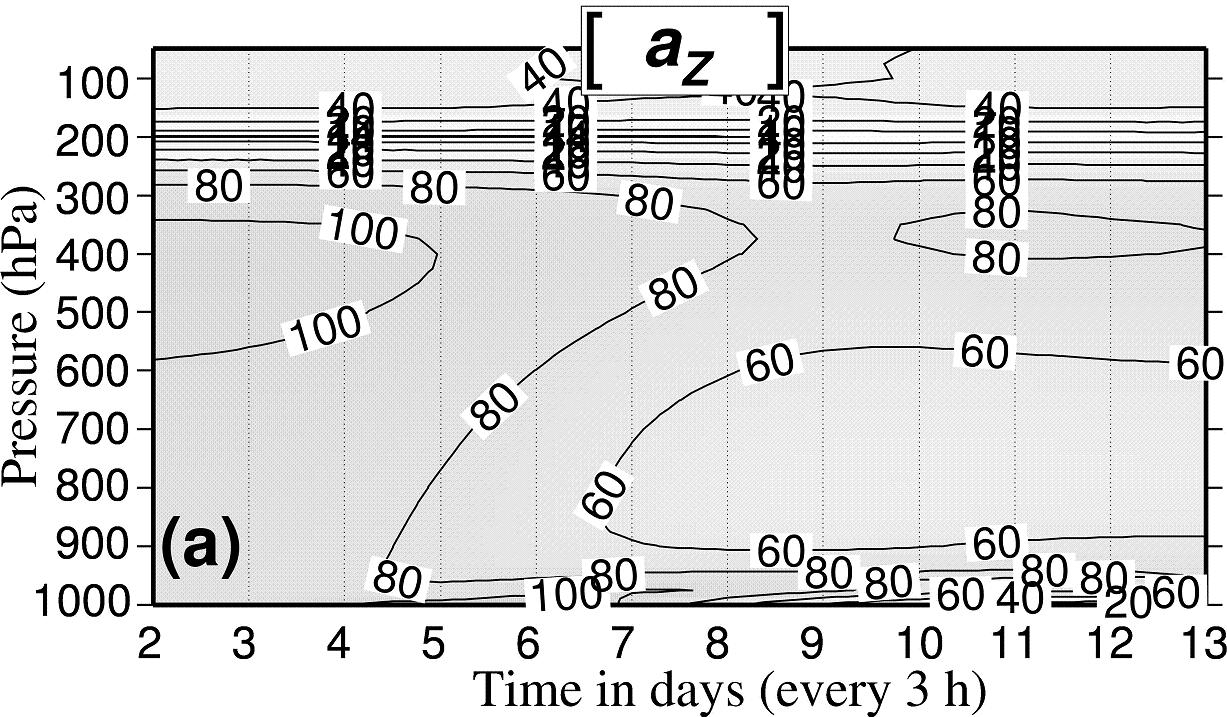}
\includegraphics[width=0.46\linewidth,angle=0,clip=true]{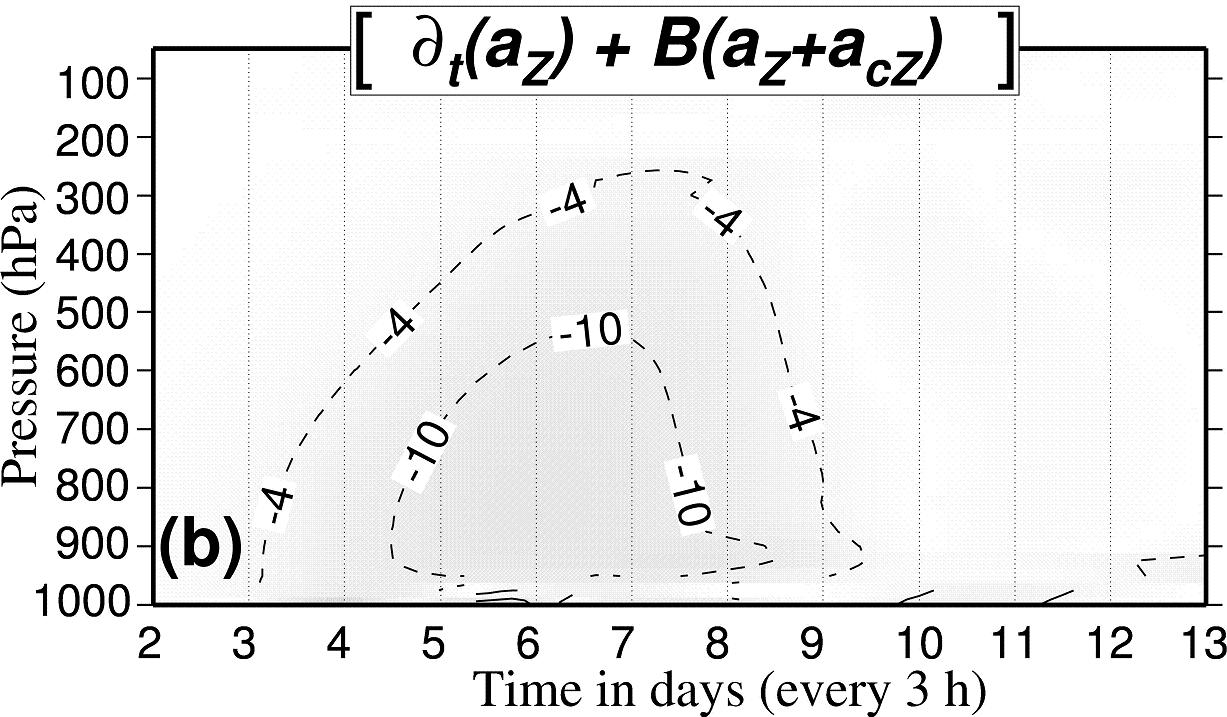}\\
\includegraphics[width=0.46\linewidth,angle=0,clip=true]{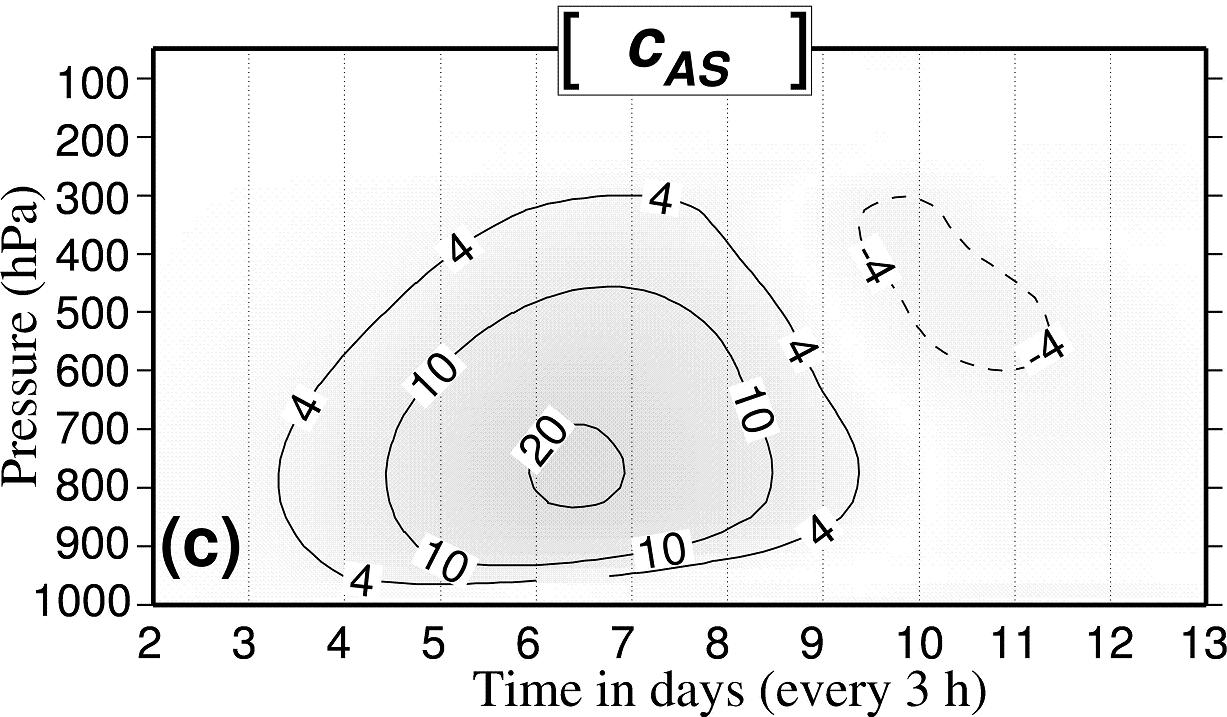}
\includegraphics[width=0.46\linewidth,angle=0,clip=true]{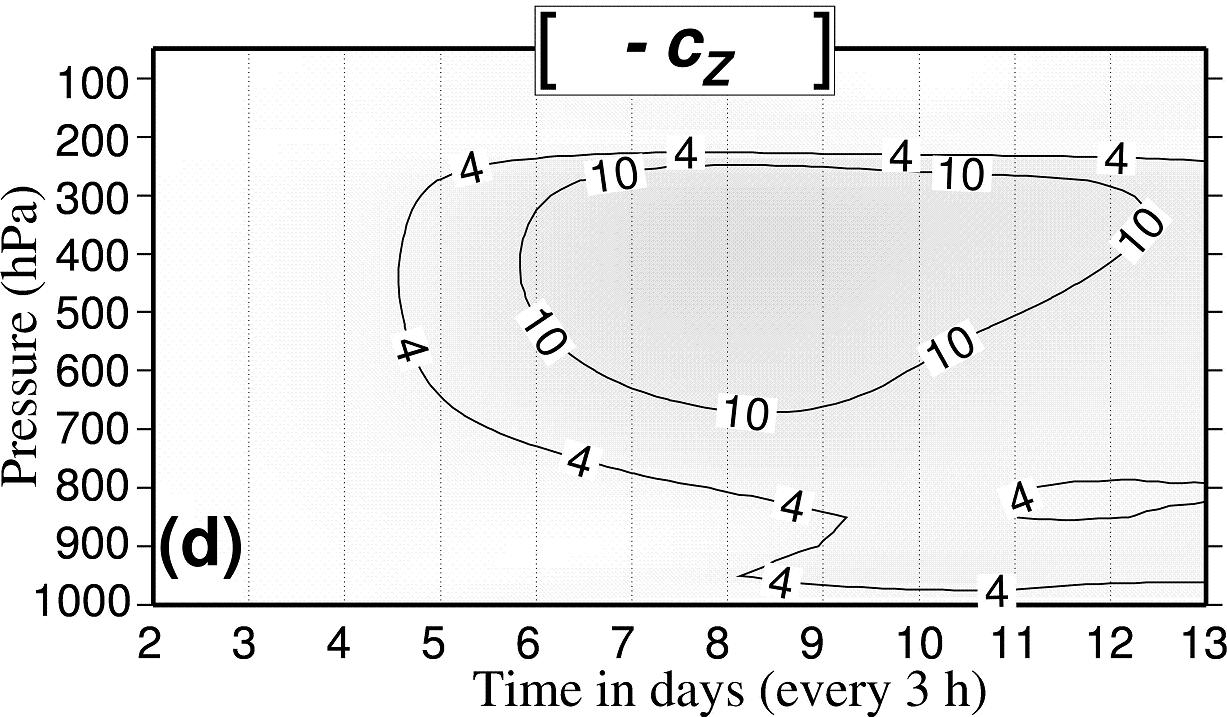}\\
\includegraphics[width=0.46\linewidth,angle=0,clip=true]{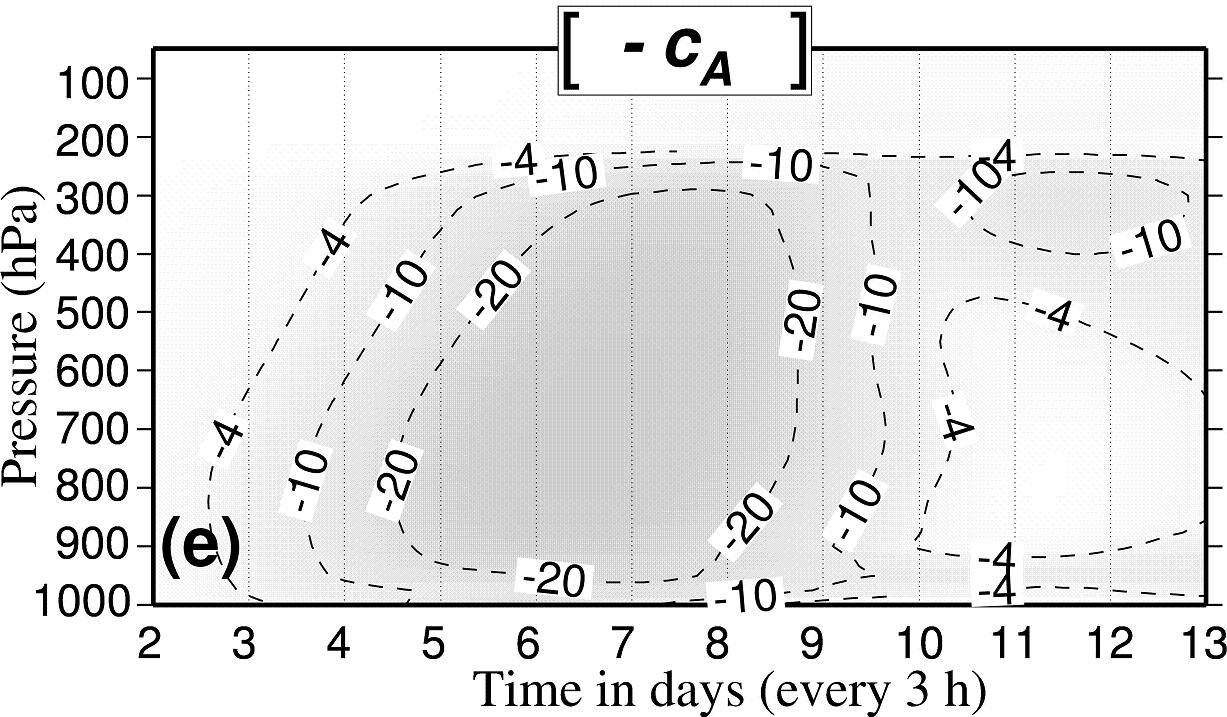}
\includegraphics[width=0.46\linewidth,angle=0,clip=true]{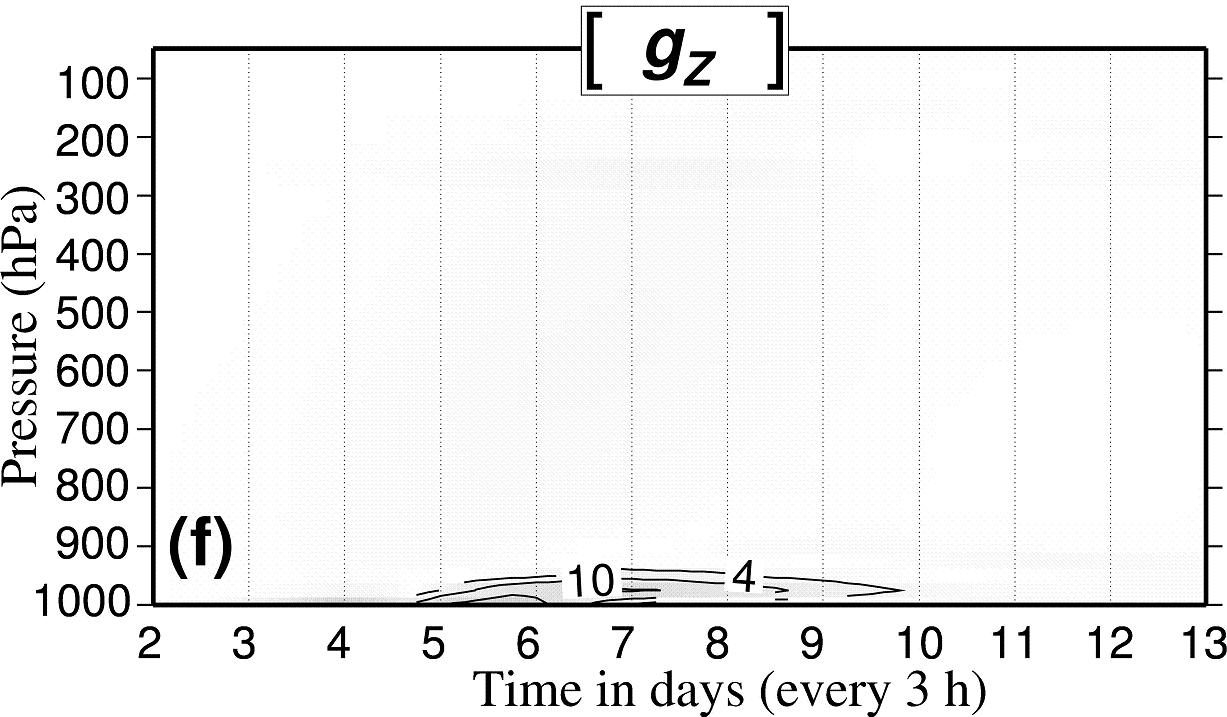}
\vspace*{-2mm}
\caption{\it \small 
As Fig.~\ref{Fig21kz}, with the same large computational
domain, with the same units and still for the diabatic simulation 
EXP-HV, but for the zonal available enthalpy component $\overline{a_Z}$.
The annotated isopleths are $\pm 4$, $\pm 10$, $\pm 20$,
$\pm 40$, $\pm 60$, $\pm 80$ and $\pm 100$.
(a) The component $\overline{a_Z}$. 
(b) The total budget of $\overline{a_Z}$ writes
$\overline{ {\partial}_t ( a_Z ) } + \overline{ B( a_{Z} )}
+ \overline{ B( a_{cZ} )}$:
it is the sum of the local time derivative 
plus the divergence of boundary fluxes. It 
is equal to (c)$\:+\:$(d)$\:+\:$(e)$\:+\:$(f).
(c) The conversion $\overline{c_{AS}}$.
(d) The conversion $-\overline{c_{Z}}$.
(e) The conversion $-\overline{c_{A}}$.
(f) The generation $\overline{g_Z}$. 
\label{Fig21az}}
\end{figure}

The same method used for $\overline{k_E}$ is applied to 
$\overline{k_Z}$ in order to define and give explanations
to a non-barotropic part for $\overline{{\partial}_t(k_Z)}$.
The dissipation $-\overline{d_Z}$, the conversion with the potential 
energy $-\overline{{B( \phi )}_Z}$ 
and conversions terms $\overline{c_Z}$ and $\overline{c_{KS}}$ 
are added altogether to form the non-barotropic term 
depicted in Fig.~\ref{Fig21kz}~(d). 
The conversion $\overline{c_{KS}}$ 
(not shown) is very small in comparison with other terms.
Large negatives values observed in 
Fig.~\ref{Fig21kz}~(d) for the
jet are due to $-\overline{{B( \phi )}_Z}$ (not shown). 
They correspond to positive values of $-\overline{c_K}$ for the 
jet, and the balance between these opposite terms explains 
why the growth in barotropic conversion is inhibited before day $8$ 
from $500$ to $150$~hPa. 
In the boundary layer, the non-barotropic 
term explains entirely the growth in $\overline{k_Z}$, due to a 
balance between large negative values for $-\overline{d_Z}$ and 
large positive values for $-\overline{{B( \phi )}_Z}$ 
(not shown). These large opposite terms close to the surface can
logically be explained by use of an idealized Ekman spiral, in a way
similar to what has been done for $\overline{k_E}$ in terms of
a balance between the zonal component of dissipation and the 
energetic impact of ageostrophic circulations in zonal average.

      \subsection{Local results for $\overline{a_Z}$ 
        (diabatic simulation).} 
      \label{subsection_4.9}

Figure~\ref{Fig21az}~(a) shows the zonal available enthalpy 
component $\overline{a_Z}$ where initial values
at day $2$ are large in the troposphere (more than $80$~J~{kg}${}^{-1})$,
with maximum values greater than $100$~J~{kg}${}^{-1}$
below the jet (from $350$ to $600$~hPa) corresponding to strong 
north/south gradients of ${T}^{\lambda}$ and to large values for
${T}^{\lambda}_{\varphi}$ in Fig.~\ref{FigUTZON}~(a).
\begin{figure}[t]
\centering
\includegraphics[width=0.46\linewidth,angle=0,clip=true]{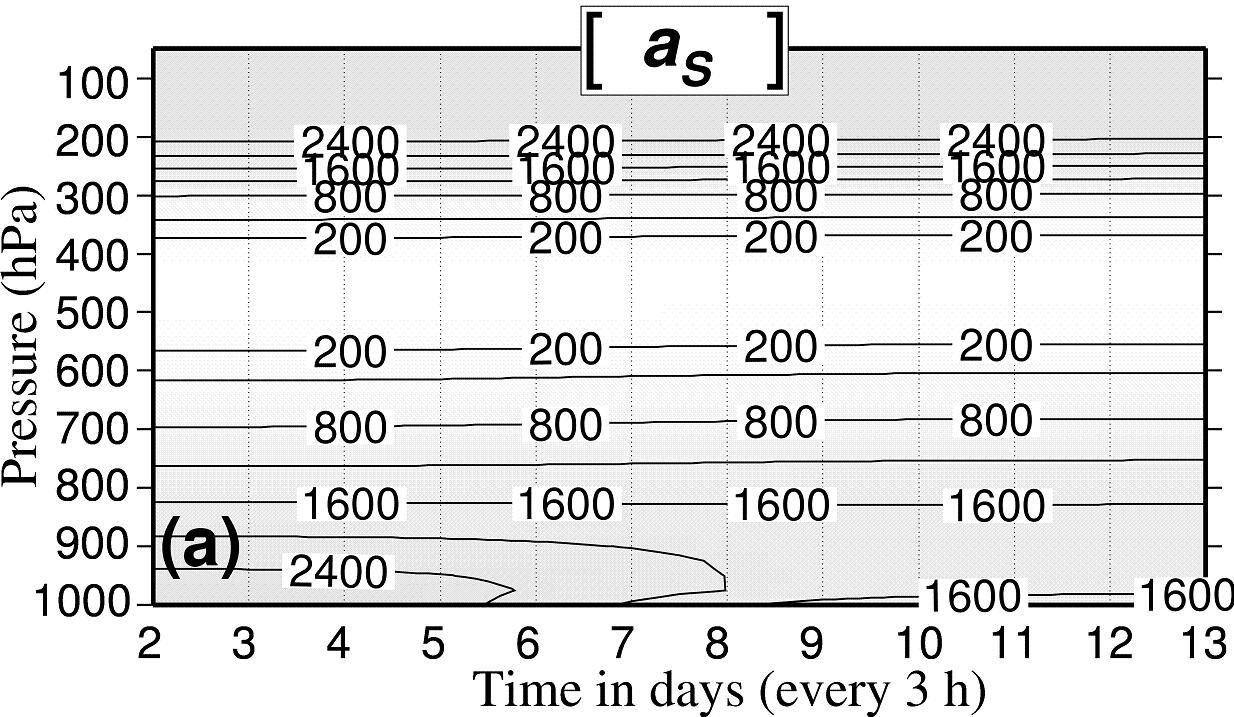}
\includegraphics[width=0.46\linewidth,angle=0,clip=true]{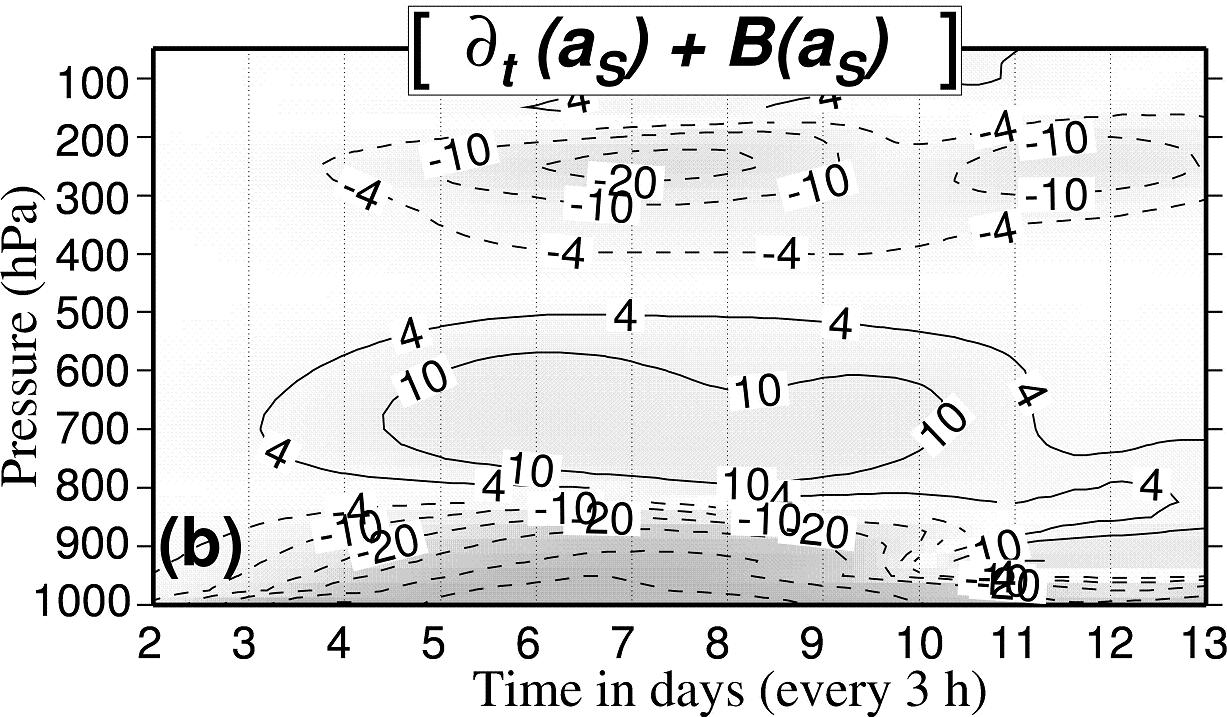}\\
\includegraphics[width=0.46\linewidth,angle=0,clip=true]{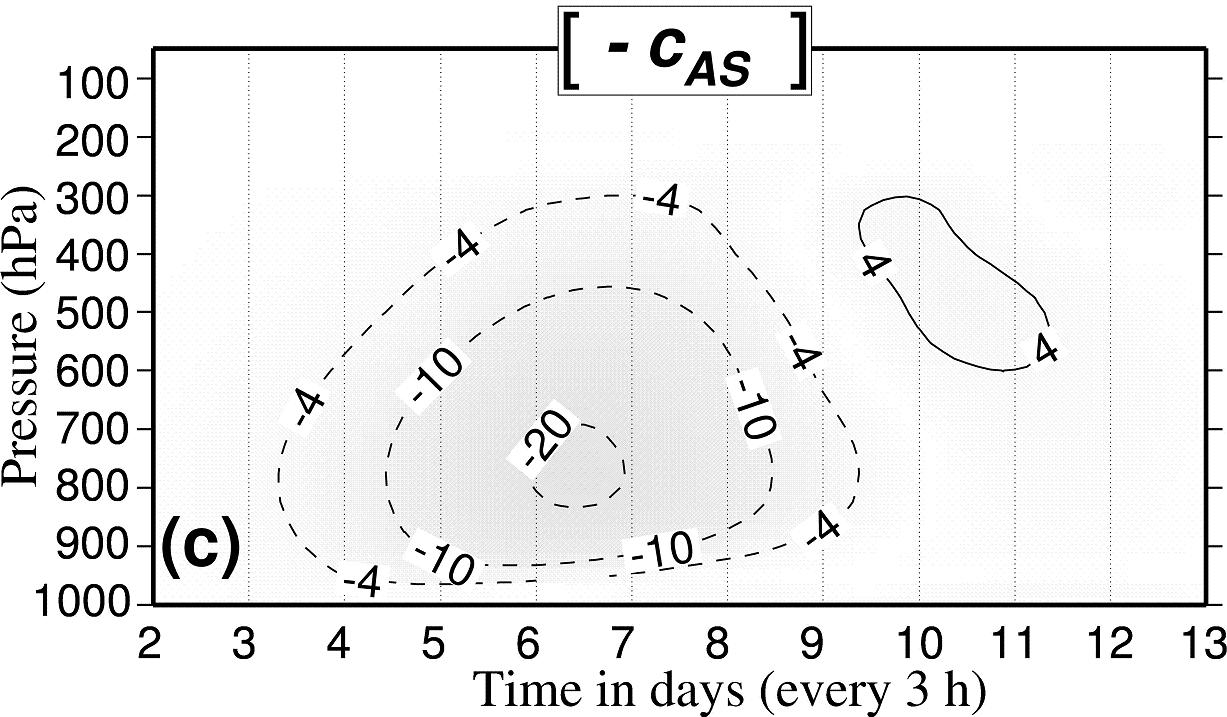}
\includegraphics[width=0.46\linewidth,angle=0,clip=true]{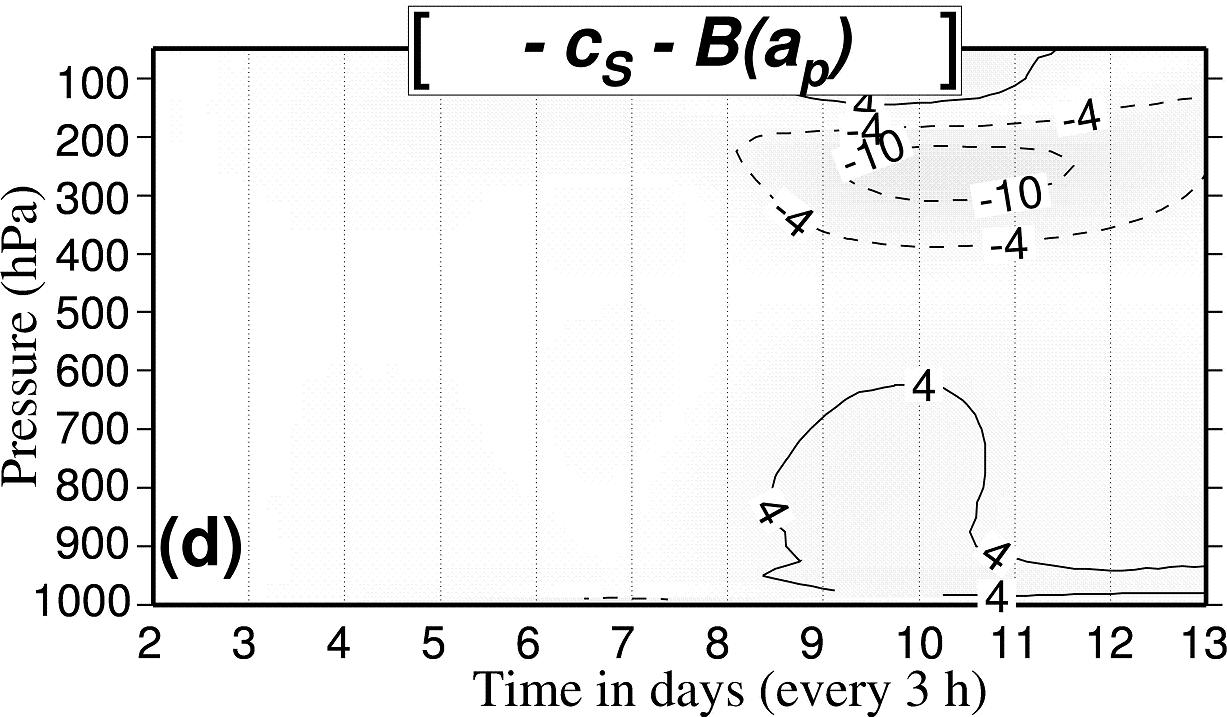}\\
\includegraphics[width=0.46\linewidth,angle=0,clip=true]{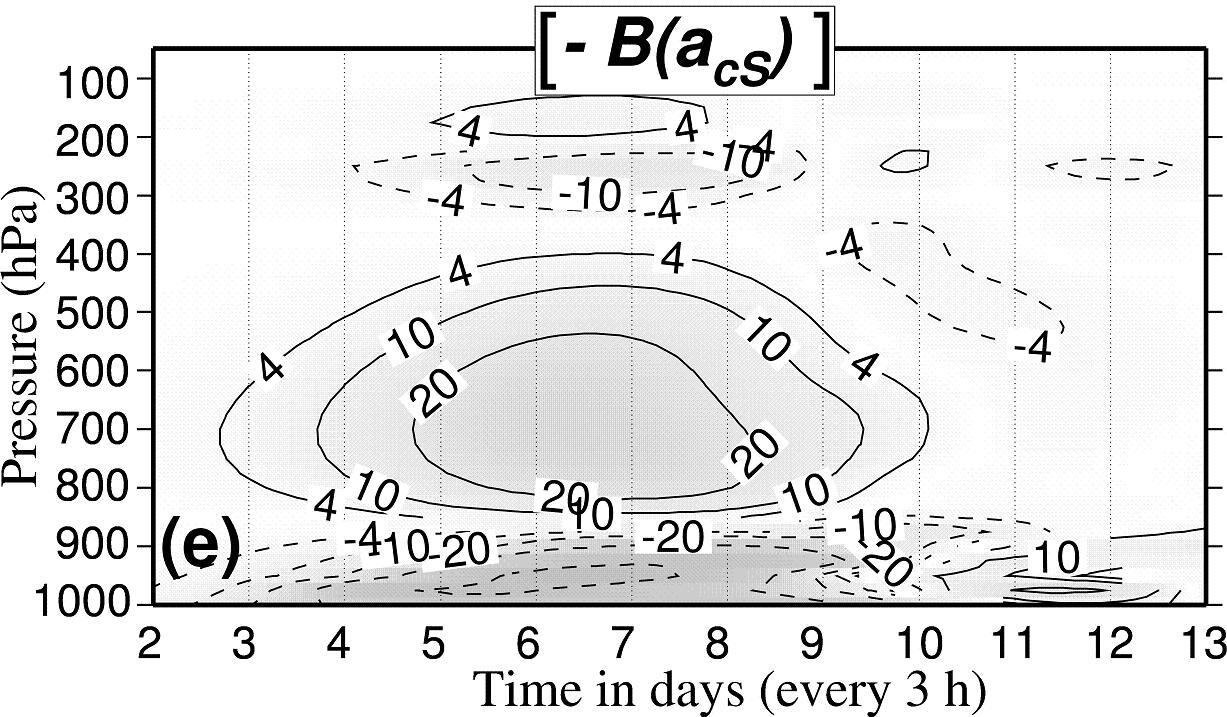}
\includegraphics[width=0.46\linewidth,angle=0,clip=true]{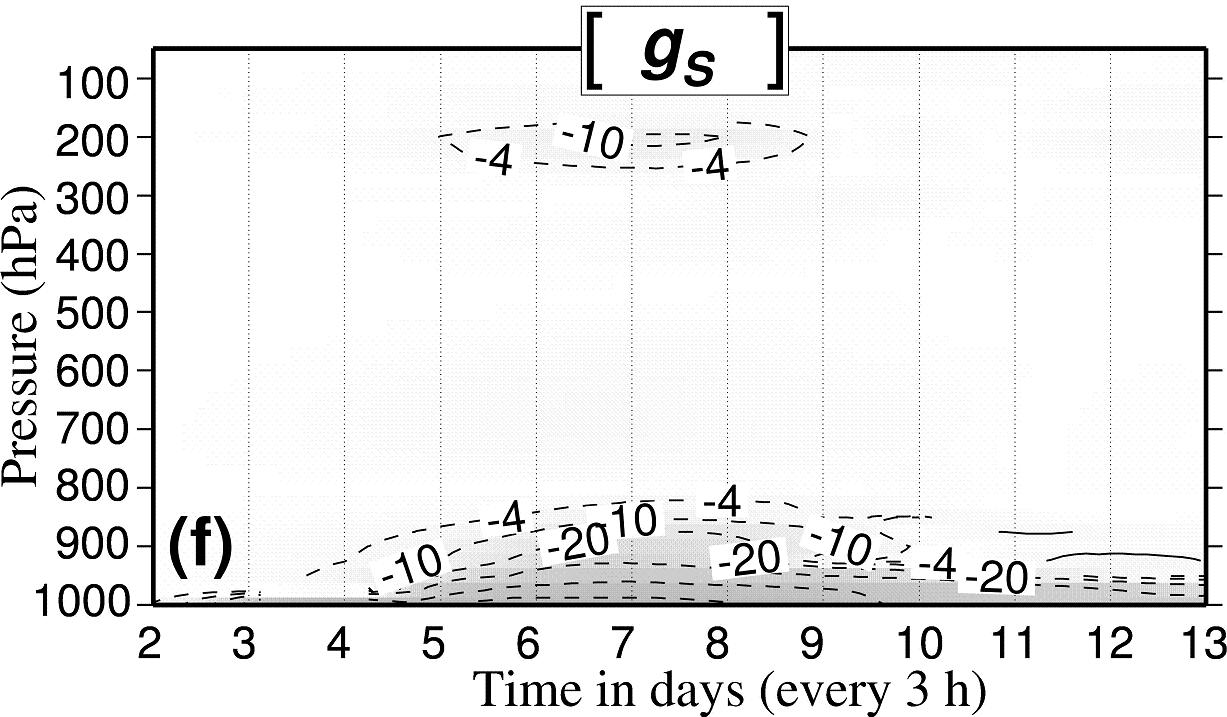}
\vspace*{-2mm}
\caption{\it \small 
As Fig.~\ref{Fig21kz}, with the same large computational
domain, with the same interval and units in (b) to (f) 
and still for the diabatic simulation EXP-HV, 
but for the static stability component $\overline{a_S}$.
Isopleths in (a) are $200$, $400$, $800$,
$1200$, $1600$, $2000$ and $2400$ J~{kg}${}^{-1}$.
For the other panels the annotated isopleths are 
$\pm 4$, $\pm 10$ and $\pm 20$, followed by the 
contoured isopleths $\pm 40$, $\pm 100$ and 
$\pm 200$~$10^{-5}$~W~{kg}${}^{-1}$.
The budget of $\overline{a_S}$ 
corresponds to (b)$\:=\:$(c)$\:+\:$(d)$\:+\:$(e)\-$\:+\:$(f).
(a) The component $a_S$. 
(b) The total budget 
$\overline{ {\partial}_t ( a_S ) } + \overline{ B( a_{S} )}
+ \overline{ B( a_{cS} )}$:
it is the sum of the local time derivative 
plus the divergence of boundary fluxes
(c) The conversion $-\overline{c_{AS}}$.
(d) The external path budget $-\overline{c_{S}}-\overline{B(a_p)}$.
(e) The boundary term $-\overline{B(a_{cS})}$.
(f) The generation $+\overline{g_S}$.
The numerous isopleths located near the surface in (b), (e)
and (f) are not artefacts. 
They correspond to real large positive or negative 
values discussed in the text.
\label{Fig21as}}
\end{figure}

The total change in $\overline{a_Z}$ is depicted in 
Fig.~\ref{Fig21az}~(b), including the boundary fluxes of
$a_Z$ and $a_{cZ}$ (they cannot be neglected in the boundary layer,
not shown). 
The total change in Fig.~\ref{Fig21az}~(b) 
is negative in the growing stage of 
the unstable mode from days $3.5$ to $8.5$, with a 
maximum depletion at day $6$ in the lower troposphere 
(from $600$ to $800$~hPa). 
The total change is zero at 
the top of the boundary layer ($920$~hPa). There is a
positive event at $1000$~hPa between days $4.5$ and $7.5$
(more clearly shown with a zoom over this region, not shown here). 
It is created by a positive part of the generation $\overline{g_Z}$
in Fig.~\ref{Fig21az}~(f) where values up to $25$~$10^{-5}$~W~{kg}${}^{-1}$ 
are observed from days $5$ to $9.5$ and between $1000$ and $950$~hPa.

Associated with negative tendencies in Fig.~\ref{Fig21az}~(b) 
are minimum values for $\overline{a_Z}$ in 
Fig.~\ref{Fig21az}~(a) after day $8$, 
with values less than $70$~J~{kg}${}^{-1}$ from $500$ to 
$900$~hPa. The behaviour of $\overline{a_Z}$ close to the 
surface is more complex. Values are as low as $20$~J~{kg}${}^{-1}$ 
at $1000$~hPa and after day $8$, with a region of moderate values 
(more than $70$~J~{kg}${}^{-1}$) observed just at the top of the
boundary layer (between $975$ and $875$~hPa, best shown with
a zoom over this region, not shown here).

It is explained in section 4.2 how global increases in
$K_E$ and $K_Z$ are obtained at the expense of global 
components $A_Z$ or $A_S$. 
However, there is no 
one-to-one local connection between $\overline{a_Z}$ or $a_S$ 
with $\overline{k_E}$ or $\overline{k_Z}$. The patterns
are clearly out of phase because maxima are located at the surface 
and near the jet for kinematic components, whereas maxima
occur in the middle troposphere for the temperature 
components. 

The decrease in $\overline{a_Z}$ can be explained by the balance between
the positive conversion $\overline{c_{AS}}$ and the negative convection
$-\overline{c_{A}}$ in Figs.~\ref{Fig21az}~(c) and (e).
The positive conversion $-\overline{c_{Z}}$ exports energy
from $\overline{k_Z}$ to $\overline{a_Z}$ but it does not correspond 
to change in $\overline{a_Z}$. In fact, it is connected to associated 
negative values for $-\overline{c_{A}}$. For the jet and the
middle troposphere, $-\overline{c_{Z}}$ and $-\overline{c_{A}}$
act as direct transfers of energy from $\overline{k_Z}$ to 
$\overline{a_E}$.

      \subsection{Local results for $\overline{a_S}=a_S$ 
                  (diabatic simulation).} 
      \label{subsection_4.10}

The static stability component $a_S$ in Fig.~\ref{Fig21as}~(a)
is maximum in the lower troposphere and in the stratosphere, 
with minimum values near the  $500$~hPa level.
These observations are in agreement with local values of 
$(\overline{T}-T_r){}^2 \propto a_S $ and with $T_r \equiv 250$~K.

The equation for $a_S$  is equivalent in Fig.~\ref{Fig21as} to 
 (b)$\:=\:$(c)$\:+\:$(d)$\:+\:$(e)$\:+\:$(f).
The development of the mode is mainly associated in 
Fig.~\ref{Fig21as}~(a) with a 
decrease of $a_S$ below the level $800$~hPa. It corresponds
in Fig.~\ref{Fig21as}~(b) to large negative values for the total change in 
$a_S$, including the boundary fluxes.
Associated with them, large negative values for the generation 
$\overline{g_S}$ in Fig.~\ref{Fig21as}~(f) 
are the main cause for the loss of energy
in the system, with observed decreasing gradients of temperature 
below the  $800$~hPa level.

Clearly, possible one to one local conversions between 
$a_S$ and $\overline{k_E}$ or $\overline{k_Z}$, 
suggested in section 4.2 from global results, cannot be 
established.

Values for the budget $-\overline{c_S}-\overline{B(a_p)}$ 
are small in Fig.~\ref{Fig21as}~(d). 
It is a confirmation that the external path 
does not contribute greatly to the energetics of the 
available-enthalpy cycle.
As mentioned in the previous section, the conversion 
$-\overline{c_{aS}}$ in 
Fig.~\ref{Fig21as}~(c) can explain the growth in 
$\overline{a_Z}$, although it cannot describe the pattern
of the total change in 
Fig.~\ref{Fig21as}~(a). It means that the other
boundary flux $-\overline{B(a_{cS})}$ is an important
feature for $a_S$, acting as vertical redistributions 
of energy with large positive and negative values but
with small global integral (not shown).

 \section{\Large \underline{A new $A3+K3+\phi$ available 
          enthalpy cycle}.} 
 \label{section_5}

Investigations of the energetics of the diabatic simulation have
shown that observed large and positive values for dissipation
in the boundary layer correspond to large and opposite values 
for the three potential energy conversion terms 
$-\overline{{B(\phi)}_S}$, $-\overline{{B(\phi)}_Z}$ and 
$-\overline{{B(\phi)}_E}$.
An attempt will be presented in this section to modify
(36) and Fig.~5~(b) of Part~I in order to
take into account these balanced terms which are interpreted, 
according to OS95 and section 4.6, as vertical redistributions 
of energy via work done by pressure forces and by ageostrophic 
circulations.

Ageostrophic conversions are denoted by $\overline{(c_{ag}){}_X}$,
for subscripts $X=(S,Z,E)$. They are written
\begin{eqnarray}
  \overline{ (c_{ag}){}_S } 
        \! \!  & = &  \! \! 
           \overline{c_S} \:
      - \: \overline{{B( \phi )}_S}
        \; = \;
      - \: \overline{{\bf U}_h} \: . \: 
           \overline{{\bf \nabla}_{\!p} \, \phi } \mbox{\hspace{1.0 cm}}
        \; = \;
              {\bf k} \: . \: 
                       \overline{(f \: {\bf U}_g)}
               \times  \overline{(  {\bf U}_a)} \: ,
        \label{eq:cagterms1}
        \vspace*{-0.4 cm} \\
  \overline{ (c_{ag}){}_Z } 
        \! \!  & = &  \! \! 
           \overline{c_Z}  \:
      - \: \overline{{B( \phi )}_Z}
        \; = \;
      - \: \overline{
            {({\bf U}_h)}^{\lambda}_{\varphi} \: . \: 
            {({\bf \nabla}_{\!p} \,\phi)}^{\lambda}_{\varphi} }
        \; = \;
              {\bf k} \: . \: 
              \overline{ {(f \: {\bf U}_g)}^{\lambda}_{\varphi}
                       \times {({\bf U}_a)}^{\lambda}_{\varphi} 
                       } \: ,
        \label{eq:cagterms2}
        \vspace*{-0.4 cm} \\
  \overline{ (c_{ag}){}_E } 
        \! \!  & = &  \! \! 
           \overline{c_E}  \:
      - \: \overline{{B( \phi )}_E}
        \; = \;
      - \: \overline{{({\bf U}_h)}_{\lambda} \: . \: 
           {({\bf \nabla}_{\!p} \, \phi)}_{\lambda}}
        \; = \;
              {\bf k} \: . \: 
              \overline{ f \: {({\bf U}_g)}_{\lambda}
                       \times {({\bf U}_a)}_{\lambda} 
                       } \: .
\label{eq:cagterms3}
\end{eqnarray}

The results obtained in sections 4.3, 4.4 and 4.8 show 
that local changes in kinetic-energy components 
are out of phase with baroclinic conversions, whereas they 
are in one-to-one relationships with ageostrophic conversions.
It is thus necessary to reorganize (36) 
and Fig.~5~(b) of Part~I so that the 
barotropic conversions $\overline{c_E}$, $\overline{c_Z}$ 
and $\overline{c_S}$ do not directly supply energy to
$\overline{k_E}$, $\overline{k_Z}$ and $\overline{k_S}$. 
Discussions presented in Johnson and Downey (1982) are
suitable to solve this problem. They suggest maintaining an
explicit degree of freedom for boundary work and to avoid
concept of direct conversion between mechanical and thermodynamic
energies. This program will be partly retained, leading to
modifications in the Lorenz internal cycle (encompassing a
hatched area in Fig.~5~(b) of Part~I).

\begin{figure}[t]
\centering
\includegraphics[width=0.97\linewidth,angle=0,clip=true]{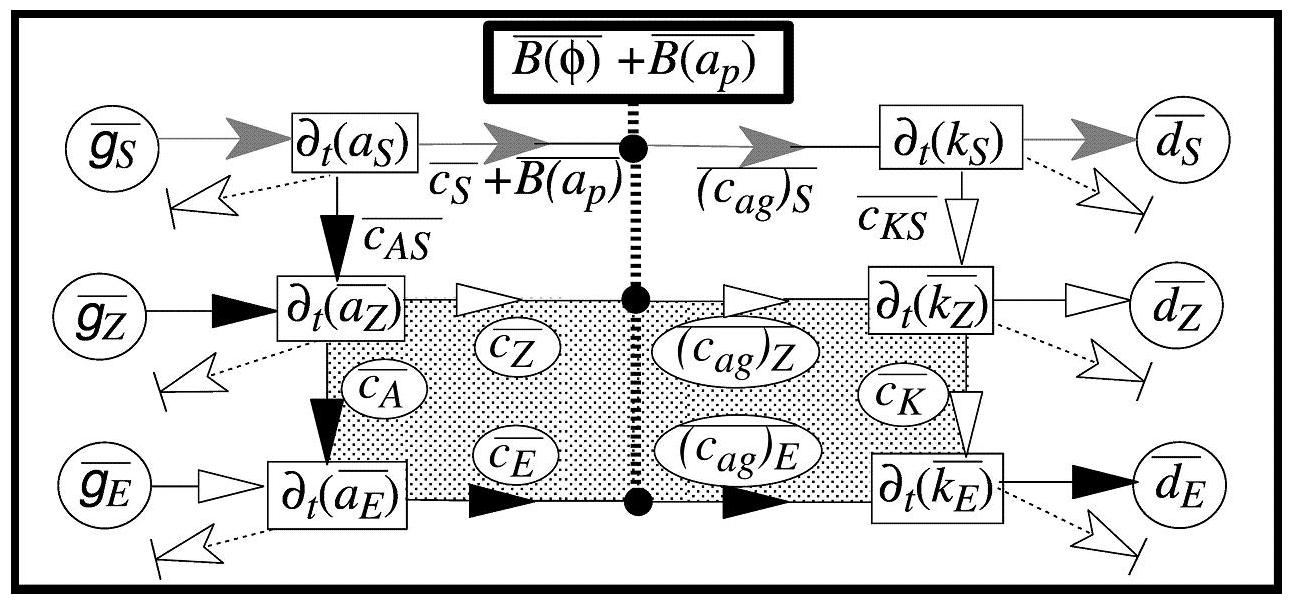}
\vspace*{-2mm}
\caption{\it \small 
A new proposal for the limited-area enthalpy cycle.
Conversion terms with potential energy 
$-\overline{{B(\phi)}_S}$, $-\overline{{B(\phi)}_Z}$
and $-\overline{{B(\phi)}_E}$ in Fig.~5~(b) of Part~I are
combined with $\overline{c_S}$, $\overline{c_Z}$ and $\overline{c_E}$
to form ageostrophic conversions (\ref{eq:cagterms1}) to
(\ref{eq:cagterms3}). The potential-energy equation is inserted between
available-enthalpy and kinetic-energy components. It is depicted
by vertical heavy dotted lines and dark points and the associated 
non-partitioned generation term is 
$\overline{B{(\phi)}} + \overline{B(a_p)}$.
The six non-labelled outgoing white arrows represent
the boundary fluxes for each of the six energy 
components. The formulations are given in the first
terms on the right-hand sides of (\ref{eq:cycle2new})).
\label{FigAHNEWCYCLEb}}
\end{figure}

The proposal for the new limited-area available-enthalpy
cycle is represented by (\ref{eq:cycle2new}) and
Fig.~\ref{FigAHNEWCYCLEb}.
\begin{equation}
\left.
\begin{aligned}
    \mbox{\hspace*{0.7 cm}} \overline{ {\partial}_t ( a_S )  }  
      \:  & =   \:
            \; - \; \overline{  B( a_S + a_{cS} ) } 
            \; - \; \overline{  c_{AS} } 
            \; - \; \{ \: \overline{  c_S  } + \overline{  B(a_{p}) }  \: \}
                  \mbox{\hspace{1.15 cm}}
            \; + \; \overline{g_S} \vspace*{-0.4 cm} \\
    \mbox{\hspace*{0.7 cm}} \overline{ {\partial}_t ( a_Z ) } 
       \:  & =   \:
            \; - \; \overline{  B( a_Z + a_{cZ} ) } 
            \; + \; \overline{  c_{AS}  }
            \; - \; \overline{  c_Z  }       \mbox{\hspace{0.1 cm}}
            \; - \; \overline{  c_A  }       \mbox{\hspace{1.86 cm}}
            \; + \; \overline{g_Z} \vspace*{-0.4 cm} \\
    \mbox{\hspace*{0.7 cm}} \overline{ {\partial}_t ( a_E ) } 
      \:  & =   \:
            \; - \; \overline{  B( a_{E} ) } \mbox{\hspace{2.3 cm}}
            \; - \; \overline{  c_E  }       \mbox{\hspace{0.1 cm}}
            \; + \; \overline{  c_A  }       \mbox{\hspace{1.86 cm}}
            \; + \; \overline{g_E} \vspace*{-0.4 cm} \\
    \overline{B{(\phi)}} + \overline{B{(a_p)}}
      \:  & =   \:
            \: + \: \{ ( \overline{ c_S  } + \overline{B{(a_p)}} )+
                       \overline{ c_Z  } +
                       \overline{ c_E  } \}
            \: - \: \{ \overline{ (c_{ag}){}_S } + 
                       \overline{ (c_{ag}){}_Z } +
                       \overline{ (c_{ag}){}_E } \}
            \vspace*{-0.4 cm} \\
    \mbox{\hspace*{0.7 cm}} \overline{ {\partial}_t ( k_S ) } 
      \:  & =   \:
            \; -  \; \overline{ B( k_{S} + k_{cS} )} 
            \; -  \; \overline{ c_{KS} }       \mbox{\hspace{0.05 cm}}
            \; + \; \overline{ (c_{ag}){}_S } \mbox{\hspace{2.4 cm}}
            \; -  \; \overline{d_S} \vspace*{-0.4 cm} \\
    \mbox{\hspace*{0.7 cm}} \overline{ {\partial}_t ( k_Z ) }  
      \:  & =   \:
            \; - \; \overline{ B( k_Z + k_{cZ} )} 
            \; + \; \overline{ c_{KS} }
            \; + \; \overline{ (c_{ag}){}_Z } 
            \; - \; \overline{ c_K }          \mbox{\hspace{1.3 cm}}
            \; - \; \overline{d_Z} \vspace*{-0.4 cm} \\
    \mbox{\hspace*{0.7 cm}} \overline{ {\partial}_t ( k_E ) } 
      \:  & =   \:
            \; - \; \overline{ B( k_{E} )}   \mbox{\hspace{2.35 cm}}
            \; + \; \overline{ (c_{ag}){}_E }           
            \; + \; \overline{ c_K }         \mbox{\hspace{1.3 cm}}
            \; - \; \overline{d_E}
\end{aligned}
\; \;
\right\} \; .
\label{eq:cycle2new}
\end{equation}

Ageostrophic conversions $\overline{(c_{ag}){}_X }$ for subscripts $X=(S,Z,E)$
appear in the three kinetic-energy equations in
(\ref{eq:cycle2new}). They also appear in the equation for 
$\overline{B(\phi})$ which is partitioned into the three 
incoming terms $\overline{c_X}$ and
the three outgoing terms $\overline{ (c_{ag}){}_X }$,
plus the term $\overline{B(a_p)}$ added to the two sides of 
the equation. These transformations between baroclinic 
and ageostrophic components are equivalent to the vertical 
redistribution of energy described in OS95.

However, the question as to how values $\overline{c_X}$
are transformed into values $\overline{ (c_{ag}){}_X }$ will not be
tackled in this paper, owing to a lack of possible separation of 
$\overline{\phi}$ into zonal, eddy and static-stability components.
Indeed, $\overline{\phi}$ is not a quadratic function and the local
separation $\phi = \overline{\phi} + 
{\phi}^{\lambda}_{\varphi} + {\phi}_{\lambda}$
disappears for averaged values, because
$\overline{{\phi}^{\lambda}_{\varphi}}=0$
and $\overline{{\phi}_{\lambda}}=0$.
In fact, this question is equivalent to the other
difficulty in understanding the real physical 
meanings for $- \overline{{B( \phi )}_X}$ in
Eqs.(36) of Part~I.

Graphically, Fig.~\ref{FigAHNEWCYCLEb} is obtained from
Fig.~5~(b) of Part~I by folding back $\overline{{B( \phi )}_X}$ 
arrows to $\overline{c_X}$ arrows on the right part of the inner 
Lorenz cycle. The left part of the Lorenz cycle is unchanged and
still involves the usual baroclinic conversions. The vertical separation 
between the two parts corresponds to the equation for $\phi$, depicted by
vertical heavy dotted lines and dark branching points. 
The external path of energy is also modified. The boundary term 
$\overline{B( a_p )}$ is added to $\overline{c_S}$ in order to
avoid large terms in the budget of $\overline{a_S}$.

As suggested by Johnson and Downey (1982), connections between 
potential-energy components and others components do not occur through direct 
conversions toward kinetic-energy components. Energy coming from $\phi$ 
rather enters in the middle of horizontal branches of the internal 
Lorenz cycle in Fig.~\ref{FigAHNEWCYCLEb} (dark points). 
Transformations leading to (\ref{eq:cycle2new}) are obtained without loss of generality and
a gain in simplicity is observed, since there are fewer terms to manage 
in the kinematic part.

 \section{\Large \underline{An application to the IOP15 
          of FASTEX}.} 
 \label{section_6}

      \subsection{Time--Pressure diagrams 
       for IOP15.} 
      \label{subsection_6.1}

It is not the aim of this section to show a complete study of the
energetics of IOP15 during the FASTEX experiment (Joly {\it et al.}, 
1997). Only preliminary results will be presented, in order to assess 
the realistic aspect of the idealized diabatic simulation EXP-HV. 
Both FASTEX and EXP-HV cases will be investigated with the
local available-enthalpy diagnostic package, though with a different 
limited-area domain.

\begin{figure}[t]
\centering
\includegraphics[width=0.49\linewidth,angle=0,clip=true]{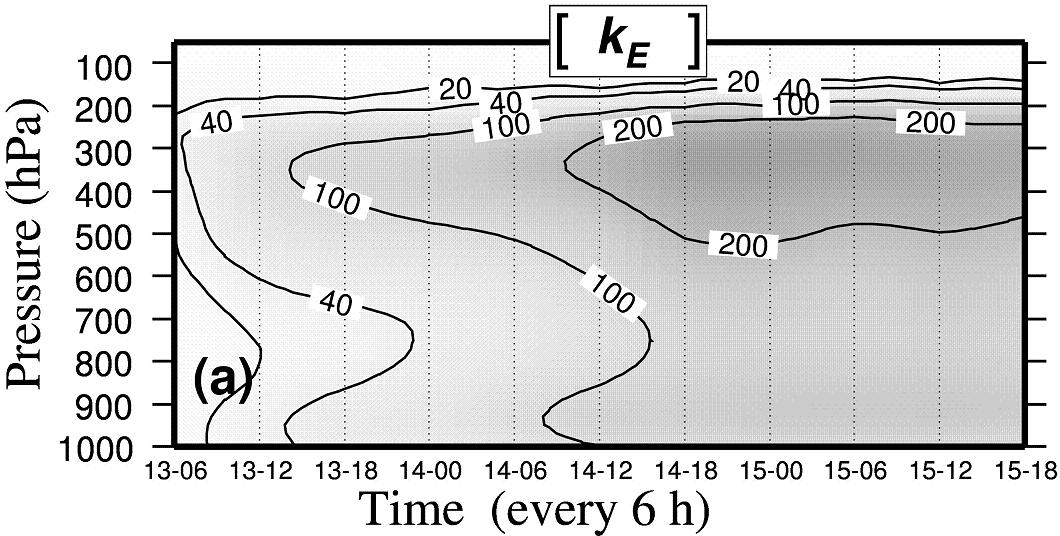}
\includegraphics[width=0.49\linewidth,angle=0,clip=true]{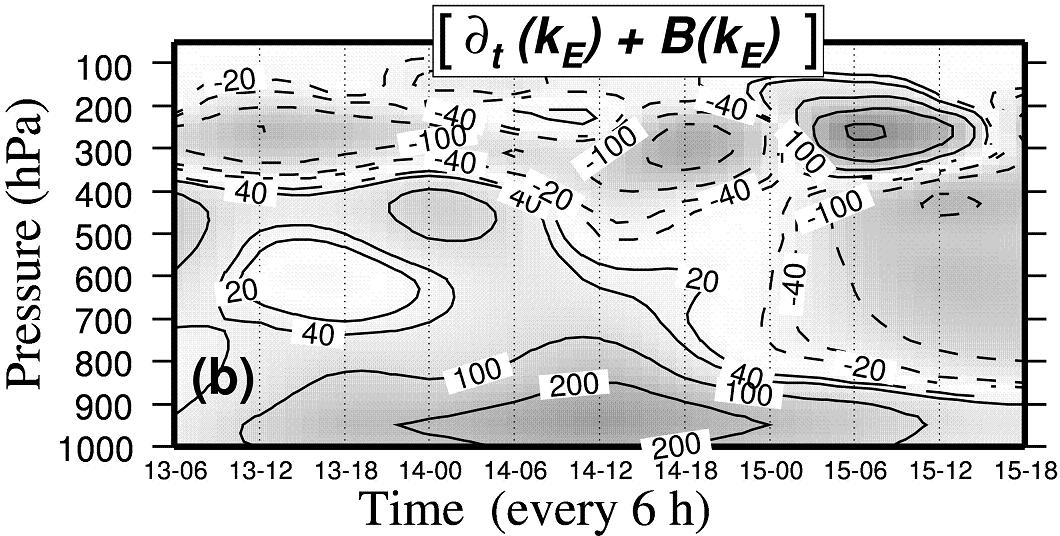}\\
\includegraphics[width=0.49\linewidth,angle=0,clip=true]{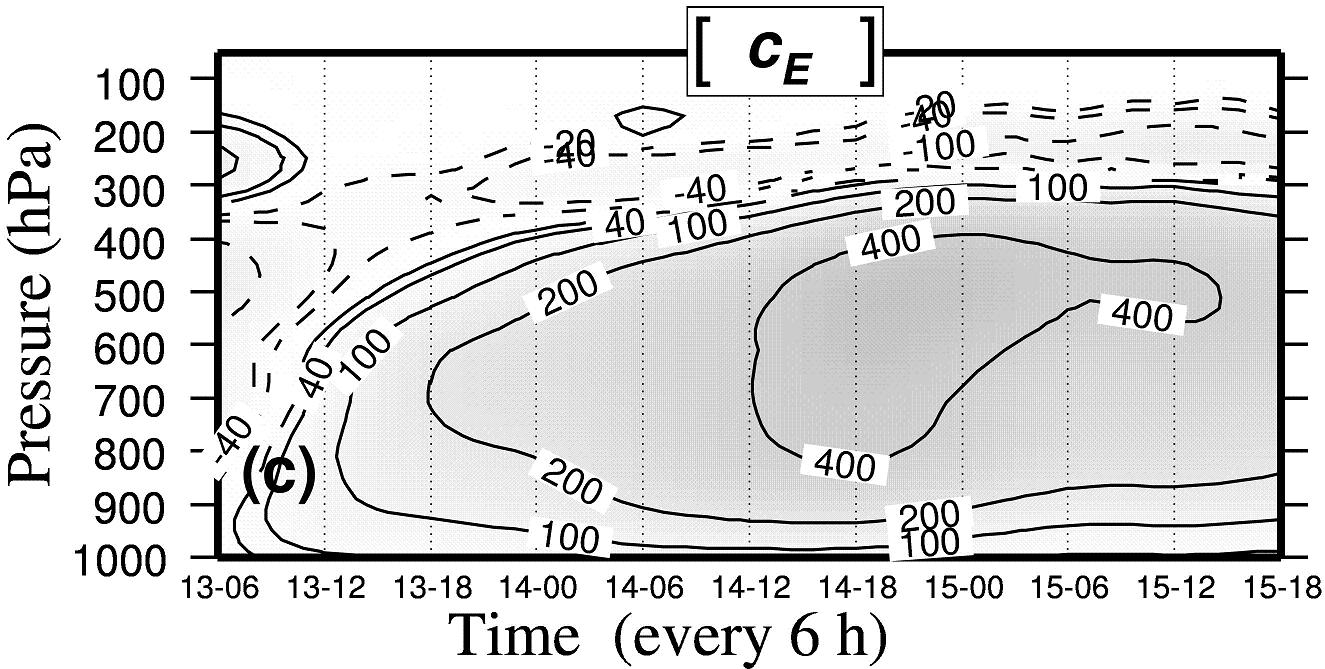}
\includegraphics[width=0.49\linewidth,angle=0,clip=true]{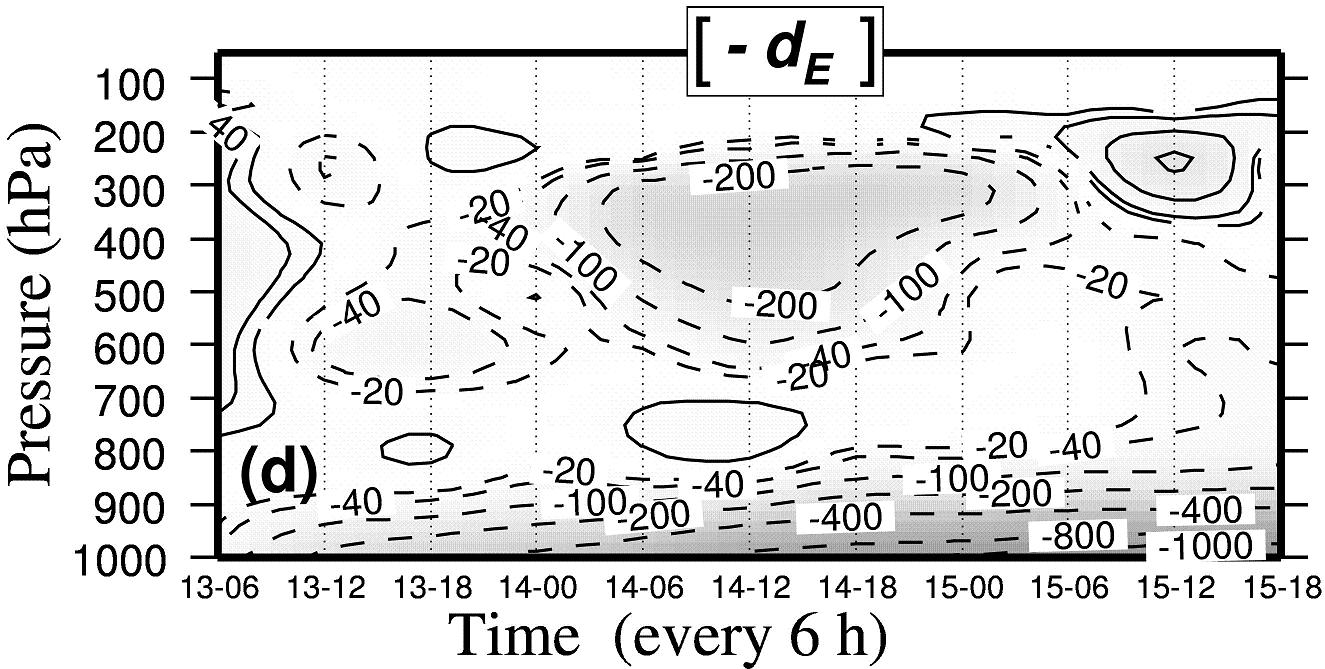}\\
\includegraphics[width=0.49\linewidth,angle=0,clip=true]{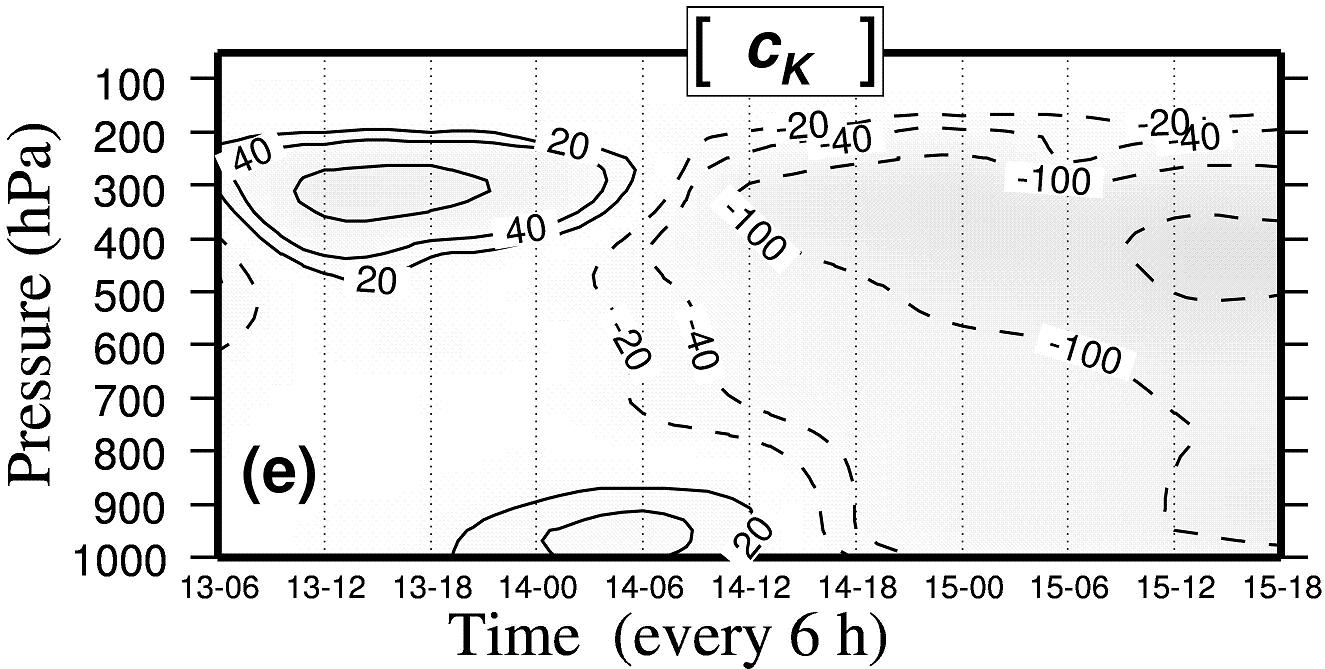}
\includegraphics[width=0.49\linewidth,angle=0,clip=true]{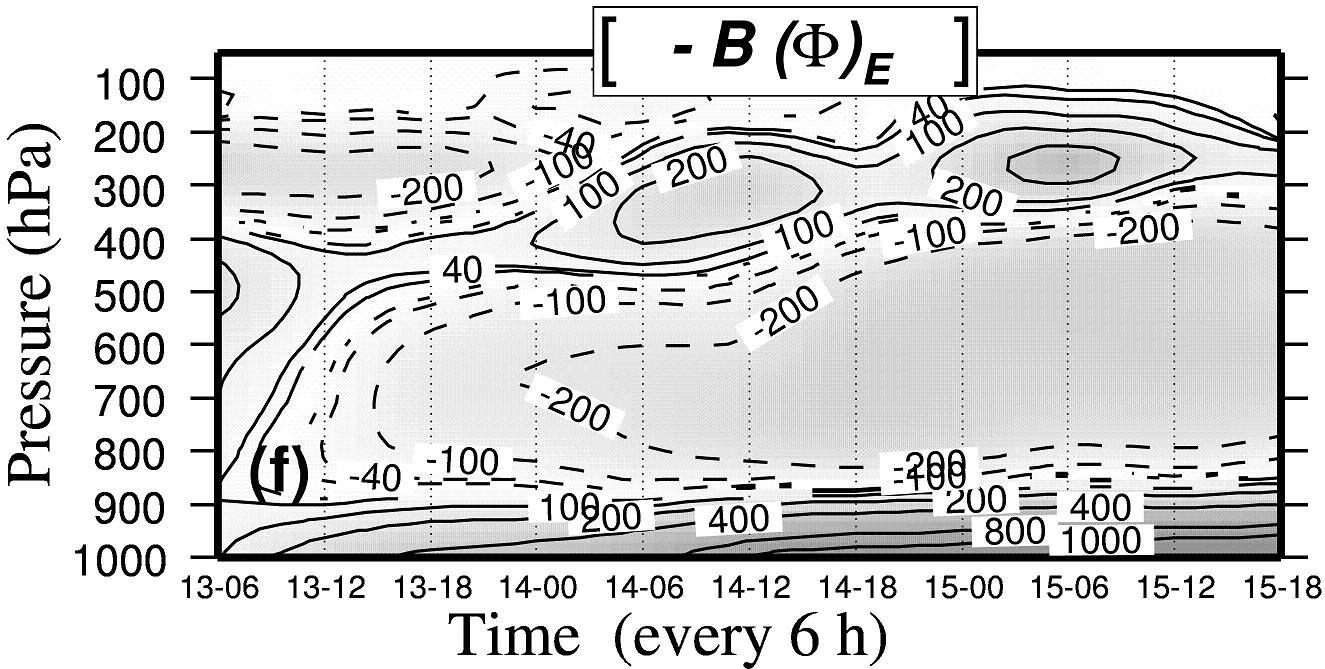}\\
\includegraphics[width=0.49\linewidth,angle=0,clip=true]{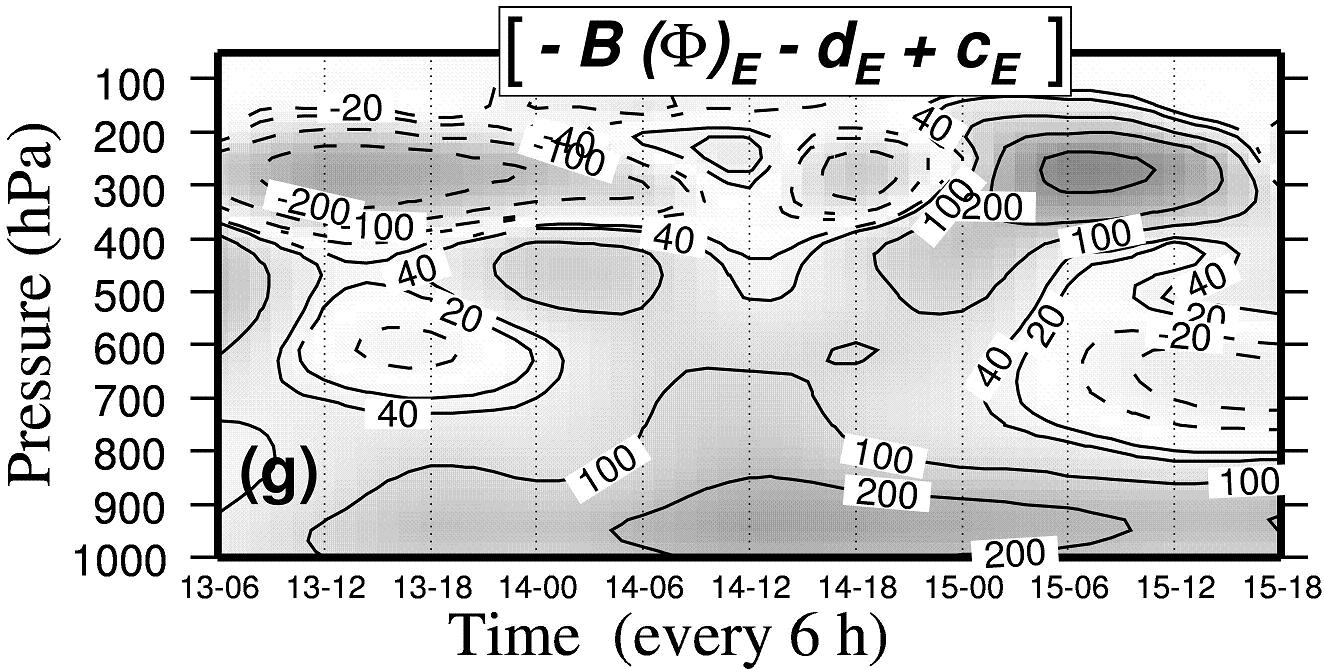}
\includegraphics[width=0.49\linewidth,angle=0,clip=true]{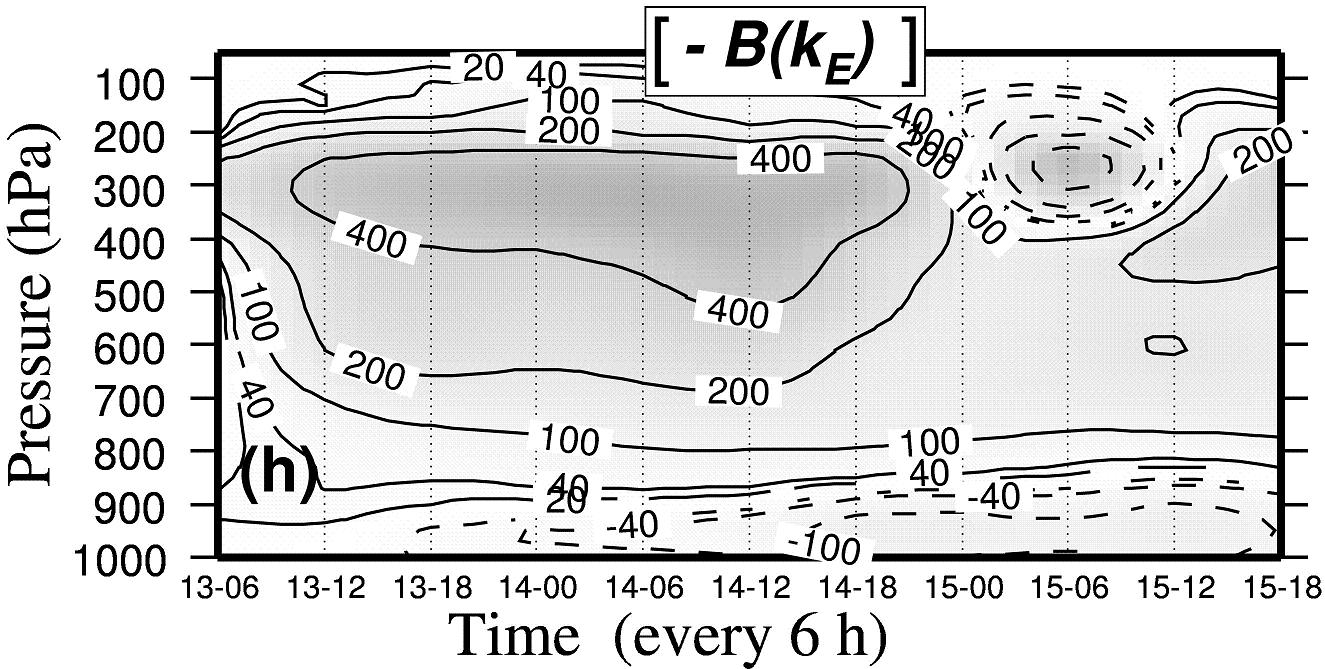}
\vspace*{-2mm}
\caption{\it \small
Time--Pressure diagrams for the IOP15 of FASTEX 
(13--16 February 1997). Data are available every 
$6$~h and results have been omitted for $1300$ and $1600$ UTC,
because only centred schemes are considered and both initial and
final dates cannot be computed. All diagrams can be compared with
Figs.~{\ref{FigEXP20}}, {\ref{FigEXP21a}} and {\ref{FigEXP21b}}.
for the other diagrams (b) to (h). 
The annotated isopleths are $\pm 20$, $\pm 40$, $\pm 100$,
$\pm 200$, $\pm 400$, $\pm 800$, $\pm 1000$ and $\pm 1500$.
The equation for $\overline{k_E}$
corresponds to (b)$\;=\:$(c)$\:+\:$(e)$\:+\:$(f)$\:+\:$(d)
or equivalently to (b)$\;=\:$(e)$\:+\:$(g).
Units are J~{kg}${}^{-1}$ for (a) 
The component $\overline{k_E}$. 
Units are $10^{-5}$~W~{kg}${}^{-1}$ for (b)--(g).
(b) The total budget 
$\overline{ {\partial}_t ( k_E ) } + \overline{ B( k_{E} )}$
(c) The baroclinic conversion $\overline{ c_E }$.
(d) The dissipation term $- \overline{d_E}$.
(e) The barotropic conversion $\overline{ c_K }$.
(f) The conversion term $- \overline{{B( \phi )}_E}$.
(g) The non-barotropic term 
$- \overline{{B( \phi )}_E} - \overline{d_E} + \overline{c_E}$.
(h) The boundary term $- \overline{ B( k_{E} )}$.
\label{FigFASTEX}}
\end{figure}

\clearpage

A three-day forecast has been simulated with an old operational version 
of the French Arpege model (triangular truncation T$149$, stretching 
factor of $3.5$, $27$ hybrid vertical levels). The operational suite
of analyses had used some non-conventional observations and the
quality of the forecast was good. The modified time scheme
(\ref{eq:bilanINT}) has been used with moving limited-area diagnostic 
domains following the storm along its trajectory. This method is close 
to the quasi-Lagragian method described in Michaelides {\it et al.}, 1999.
However, in this paper, the scheme (\ref{eq:bilanINT}) is evaluated at a 
given diagnostic time $t_0$ with the same limited-area for
$ t_{(-)}$, $t_0$ and $t_{(+)}$. And when passing from $t_0$ to $t_0+\Delta t$,
the size and location of the common diagnostic domains change at the same time 
for the new $ t_{(-)}+\Delta t $, $t_0+\Delta t $ and $t_{(+)}+\Delta t $
diagnostic times in the scheme (\ref{eq:bilanINT}).

Comparisons of diagrams between Figs.~(\ref{FigFASTEX})~(a) and 
(\ref{FigEXP21a})~(a) show that the idealized diabatic experiment EXP-HV 
can reproduce some of
energetic features observed in IOP15 of FASTEX. In the two cases, the eddy 
component $\overline{k_E}$ is a maximum close to the jet and in the boundary 
layer at a level above the surface. The main differences concern the values
of $\overline{k_E}$ which are enlarged, going from $30$-$50$ units
for EXP-HV to $200$-$300$ units for FASTEX. Another difference is that 
the maximum close to the surface occurs $12$~h later than for the jet
in the FASTEX experiment. The contrary is observed in EXP-HV.

Patterns of total budget in Figs.~(\ref{FigFASTEX})~(b) 
and close to the surface are similar, 
with a maximum of development of $\overline{k_E}$ on $14-12$~h 
and at level $950$~hPa for FASTEX. However, total budgets close to 
the jet are very different. The explanation is given by 
Fig.~(\ref{FigFASTEX})~(h) where the boundary 
flux is equivalent and opposite to the local 
tendency, leading to weak values of the sum 
$\overline{ {\partial}_t ( k_E ) } 
+ \overline{ B( k_{E} )}$ depicted on 
Fig.~(\ref{FigFASTEX})~(b). As a consequence, 
advection processes seem to control energetics of the jet in
FASTEX, whereas local developments appear to be the 
prevailing sources of $\overline{k_E}$ in the boundary layer.
Such advection processes for the jet could not appear in EXP-HV 
because of the eight waves surround the earth. Even for a small
limited area, the incoming and outgoing energy is equal for EXP-HV.

Patterns for the baroclinic and barotropic conversions, 
the dissipation term and the conversion term with potential 
energy are to a large extent similar for idealized EXP-HV and 
for the real FASTEX experiments 
(see the Figs.~(\ref{FigFASTEX})~(c)--(f)).
In particular, the dissipation is generally weak above the
boundary layer, though with surprising negative large values
close to $400$~hPa and centred on day $14$ at $12$~h.

The non-barotropic term in 
Fig.~(\ref{FigFASTEX})~(g) can be compared with 
{\ref{FigEXP21b}}~(b). As for the total budget in 
Fig.~(\ref{FigFASTEX})~(b),
there are large differences for the jet region of
FASTEX, where increase in $\overline{k_E}$ is
mainly due to advection processes. For FASTEX as for
EXP-HV, the ageostrophic conversion,
equivalent to (g)--(d)
in Fig.~(\ref{FigFASTEX}), seems to be the relevant 
term (not shown) to explain the forcing coming 
from $\overline{a_E}$, in place of the usual
baroclinic conversion $\overline{c_E}$.

      \subsection{The available-enthalpy cycles for 
        IOP15.} 
      \label{subsection_6.2}

\begin{figure}[t]
\centering
\includegraphics[width=0.49\linewidth,angle=0,clip=true]{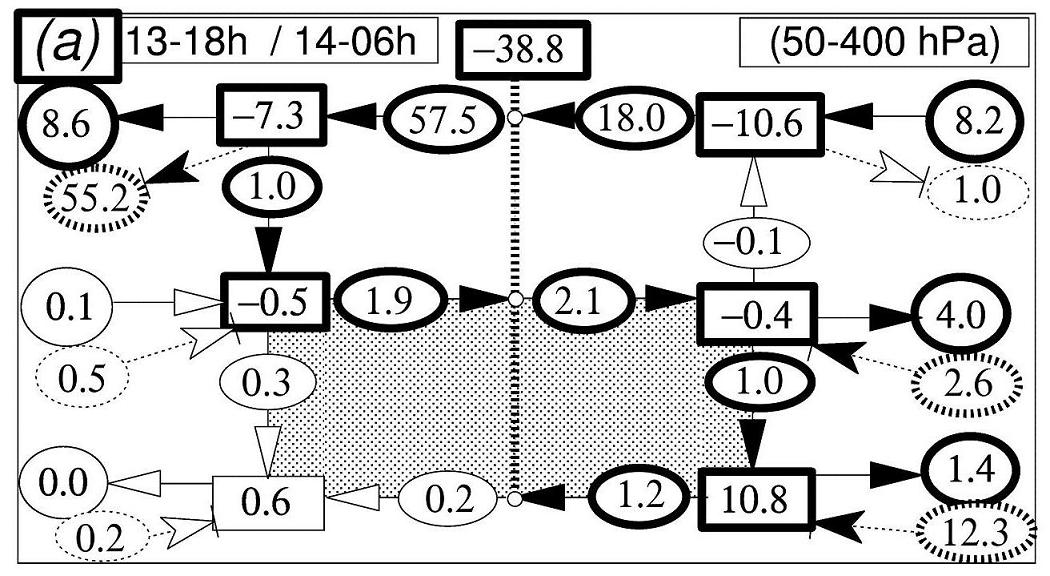}
\includegraphics[width=0.49\linewidth,angle=0,clip=true]{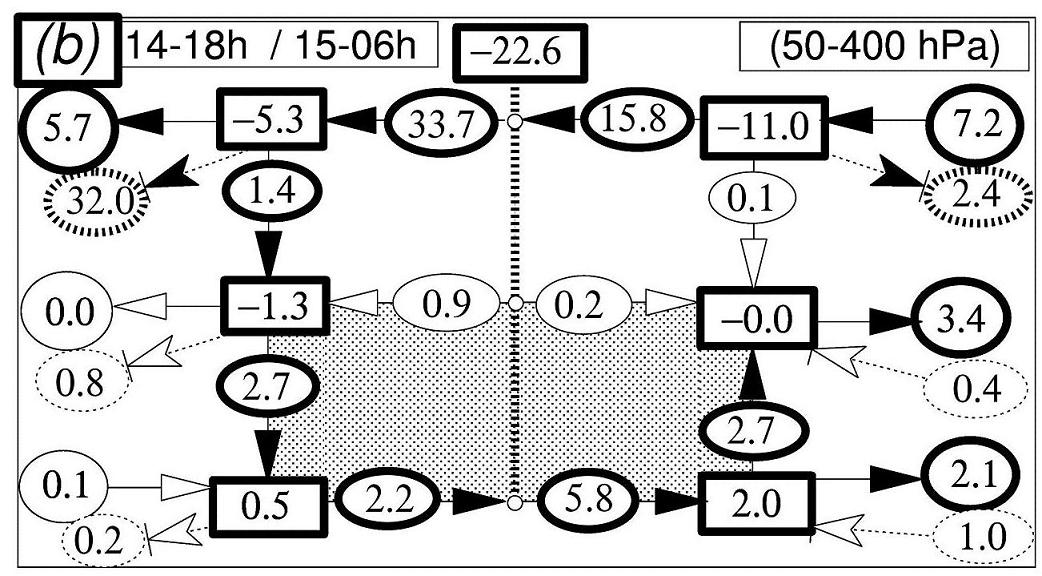}\\
\includegraphics[width=0.49\linewidth,angle=0,clip=true]{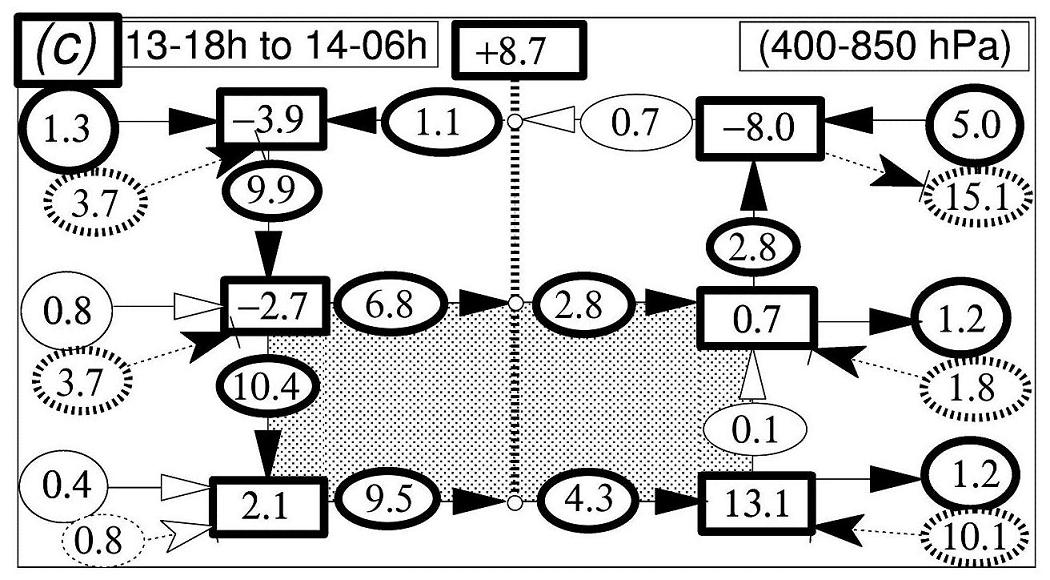}
\includegraphics[width=0.49\linewidth,angle=0,clip=true]{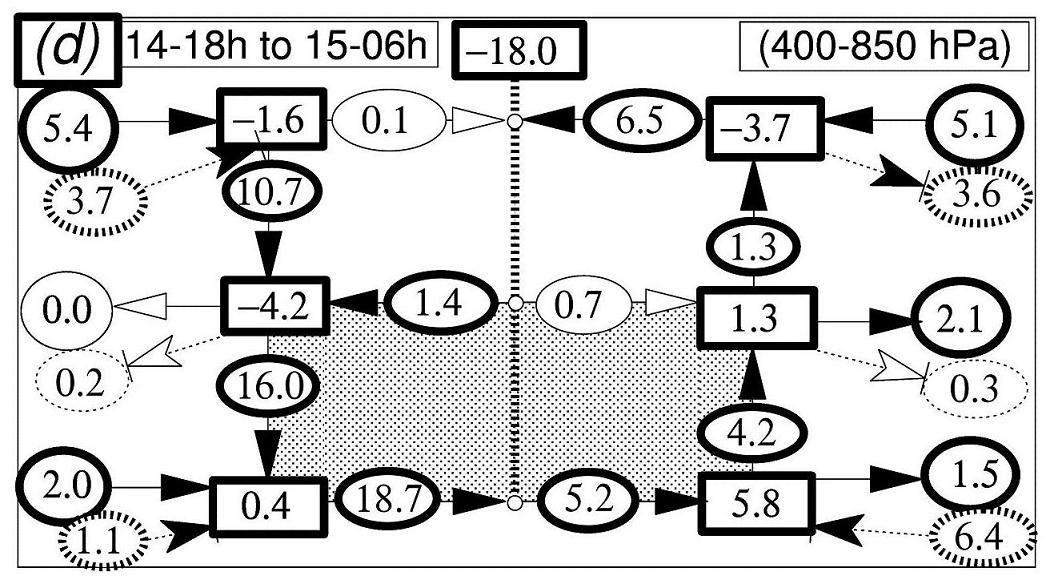}\\
\includegraphics[width=0.49\linewidth,angle=0,clip=true]{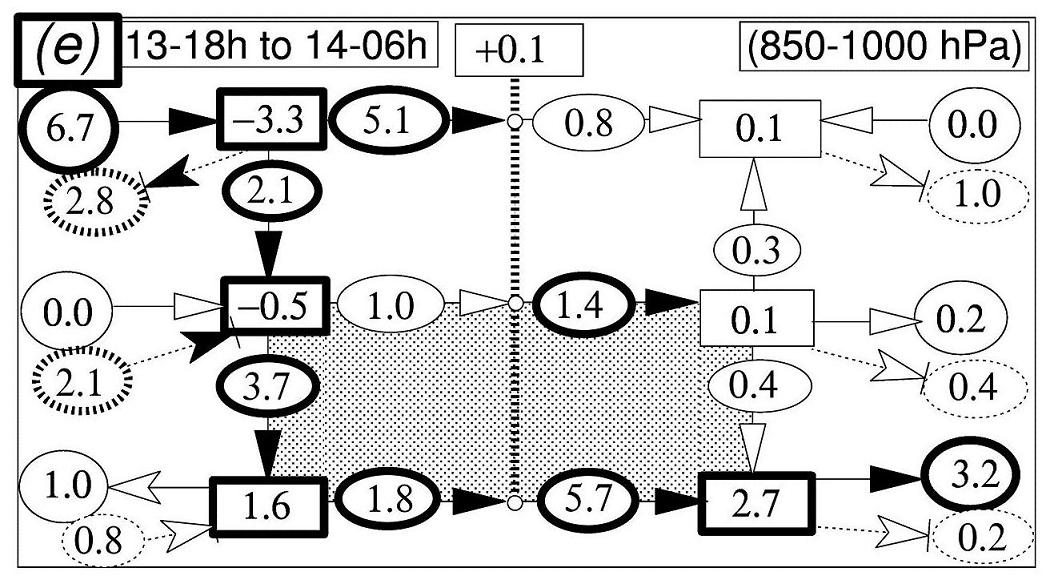}
\includegraphics[width=0.49\linewidth,angle=0,clip=true]{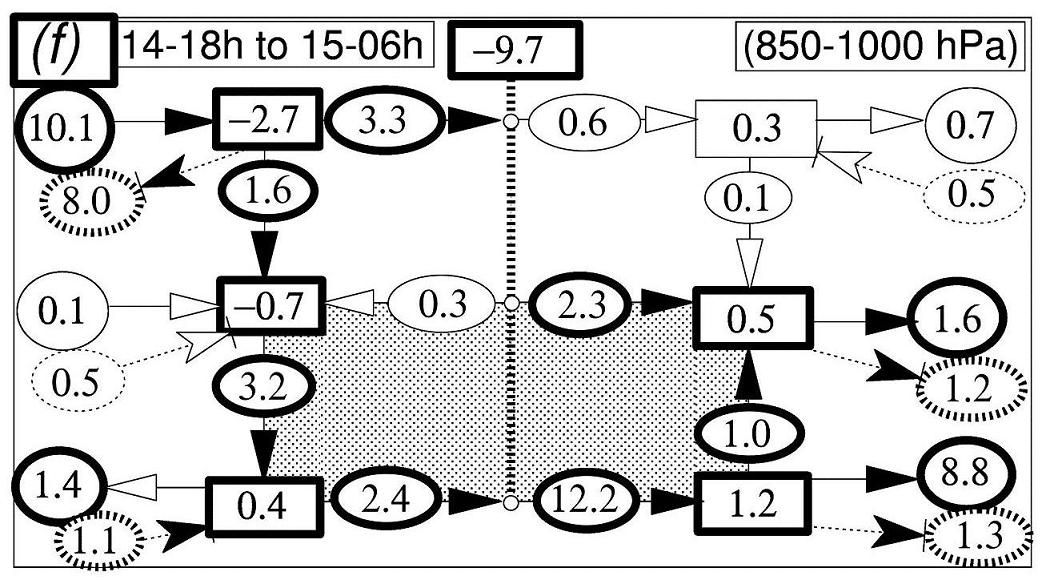}
\vspace*{-2mm}
\caption{\it \small 
The available-enthalpy cycles for the IOP15 of FASTEX.
The boxes, circles and arrows correspond to the cycle
(\ref{eq:cycle2new}) and to Fig.~\ref{FigAHNEWCYCLEb}.
Left column: the first growing stage from 
the 18 UTC 13 February to the 06 UTC 
14 February.
Right column: the last mature and decaying stages 
from 18  UTC 14 February to the 06 UTC
15 February. 
Units are W~{m}${}^{-2}$.
(a) and (b) Upper-troposphere and stratosphere region 
     ($50$ to $400$~hPa).
(c) and (d)  Middle-troposphere region ($400$ to $850$~hPa).
(e) and (f) Lower-troposphere and boundary-layer region 
      ($850$ to $1000$~hPa).
\label{FigFASTEXcycles}}
\end{figure}

The available-enthalpy cycles presented in Fig.~\ref{FigFASTEXcycles}
correspond to the storm investigated during the IOP15 of FASTEX and 
already described in the previous section (see Fig.~\ref{FigFASTEX}).
The cycles are computed for three 
vertical layers (upper, middle and lower troposphere) 
and for the two growing and decaying 
stages of development of the storm. The vertical integral of
any term $X$ is equal to the sum of $X dp/g$, where
$g$ is the acceleration due to gravity. 
Units are W~{m}${}^{-2}$.
The contributions to the global budget of upper, 
middle and lower pressure layers are equal to $35$, 
$45$ and $15$\%, respectively, according to the differences 
in pressure for the layers.

The main objective of this section is to determine how far
the ageostrophic conversions $\overline{(c_{ag})_X}$ 
acting on $\overline{k_X}$ can differ from the baroclinic 
conversions $\overline{c_X}$ acting on $\overline{a_X}$,
for the subscripts $X$ = ($S$, $Z$, $E$). Clearly,
 Figs.~\ref{FigFASTEXcycles}~(a)--(e) show 
that the differences between the two kinds
of conversion terms can be large, for all layers and for 
the two stages of the storm.

An important example is the low-level growing stage cycle
shows in Fig.~\ref{FigFASTEXcycles}~(e). It corresponds to an 
ageostrophic conversion of $+5.7$ units and to a 
baroclinic conversion of only $+1.8$ units. The difference of
$-3.9$ units for ($\overline{a_E}$, $\overline{k_E}$)
represents the eddy contribution for
$\overline{B{(\phi)}} + \overline{B(a_p)}$, with the
contributions of $-0.4$ units for ($\overline{a_Z}$, 
$\overline{k_Z}$) and $+4.3$ units for 
($\overline{a_S}$, $\overline{k_S}$),
leading to the observed small total change of 
$+0.1$ units.

This low-level growing-stage cycle corresponds to 
typical features of baroclinic developments for midlatitude 
storms. The generation term $\overline{g_S} =+6.7$
and the conversion term $\overline{c_S}=+5.1$
are associated with a release of $\overline{a_S}$ reaching 
$-3.3$ units, with the sequence of positive conversion
terms $\overline{c_{AS}}=+2.1$, $\overline{c_{A}}=+3.7$,
$\overline{c_{E}}=+1.8$ and $\overline{(c_{ag})_E}=+5.7$
units. The budget of $\overline{k_E}$ is closed
by a barotropic instability $\overline{c_{K}}=+0.4$, with
a growing of $+2.7$ units for $\overline{k_E}$ and
a realistic dissipation term $\overline{d_{E}}=3.2$ 
units.

The main difference between the low-level growing 
stage in Fig.~\ref{FigFASTEXcycles}~(e) and the 
mature or decaying stage in Fig.~\ref{FigFASTEXcycles}~(f)
is the expected change of sign for the barotropic 
conversion ($\overline{c_{K}}=-1.0$).
This barotropic stabilization
corresponds to a growth of $\overline{k_{Z}}$ 
($+0.5$ units). The dissipation terms increase
for the mature stage,
reaching $1.6$ units for $\overline{d_{Z}}$
and $8.8$ units for $\overline{d_{E}}$. The large
values for the eddy dissipation balance the large 
ageostrophic conversion ($+12.2$ units),
giving an explanation to the observed moderate growth of 
$\overline{k_{E}}$ (only $+1.2$ units).

The total budget for $\overline{B{(\phi)}} + \overline{B(a_p)}$
is small ($+0.1$ units) only for the low-level growing stage in 
Fig.~\ref{FigFASTEXcycles}~(e). Larger values 
from $-38.8$ to $+8.7$ units are observed for all other
cases(Figs.~\ref{FigFASTEXcycles}~(a) to (d), and (f)). 
It means that the budget of 
potential energy plays an important role in the 
available-enthalpy cycle by supplying (extracting) energy to (from) 
other forms of energy. There is a need to elucidate 
the uncertain involved processes.

The conversion terms $\overline{c_{Z}}$ and 
$\overline{(c_{ag})_Z}$ have opposite signs 
for the three decaying stage cases 
(Figs.~\ref{FigFASTEXcycles}~(b), (d) and (f)), 
whereas they have the same signs for the 
three growing-stage cases
 (Figs.~\ref{FigFASTEXcycles}~(a), (c) and (e)).
Other comparisons between the eddy ageostrophic and
baroclinic conversions show that $\overline{(c_{ag})_E}$ 
is the governing term for the upper-level and for the 
lower-level cases (Figs.~\ref{FigFASTEXcycles}~(a), 
(b), (e) and (f)). The governing term
is $\overline{c_{E}}$ for the mid-troposphere cases
(Figs.~\ref{FigFASTEXcycles}~(c) and (d)).

These comparisons can be understood by analysing the vertical and time
distributions for $\overline{c_{E}}$ and $- \overline{{B( \phi )}_E}$, 
as depicted in Fig.~\ref{FigFASTEX}~(c) and (f), respectively. 
According to (\ref{eq:cagterms3}), the ageostrophic conversion is 
the sum of the two terms: $\overline{c_{E}} - \overline{{B( \phi )}_E}
= \overline{(c_{ag})_E}$. For the mid-tropospheric region
they have opposite sign, leading to the observed small values of 
$\overline{(c_{ag})_E}$. 
For the low-level troposphere, the two terms are positive, 
leading to the large positive values for $\overline{(c_{ag})_E}$. 
For the upper-levels region the sign of $- \overline{{B( \phi )}_E}$ 
is somewhat irregular close to the jet. The observed increase 
of $\overline{(c_{ag})_E}$ is less easy to explain.

\section{\Large \underline{Conclusions}.} 
 \label{section_7}

The aim of this paper was to investigate the local energetics of
idealized simulations of adiabatic and diabatic versions
of baroclinic waves. The final diagnostic tool is the
fully symmetric limited area available enthalpy cycle 
with $A3+K3+\phi$ components, defined by (\ref{eq:cycle2new}) 
and Fig.~\ref{FigAHNEWCYCLEb}. 

As stated in McIntyre (1980) and Plumb (1983), 
the transformed Eulerian-mean systems or 
the generalized Eliassen-Palm fluxes
can lead to better ways of analysing wave and
mean-flow interactions.

Furthermore, the partition of Lorenz into
{\em zonal mean\/} and {\em eddy \/} components of
the flow could be improved in many
ways, for instance by following variational processes
as suggested by Van Mieghem (1956), Plumb (1983) 
and Kucharski (1997). 
An example of the use of a flow-dependent
reference state is given in Kucharski 
and Thorpe (2000a), where {\em rotated zonal means\/}
are defined. But in this two-part paper,
the conventional partition of Lorenz has been
considered and the same choice has been made
in the most recent papers 
of Michaelides et al. (1999), Kucharski and Thorpe 
(2000b, 2001) or Mishra and Rao (2001).

As for the non-uniqueness of energy cycles 
widely discussed in Johnson and Downey (1982) and
in Plumb (1983), the definition of the
limited-area available enthalpy cycle 
(\ref{eq:cycle2new}) suffers from the same
general problem of uncertainty in
the interpretations of conversion terms.
It is, for instance, always possible from a mathematical
point of view to cancel out any branch of a cycle,
by adding a common term to each part of a closed loop. 
The simplest example is
a triad of energy components ($e_1$, $e_2$, $e_3$) with
the conversions $C_{(1,2)}$, $C_{(2,3)}$ and $C_{(3,1)}$ acting
between them. If the quantity $-C_{(3,1)}$ is added to the
three branches, the modified cycle is still valid and
both the energy and net tendencies are the same
before and after the modification. The impact on the
diagram for the triad-cycle would be of a
cancellation of the direct conversion between $e_1$ and $e_3$,
the other conversions becoming $C_{(1,2)}-C_{(3,1)}$ 
and $C_{(2,3)}-C_{(3,1)}$.

It is thus important to verify that the
physical interpretations of the modified
conversion and flux terms are still valid
in the available-enthalpy cycle (\ref{eq:cycle2new}), 
as stated in McIntyre (1980), Johnson and Downey (1982),
and Plumb (1983).

Firstly, it has been verified that 
there are no approximations and no missing terms, 
by showing that observed dissipation and generation residuals 
are small in case of adiabatic simulations in both global 
and local cases.
Furthermore, there is a real gain in physical basis in the 
definition of the new available enthalpy cycle. 
In (\ref{eq:cycle2new}) and Fig.~\ref{FigAHNEWCYCLEb},
Lorenz's internal cycle has been modified in order to
highlight the ageostrophic conversion
$-{\bf U}_h \: . \: {\bf \nabla}_{\!p} \:(\phi)$.
It acts as a forcing directly applied to the kinetic-energy 
components, in place of the usual baroclinic
conversion $- R \: \omega \: T / {p}$.

The potential-energy component plays the role of vertical 
redistribution of energy as described in OS95, by transforming 
baroclinic into ageostrophic conversions, via unknown 
processes to be discovered.
The use of ageostrophic conversion terms in the available-enthalpy 
cycle is the main modification brought to the study of Pearce (1978). 
It appears to be in agreement with the results obtained 
with the numerical simulation of the IOP15 of FASTEX, 
as presented in section 6.2. The magnitudes of the baroclinic and 
ageostrophic conversions are indeed very different. Even their 
sign can change. When the potential-energy component is included,
the values of the generation and dissipation terms
computed as residuals of the cycle (\ref{eq:cycle2new})
turn out to be realistic.

Other comparisons of results from the idealized case EXP-HV and from
the IOP15 real case of FASTEX show that surface patterns are
almost the same, although the jet energetics are very
different. It could be worthwhile to continue this study on
IOP15 of FASTEX, with special attention paid to the impact of 
surface energy fluxes on the atmospheric boundary-layer front, 
as already stated in Giordani and Planton (2000). 

The limited area available enthalpy cycle (\ref{eq:cycle2new})
must be particularly suitable for studying the energetics of isolated
structures like frontal waves, for which FASTEX has been
organized. To do so, the potential-vorticity inversion mechanism could be used
to make simulations including and not including some localized 
small patterns, in order to understand their dynamical consequences.

Other idealized cases could be investigated in the future,
with possible enhanced horizontal or vertical resolutions. 
It could also be worthwhile to define more 
realistic simulations, by including humidity
and other diabatic processes. Results from Marquet (1993) 
could serve as a starting point for defining a moist
local available-enthalpy cycle.

\vspace{5mm}
\noindent{\Large\bf \underline{Acknowledgements}.}
\vspace{2mm}

The author is most grateful to S. Malardel, Ph. Arbogast and 
C. Freydier for their support and useful discussions about 
applications of available-enthalpy energetics to idealized 
simulations and for the preparation of the basic state with 
the most unstable mode. I also thank R. Clark and the
two referees who suggested many clarifications and modifications 
to the manuscript.

\vspace{6mm}
\noindent
{\Large\bf Appendix A. Computations of the pseudo-geostrophic wind.}
             \label{appendix_A}
\renewcommand{\theequation}{A.\arabic{equation}}
  \renewcommand{\thefigure}{A.\arabic{figure}}
   \renewcommand{\thetable}{A.\arabic{table}}
      \setcounter{equation}{0}
        \setcounter{figure}{0}
         \setcounter{table}{0}
\vspace*{1mm}
\hrule

\vspace*{2mm}
The stationary jet depicted in Fig.~\ref{FigUTZON}~(a) is 
defined by the pseudo-geostrophic wind 
($u_g^{\ast}, v_g^{\ast}=0$). It is obtained from
(29) of Part~I, together with the hypotheses 
that all the zonal and meridian tendencies are zero 
and that all the variables ($T, \phi, u_g^{\ast}$) are zonally 
symmetric, with a constant surface pressure and without friction.
The result is given by (\ref{defevolUv}) .
\vspace{-0.15cm}
\begin{eqnarray}
    \frac{d v_g^{\ast}}{dt}     & = & 
                  - \: [{\bf \nabla}_{\!p}(\phi)]{}_{y}
               \; - \; f^{\ast} u_g^{\ast}  
               \; \;  = \; \; 0   \: . \label{defevolUv}  
\end{eqnarray}

If the usual geostrophic wind is denoted by 
$u_g =$ $ - [{\bf \nabla}_{\!p}(\phi)]{}_{y} / f$,
the pseudo-\-geos\-tro\-phic version is obtained
by solving an equation of the second degree,
derived by inserting 
$ f^{\ast} = f + u_g^{\ast} \tan(\varphi) / {\cal R}$,
into (\ref{defevolUv}), to give
\vspace{-0.15cm}
\begin{eqnarray}
(u_g^{\ast}){}^2 \; \frac{\tan(\varphi)}{\cal R}
\: + \:  f \: u_g^{\ast} 
\: + \: [{\bf \nabla}_{\!p}(\phi)]{}_{y} 
\: = \: 0
\hspace{4mm} 
\Longrightarrow
\hspace{2mm} 
u_g^{\ast}   \: = \: 
{\cal V}_T \: \left\{ \sqrt{ 1+2 u_g / {\cal V}_T } \: - 1 \: \right\}
. \label{AppB:defJETug}    
\end{eqnarray}
The term ${\cal V}_T = \Omega \: {\cal R} \cos(\varphi)$ 
is the velocity due to the rotation of
the earth with $\Omega \: {\cal R} \approx 464$~m~s${}^{-1}$.
The limit of (\ref{AppB:defJETug}) 
for small values of $u_g / {\cal V}_T$ 
and with $\sqrt{1+2X} -1 \approx X$ for 
small $X$ is the usual geostrophic wind $u_g$.
This approximation is valid in the tropical
region where $\cos(\varphi) \approx 1$ and
$u_g \ll 464$~m~s${}^{-1}$. But
${\cal V}_T$ decreases with increasing latitude where
$\cos(\varphi) \approx 0$, leading to possible large
differences between $u_g^{\ast}$ and $u_g$.
A significant departure from classic 
geostrophic conditions (up to $4$~m~s${}^{-1}$) is also 
obtained for the mid-latitude jet. 
The obvious singularities for 
$\varphi =0$ and $\varphi=\pm\pi/2$ in
(\ref{AppB:defJETug}) have no practical 
impact for a Gaussian grid where there is 
no point located at the poles or at the exact equator.

There is a need to take into account the formulation 
(\ref{AppB:defJETug}) to avoid significant time oscillations 
of the jet due to imbalanced wind, as clearly observed for
some spectral coefficients in a first attempt to
construct the zonal jet. The method of computing
``$[{\bf \nabla}_{\!p}(\phi)]{}_{y}$'' is also important. 
The method used in Arpege is a spectral
computation of the gradient of temperature for
``$\cos(\varphi){\partial}_{\varphi}(T)$'',
with a transformation onto the Gaussian grid.
The meridian component ``$({\cal R}){}^{-1}{\partial}_{\varphi}(\phi)$''
is finally obtained by computing the vertical integral 
of the hydrostatic equation and by a division by 
``${\cal R} \cos(\varphi)$'' on the Gaussian grid.



\vspace{5mm}
\noindent{\Large\bf \underline{References}.}
\vspace{2mm}

\noindent{$\bullet$ Brennan,~F.~E. and Vincent,~D.~G.} {1980}.
{Zonal and eddy components of the synoptic-scale energy
budget during intensification of hurricane Carmen (1974).}
{\it Mon. Weather Rev.\/}
{\bf 108,}
p.954--965.

\noindent{$\bullet$ Courtier,~J.~A., 
Freydier,~C., Geleyn,~J.F., 
Rabier,~F. and Rochas,~M.} {1991}.
{The Arp\`ege project at M\'et\'eo-France.
{\it ECMWF Seminar Proceedings.\/},
Reading, 9-13 Sept. 1991, Volume II,
p.193--231.}

\noindent{$\bullet$ Giordani,~H., Planton,~S.} {2000}.
{Modeling and analysis of ageostrophic circulation
over the Acores oceanic front during the
Semaphore experiment.
{\it Mon. Weather Rev.\/},
{\bf 128,}
p.2270--2287.}

\noindent{$\bullet$ Hoskins,~B.J. and Simmons,~A.J.} {1975}.
{A multi-layer spectral model and the semi-implicit method.
{\it Q. J. R. Meteorol. Soc.\/},
{\bf 101,}
p.637--655.}

\noindent{$\bullet$ Johnson,~R.J. and Downey,~W.K.} {1982}.
{On the energetics of open systems.
{\it Tellus\/},
{\bf 34,}
(2),
p.458--470.}

\noindent{$\bullet$ Joly~A., Jorgensen~D., Shapiro~M.~A., 
Thorpe~A., Bessemoulin~P., Browning~K.~A., 
Cammas~J.-P., Chalon~J.-P., Clough~S.~A., Emmanuel~K.~A.,
Eymard~L., Gall~R., Hildebrand~P.~H., Langland~R.~H., Lemaitre~H.,
Lynch~P., Moore~J.~A., Persson~P.~Ola~G., Snyder~C. and Wakimoto~R.~M.}
 {1997}.
{The Fronts and Atlantic Storm-Track Experiment (FASTEX):
scientific objectives and experimental design.
{\it Bull. Amer. Meteorol. Soc.\/}, 
{\bf 78,} (9), 
p.1917--1940.}

\noindent{$\bullet$ Kucharski,~F.} {1997}.
{On the concept of exergy and available potential energy.
{\it Q. J. R. Meteorol. Soc.\/}
{\bf 123},
p.2141--2156.} 

\noindent{$\bullet$ Kucharski,~F. and Thorpe,~A.,~J.} {2000a}.
{Upper-level barotropic growth as a precursor to 
cyclogenesis during FASTEX.
{\it Q. J. R. Meteorol. Soc.\/},
{\bf 126,}
p.3219--3232.}

\noindent{$\bullet$ Kucharski,~F. and Thorpe,~A.,~J.} {2000b}.
{Local energetics of an idealized baroclinic 
wave using extended exergy.
{\it J. Atmos. Sci.\/},
{\bf 57,}
p.3272--3284.}

\noindent{$\bullet$ Kucharski,~F. and Thorpe,~A.,~J.} {2001}.
{The influence of transient upper-level barotropic growth 
on the development of baroclinic waves.
{\it Q. J. R. Meteorol. Soc.\/},
{\bf 127,}
p.835--844.}

\noindent{$\bullet$ Lorenz,~E.~N.} {1955}.
{Available potential energy and the 
 maintenance of the general circulation.
{\it Tellus.\/}
{\bf 7,} (2),
p.157--167.}

\noindent{$\bullet$ Louis,~J.F.} {1979}.
{A parametric model of the vertical eddy fluxes
in the atmosphere.
{\it Boundary-Layer Meteorol.\/},
{\bf 17,}
p.187--202.}

\noindent{$\bullet$ Louis,~J.F., Tiedke,~M. and Geleyn,~J.F.} {1981}.
{A short history of the operational PBL
parameterization at ECMWF.
Pp. 59--79 in {\it  Proceedings of
ECMWF Workshop on Planetary
Boundary Layer Parameterization.\/},
25-27 November 1981, Reading, UK.}

\noindent{$\bullet$ McIntyre,~M.E.} {1980}.
{An introduction to the generalized
Lagrangian-mean description of wave,
mean-flow interation.
{\it Pure Appl. Geophys.\/},
{\bf 118,}
p.152--176.}

\noindent{$\bullet$ Marquet~P.} {1991}.
{On the concept of exergy and available
enthalpy: application to atmospheric energetics.
{\it Q. J. R. Meteorol. Soc.}
{\bf 117}:
p.449--475.
\url{http://arxiv.org/abs/1402.4610}.
{\tt arXiv:1402.4610 [ao-ph]}}

\noindent{$\bullet$ Marquet~P.} {1993}.
{Exergy in meteorology: definition and properties
of moist available enthalpy.
{\it Q. J. R. Meteorol. Soc.\/},
{\bf 119,}
p.567--590.}

\noindent{$\bullet$ Marquet,~P.} {1994}.
{\it Applications du concept d'exergie \`a
l'\'energ\'etique de l'at\-mosph\`ere. Les
notions d'enthalpie utilisables s\`eche
et humide \/}.
PhD-thesis of the Paul Sabatier University.
Toulouse, France.

\noindent{$\bullet$ Marquet,~P.} {1995}.
{On the concept of pseudo-energy of T. G. Shepherd.
{\it Q. J. R. Meteorol. Soc.}
{\bf 121}:
p.455--459.
\url{http://arxiv.org/abs/1402.5637}.
{\tt arXiv:1402.5637 [ao-ph]}}

\noindent{$\bullet$ Marquet,~P.} {2001}.
{\it The available enthalpy cycle. 
 Applications to idealized baroclinic 
 waves.\/}.
Note de centre du CNRM. Number 76.
Toulouse, France.

\noindent{$\bullet$ Marquet~P.} {2003a}.
{The available enthalpy cycle. Part~I : 
 Introduction and basic equations.
{\it Q. J. R. Meteorol. Soc.},
{\bf 129}, (593),
p.2445--2466.
\url{http://arxiv.org/abs/1403.5671}.
{\tt arXiv:1403.5671 [ao-ph]}
}

\noindent{$\bullet$ Marquet~P.} {2003b}.
{The available enthalpy cycle. Part~II : 
 Applications to idealized baroclinic waves.
{\it Q. J. R. Meteorol. Soc.},
{\bf 129}, (593),
p.2467--2494.}

\noindent{$\bullet$ Michaelides,~S.~C.} {1987}.
{Limited area energetics of Genoa cyclogenesis.
{\it Mon. Weather Rev.\/}
{\bf 115},
p.13--26.}

\noindent{$\bullet$ Michaelides,~S.~C., Prezerakos,~N.~G. 
 and Flocas,~H.,~A.} {1999}.
{Quasi-Lagrangian energetics of an intense
Mediterranean cyclone.
{\it Q. J. R. Meteorol. Soc.\/},
{\bf 125,}
p.139--168.}

\noindent{$\bullet$ Mishra,~S.,~K. and Rao,~V.,~B.} {2001}.
{The energetics of an upper tropospheric cyclonic vortex
 over north-east Brazil.
{\it Q. J. R. Meteorol. Soc.\/},
{\bf 127,}
p.2329--2351.} 

\noindent{$\bullet$ Muench,~H.~S.} {1965}.
{On the dynamics of the wintertime stratosphere 
circulation.
{\it J. Atmos. Sci.\/}
{\bf 22},
p.349-360.}

\noindent{$\bullet$ Orlanski,~I. and Sheldon,~J.~P.} {1995}.
{Stages in the energetics of baroclinic systems.
{\it Tellus\/},
{\bf 47A,}
(5),
p.605--628.}

\noindent{$\bullet$ Pearce,~R.~P.} {1978}.
{On the concept of available potential energy.
{\it Q. J. R. Meteorol. Soc.\/}
{\bf 104},
p.737--755.}

\noindent{$\bullet$  Plumb,~R.A.} {1983}.
{A new look at the energy cycle.
{\it J. Atmos. Sci.\/},
{\bf 40,}
p.1669--1688.}

\noindent{$\bullet$  Simmons,~A.J. and Hoskins,~B.J.} {1976}.
{Baroclinic instability on the sphere: normal modes
of the primitive and quasi-geostrophic equations. 
{\it J. Atmos. Sci.\/},
{\bf 33,} 
p.1454--1477.}

\noindent{$\bullet$  Simmons,~A.J. and Hoskins,~B.J.} {1978}.
{The life cycles of some nonlinear baroclinic waves. 
{\it J. Atmos. Sci.\/},
{\bf 35,} 
p.414--432.}

\noindent{$\bullet$  Thorncroft,~C.D. and Hoskins,~B.J.} {1990}.
{Frontal cyclogenesis. 
{\it J. Atmos. Sci.\/},
{\bf 47,} 
p.2317--2335.}

\noindent{$\bullet$  Van Mieghem,~J.} {1956}.
{The energy available in the atmosphere for
conversion into kinetic energy. 
{\it Beitr. Phys. Atmos.\/},
{\bf 29,} 
p.129--142.}

 \end{document}